\documentclass[aps,pra,
twocolumn,
superscriptaddress,
floatfix,
longbibliography
]{revtex4-2}

\usepackage{amsfonts}
\usepackage{amsmath}
\usepackage{layout}
\usepackage{amssymb} 
\usepackage{graphicx}
\usepackage{dcolumn}
\usepackage{bm}
\usepackage{hyperref}
\usepackage[mathlines]{lineno}
\usepackage[dvipsnames]{xcolor} 
\colorlet{BLUE}{blue}

\usepackage{tikz}
\usetikzlibrary{decorations.markings}
\usetikzlibrary{arrows,backgrounds,snakes,shapes}
\usepackage{float}
\usepackage{mathtools}
\usepackage{xcolor} 
\usepackage{lmodern} 

\tikzset{->-/.style={decoration={markings,mark=at position .7 with {\arrow{>}}},postaction={decorate}}}
\tikzset{-<-/.style={decoration={markings,mark=at position .7 with {\arrow{<}}},postaction={decorate}}}
\tikzset{>-/.style={decoration={markings,mark=at position .3 with {\arrow{>}}},postaction={decorate}}}
\tikzset{mydot/.style={circle,fill=white,draw,outer sep=0pt,inner sep=1.5pt}}
\tikzset{-</.style={decoration={markings,mark=at position .3 with {\arrow{<}}},postaction={decorate}}}
\tikzset{mydot/.style={circle,fill=white,draw,outer sep=0pt,inner sep=1.5pt}}
\tikzset{con/.style={draw=none, postaction={decoration={markings,mark=at position .5 with {\pgftransformscale{.7}\arrow{>}}},decorate},postaction={draw,densely dotted,black,decorate}}}
\tikzset{snake arrow/.style={->,decorate,decoration={snake,amplitude=.4mm,segment length=2mm,post length=1mm}}}
\tikzstyle{block} = [draw, fill=white, rectangle, 
    minimum height=0.685cm, minimum width=0.685cm, inner sep=.5pt]
\tikzstyle{bigblock} = [draw, fill=white, rectangle, 
    minimum height=0.685cm, minimum width=1.8cm, inner sep=.5pt]
\tikzstyle{sum} = [draw, fill=white, circle, node distance=1cm]
\tikzstyle{input} = [coordinate]
\tikzstyle{output} = [coordinate]
\tikzstyle{pinstyle} = [pin edge={to-,very thick,black}]

\usetikzlibrary{quantikz}
\usepackage{adjustbox}
\usepackage{physics}

\usepackage{algorithm}
\usepackage{algpseudocode}

\newcommand{\angamma}{\boldsymbol{\gamma}}
\newcommand{\anbeta}{\boldsymbol{\beta}}
\newcommand{\pauli}[2]{\sigma_{#1}^{#2}}

\begin{document}

\title{The Quantum Approximate Optimization Algorithm performance with low entanglement and high circuit depth}

\author{Rishi Sreedhar}
\affiliation{Department of Microtechnology and Nanoscience (MC2), Chalmers University of Technology, SE-412 96 G\"{o}teborg, Sweden}

\author{Pontus Vikst{\aa}l}
\affiliation{Department of Microtechnology and Nanoscience (MC2), Chalmers University of Technology, SE-412 96 G\"{o}teborg, Sweden}

\author{Marika Svensson}
\affiliation{Jeppesen, 411 03 Gothenburg, Sweden}
\affiliation{Department of Computer Science, Chalmers University of Technology, 412 96 Gothenburg, Sweden}

\author{Andreas Ask}
\affiliation{Department of Microtechnology and Nanoscience (MC2), Chalmers University of Technology, SE-412 96 G\"{o}teborg, Sweden}

\author{G{\"o}ran Johansson}
\affiliation{Department of Microtechnology and Nanoscience (MC2), Chalmers University of Technology, SE-412 96 G\"{o}teborg, Sweden}

\author{Laura Garc\'{i}a-\'{A}lvarez}
\affiliation{Department of Microtechnology and Nanoscience (MC2), Chalmers University of Technology, SE-412 96 G\"{o}teborg, Sweden}

\begin{abstract}
Variational quantum algorithms constitute one of the most widespread methods for using current noisy quantum computers. However, it is unknown if these heuristic algorithms provide any quantum-computational speedup, although we cannot simulate them classically for intermediate sizes. Since entanglement lies at the core of quantum computing power, we investigate its role in these heuristic methods for solving optimization problems.
In particular, we use matrix product states to simulate the quantum approximate optimization algorithm with reduced bond dimensions $D$, a parameter bounding the system entanglement. Moreover, we restrict the simulation further by deterministically sampling solutions. We conclude that entanglement plays a minor role in the MaxCut and Exact Cover 3 problems studied here since the simulated algorithm analysis, with up to $60$ qubits and $p=100$ algorithm layers, shows that it provides solutions for bond dimension $D \approx 10$ and depth $p \approx 30$. Additionally, we study the classical optimization loop in the approximated algorithm simulation with $12$ qubits and depth up to $p=4$ and show that the approximated optimal parameters with low entanglement approach the exact ones.
\end{abstract}

\maketitle

\section{Introduction}
\label{sec:Introduction}

The current era of quantum computing is hampered by excessive noise on available intermediate-scale hardware. Fault-tolerant quantum computers are predicted to solve certain problems faster than conventional computers, but decoherence limits harnessing the quantum mechanical properties enabling this computing power~\cite{preskill2018quantum}.
With the ongoing rapid developments of quantum computing hardware providing significant evidence that a classical computer cannot efficiently replicate quantum circuits~\cite{arute2019quantum,zhong2020quantum,zhong2021phase-programmable,wu2021strong}, recent work has focused on the quest for useful quantum algorithms to run on near-term devices. The availability of quantum processors allows us to evaluate the potential of quantum computation. However, it remains an open question whether accessible quantum algorithms can provide any advantage for practical applications.

Among current approaches for exploiting noisy quantum hardware are variational quantum algorithms (VQAs)~\cite{cerezo2021variational}. These heuristic techniques employ a classical optimization loop to update a parametrized quantum circuit designed to find the ground state of a problem Hamiltonian. 
This category includes the Variational Quantum Eigensolver (VQE) for quantum chemical calculations~\cite{peruzzo2014a-variational} and the Quantum Approximate Optimization Algorithms (QAOAs) for conventional optimization problems~\cite{farhi2001a-quantum,hadfield2019from}, both promising candidates due to their adaptability to different problems. 

The challenges encountered in numerical and analytical studies of quantum circuits have driven the improvement of classical simulation techniques for quantum computing~\cite{pednault2019leveraging,huang2020classical,pan2020contracting,gray2021hyper-optimized,pan2021simulating,medvidovic2021classical}. On the one hand, we need advanced numerical methods to reproduce quantum algorithms and benchmark their performances. On the other hand, the heuristic nature of VQAs raises questions on how they might facilitate solving optimization problems, and classical techniques help develop an intuition on what quantum properties may constitute a resource for practical advantage. Ongoing research addresses the latter question, focusing on identifying what underlying mechanisms may yield the success of quantum heuristics or restrict their effectiveness. In particular, a few studies assess how entanglement impacts the trainability and performance of VQAs. More specifically, the optimization and initialization properties of different VQAs have been studied in terms of their entanglement spectra and entangling gates' structure~\cite{wiersema2020exploring,diez-valle2021quantum,mcclean2021low-depth,chen2022how-much,dupont2022an-entanglement}.

In this work, we numerically explore the role of entanglement in the performance of QAOA applied to two canonical optimization problems: Exact Cover $3$ (EC3), and MaxCut. The standard QAOA quantum circuits are built from two parametrized unitaries, applied $p$ times sequentially. The first unitary encodes the optimization problem and may generate entangled states while the second---encompassing only single-qubit gates---cannot increase entanglement. Current research suggests that the QAOA needs a circuit depth of $p > 1$ to compete with classical algorithms for optimization problems~\cite{hastings2019classical, farhi2020the-quantum, bravyi2021classical, basso2022the-quantum}. However, the achievable depth $p$ on recent experimental runs of the algorithm is hampered by coherence times, compromising its implementation~\cite{harrigan2021quantum}. 
On the one hand, decoherence degrades entanglement~\cite{yu2004finite-time,almeida2007environment-induced} but reducing the QAOA circuit depth limits the number of entangling gates too. Then, it is natural to ask how fewer entangling operations relate to entanglement and performance.
Furthermore, previous analysis of QAOA with $p=1$ for other problem cases---the bush of implications and the Hamming weight with a spike, studied in adiabatic quantum optimization~\cite{farhi2002quantum}---reveal that entanglement can hinder the algorithm performance~\cite{mcclean2021low-depth}. 
For MaxCut problems, a recent study suggests that removing excess entanglement generated by intermediate layers of the QAOA may lead to better results~\cite{chen2022how-much}. Additionally, large-depth QAOA circuits exhibit an entanglement barrier between the initial state and the final one that complicates its classical simulation and benchmarking~\cite{dupont2022calibrating,dupont2022an-entanglement}. 
These different results motivate further investigation to determine what problem instances may benefit from entanglement in QAOA.

To analyze how different degrees of entanglement affect QAOA for EC3 and MaxCut, we use matrix product states (MPSs), a key ingredient in many tensor-network-based methods~\cite{schollwock2011the-density-matrix}. 
The effectiveness of the density matrix renormalization group (DMRG) algorithm~\cite{white1992density,white1993density-matrix} in simulating quantum spin chains yielded remarkably precise results, establishing these numerical techniques among the most powerful to tackle the challenges of quantum many-body physics and impacting the way quantum systems can be treated computationally.
This success of DMRG has later been attributed to its connection to MPSs, a class of one-dimensional quantum state representations suitable to model efficiently relevant many-body wave functions with low entanglement. Gapped one-dimensional quantum systems follow an area law for entanglement entropy~\cite{hastings2007an-area}, and any state exhibiting that behavior is provably well approximated by an MPS.
In the same way, these tensor network techniques have been used widely in quantum computation~\cite{vidal2003efficient}, with recent works focused on approximately simulating imperfect quantum computers~\cite{markov2008simulating,zhou2020what,dupont2022calibrating} and benchmarking current experiments~\cite{pednault2019leveraging,huang2020classical,pan2020contracting,gray2021hyper-optimized,pan2021simulating}.
Here, we rely on the tensor product structure of MPSs to characterize QAOA circuits, describing states in terms of varying degrees of entanglement at interfaces between different parts of the system. To be precise, we adapt the so-called bond dimensions to control the expressivity of an MPS and, consequently, the allowed amount of entanglement.
Current efforts benchmarking realistic devices connect the gate fidelities in a quantum circuit to the allowed entanglement~\cite{zhou2020what,dupont2022calibrating}. 
Here, we examine the approximated QAOA performance for the specified classical problems. That is, we focus on studying the approximated quantum algorithm behavior solving the problems rather than comparing it to the exact quantum algorithm.

This paper is structured as follows.
In Sec.~\ref{sec:QAOA}, we describe the QAOA algorithm and introduce the computationally hard problems EC3 and MaxCut. Secondly, in Sec.~\ref{sec:MPS_Simulation}, we review the MPS representation and simulation methods, relating them with the entanglement analysis of the system. Additionally, we introduce a deterministic sampling method that leads to a restricted form of QAOA. Sec.~\ref{sec:QAOA_det_sam_perform} includes an extensive analysis of QAOA performances ($p\leq100$) for varying maximum bond dimensions ($D\leq 128$), with up to $60$ qubit instances. Then, in Sec.~\ref{sec:QAOA_training}, we study how the classical optimization loop behaves for QAOA states with low entanglement in their MPS representation for circuit depths up to $p=4$. Here, we also examine the performance of both an exact QAOA and our approximated QAOA with the approximate parameters derived from low entangled MPSs. Finally, we summarize our results in Sec.~\ref{sec:conclusion}.

\section{Quantum Approximate Optimization Algorithm}
\label{sec:QAOA}

\begin{figure*}[!hbtp]
    \centering
    \includegraphics[width=0.8\linewidth]{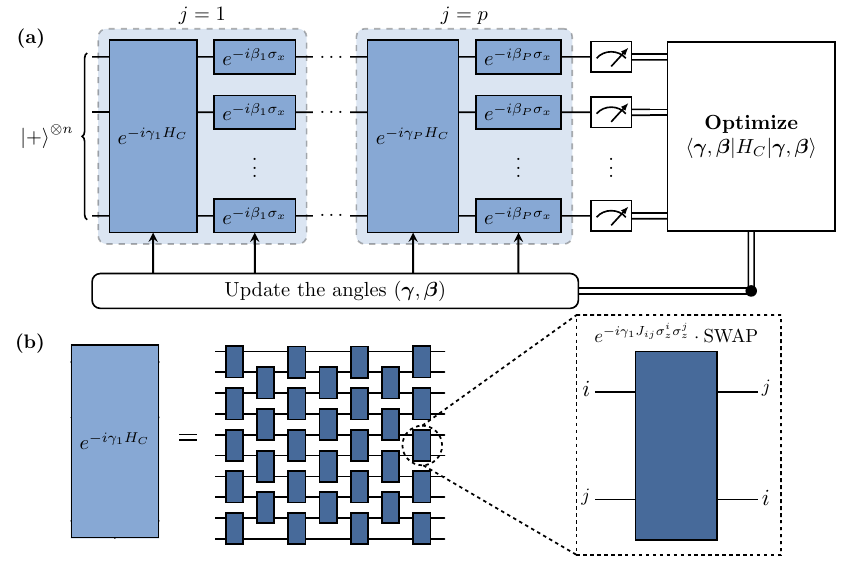}
    \caption{The QAOA circuit comprises $p$ steps with the equal superposition state $\ket{+}^{\otimes n}$ as input. Each step $j$ consists of two unitaries $e^{-i \beta_{j} H_B}$ and $ e^{-i \gamma_{j} H_C}$ based on a mixing Hamiltonian $H_B$ and cost Hamiltonian $H_C$, parametrized by two angles $\beta_j$ and $\gamma_j$, respectively. The final quantum state of the circuit is a parametrized ansatz $\ket{\angamma,\anbeta}$ expressed in terms of $2p$ independent angles $(\angamma,\anbeta)$. 
    The ansatz is classically optimized using an external feedback loop until reaching the quantum state $\ket{\angamma_{\text{opt}}, \anbeta_{\text{opt}}}$ that approaches the ground state of cost Hamiltonian $H_C$ for large $p$. $\mathbf{(a)}$ Full Schematic of $p$-depth QAOA with quantum operations (blue) and classical optimization loop. $\mathbf{(b)}$ Compilation of quantum operations using a SWAP network to adapt non-local operations to a qubit linear connectivity.}
    \label{fig:QAOAcirc}
\end{figure*}

VQAs are heuristic hybrid quantum-classical algorithms centered around the variational method in quantum mechanics. Here, we consider the standard version of QAOA~\cite{farhi2014a-quantum} to solve MaxCut and EC3 problem instances. We define each problem instance with a cost function to be optimized and apply QAOA to find the solution. 
First, we rewrite the cost function as a quantum Hamiltonian $H_C$ with a ground state encoding the solution to the optimization problem of interest.
As illustrated in Fig.~\ref{fig:QAOAcirc}, the QAOA quantum circuit comprises of a sequence of parametrized unitaries generated by the cost Hamiltonian $H_C$ and a mixing Hamiltonian $H_B$, applied alternatively $p$ times to the state $\ket{+}^{\otimes n}$ with angles $\angamma=(\gamma_1,\dots,\gamma_p)$ and $\anbeta=(\beta_1,\dots,\beta_p)$, respectively.
This circuit produces parametrized quantum ans\"{a}tze given by
\begin{equation}
    \label{eq:qaoa_state_creation}
    \ket{\angamma, \anbeta} = \prod_{j=1}^{p} e^{-i \beta_{j} H_B}e^{-i \gamma_{j} H_C} \ket{+}^{\otimes n},
\end{equation}
which one varies to minimize the expectation value of the cost
\begin{equation}
\label{eq:cost_p}
    C(\angamma,\anbeta) = \bra{\angamma,\anbeta} H_C \ket{\angamma,\anbeta}.
\end{equation}
In the standard algorithm, the mixing Hamiltonian is $H_B = \sum_{i = 1}^n \pauli{x}{i}$, where Pauli operators $\pauli{x}{i}$ acts locally on each qubit. Such choice allows the interpretation of QAOA as a Trotterized quantum annealing, with the system initialized in the ground state of $-H_B$ and evolving to the ground state of $H_C$ with an annealing schedule related to the $2p$ circuit parameters $(\angamma,\anbeta)$.
In QAOA, a classical optimization loop aims to find the set of angles $(\angamma_{\text{opt}}, \anbeta_{\text{opt}})$ minimizing $C(\angamma,\anbeta)$, which may lead to non-adiabatic mechanisms elusive in quantum annealing~\cite{zhou2020quantum}.
The performance of QAOA improves with increasing circuit depth $p$~\cite{farhi2014a-quantum}, asymptotically leading to adiabatic quantum computation. That is,  $\ket{\angamma_{\text{opt}}, \anbeta_{\text{opt}}}_p$ exactly approach the ground state of $H_C$ for $p \rightarrow \infty$. For a finite depth QAOA, one measures the final state in the computational basis and aims to sample a bitstring solving the problem approximately with high probability.

Here, we analyze the standard QAOA applied to two widely studied classically hard problems, MaxCut on randomly generated Erd\H{o}s–R\'{e}nyi undirected graphs and EC3. Given a graph $G(V,E)$ with $V$ vertices and $E$ edges, in MaxCut, we look for two complementary partitions of the set $V$ such that the number of cut edges between them is maximized. In this work, we consider unweighted graphs defined by symmetric adjacency matrices $A$ such that $A_{ij} = 1$ if there is an edge present between vertices $i$ and $j$ and $0$ otherwise. We translate the problem of finding the maximum cut of a graph with $n$ vertices to minimizing the cost Hamiltonian
\begin{equation}
    H^{\textsc{MC}}_C = -\frac{1}{2} \sum_{i,j} A_{ij} \left( 1 - \pauli{z}{i} \pauli{z}{j} \right).
    \label{eq:mxc_hamiltonian}
\end{equation}
The minus in front of $H^{\textsc{MC}}_C$ frames it as a minimization problem such that the ground state encodes the maximum cut of the graph, and low energy states constitute approximate solutions. 

Another hard problem for conventional computers---studied in adiabatic quantum computation~\cite{farhi2001a-quantum,choi2011different}---is EC3, a particular case of the Exact Cover problem. EC3 is a special case of the 3SAT satisfiability problem, which can be formulated considering $n$ bits and $m$ clauses. In EC3, each clause involves exactly $3$ bits $x_i, x_j, x_k$ such that $x_i + x_j + x_k = 1$. We construct EC3 satisfiable problem instances, such that one could find a bitstring $\mathbf{x} = x_1 \dots x_n$ that satisfies all the $m$ clauses. After mapping this problem to qubit variables, the solution is encoded in the ground state of the Ising cost Hamiltonian
\begin{equation}
    H^{\textsc{EC3}}_C = \sum_i h_i \pauli{z}{i} + \sum_{i,j} J_{ij}\pauli{z}{i} \pauli{z}{j}.
    \label{eq:exc_hamiltonian}
\end{equation}
In contrast to MaxCut, only optimal solutions satisfying all the clauses are valid in EC3. Refer to Appendix~\ref{appx:rand_inst_gen_appx} for a detailed description of the creation of MaxCut and EC3 suitable problem instances.

Finally, once the QAOA instances are created, we implement a SWAP network~\cite{kivlichan2018quantum} to describe the algorithms in terms of single-qubit operations and two-qubit nearest-neighbor interactions in a qubit chain, as shown in Fig.~\ref{fig:QAOAcirc}. In this manner, we can use the framework of matrix product states (MPS)---suitable for one-dimensional architectures---to analyze the QAOA with cost Hamiltonians of Eqs.~(\ref{eq:mxc_hamiltonian}) and (\ref{eq:exc_hamiltonian}), now rewritten in a suitable way.

\section{QAOA simulation with matrix product states}
\label{sec:MPS_Simulation}
Given a quantum algorithm comprised of single and two-qubit gates between nearest-neighbor qubits in a linear array, we consider matrix product states a convenient tool for its analysis.
First, we represent the state of a one-dimensional $n$ qubit register with an MPS. This description is adequate to analyze the role of entanglement in quantum algorithms, since we can bound the amount of entanglement between different bipartitions of the system. To illustrate the construction of an MPS, we first analyze the entanglement between two separate blocks of the register, with $m$ and $n-m$ qubits, respectively. The general form of such bipartite state is a superposition in terms of product states of the orthonormal bases $\{\ket{i_L}\}$ and $\{\ket{j_R}\}$ of the two separate Hilbert spaces $\cal{H}_L$ and $\cal{H}_R$, corresponding to the left and right partitions as given below,
\begin{equation}
\label{eq:bi_partition}
    \ket{\psi} = \sum_{i,j} \alpha_{ij} \ket{i_L} \ket{j_R}.
\end{equation}
The singular value decomposition (SVD) of the $2^m \times 2^{n-m}$ matrix representation of the complex amplitudes $\alpha_{ij}$, such that $\alpha_{ij}=\sum_k  U_{ik} \lambda_k V^{\dagger}_{kj}$, leads to the Schmidt decomposition of $\ket{\psi}$ in terms of the positive singular values $\lambda_k$,
\begin{align}
\label{eq:schmidtDeriv}
    \ket{\psi} &= \sum_{i,j} \sum_k U_{ik} \lambda_k V^{\dagger}_{kj} \ket{i_L} \ket{j_R} \nonumber  \\
    & = \sum_k \lambda_k \sum_{i}  U_{ik} \ket{i_L} \sum_{j} V^{\dagger}_{kj} \ket{j_R} \nonumber \\
    & = \sum_k \lambda_k \ket{k_L} \ket{k_R},
\end{align}
with $\{\ket{k_L}\}$ and $\{\ket{k_R}\}$ being the new orthonormal bases of $\cal{H}_L$ and $\cal{H}_R$ respectively. The Schmidt expansion in Eq.~(\ref{eq:schmidtDeriv}) exhibits explicitly the entanglement between the left $L$ and right $R$ subsystems, with the entanglement entropy related to the probabilities $\lambda^2_k$ as
\begin{align}
    S_L = S_R &= - \Tr{\rho_L\log_2\rho_L} = - \Tr{\rho_R\log_2\rho_R} \nonumber \\
    &=- \sum_k \lambda_k^2 \log_2 \lambda_k^2,
\end{align}
where $\rho_L=\Tr_R\ketbra{\psi}$ and $\rho_R=\Tr_L\ketbra{\psi}$ are reduced density matrices on the subspaces $L$ and $R$. For separable systems, there is only one $\lambda_k$ which takes the value $1$. Hence, $S_L =S_R=0$. If two or more singular values are nonzero, the entanglement entropy becomes positive and one loses mutual information by focusing on the separate subsystems independently.
The number of nonzero Schmidt coefficients $\lambda_k$ is the rank $r$ of the matrix of complex coefficients $\alpha_{ij}$ in Eq.~(\ref{eq:bi_partition}), which can be at most $\min(2^m,2^{n-m})$. A straightforward truncation for approximate simulations consists in restricting the number of nonzero Schmidt coefficients by setting the smaller singular values to zero. This simplification leads to a new representation of the state retaining a reduced number of parameters to approximately describe the correlation between two separate subsystems.

\begin{figure}[!hbtp]
    \centering
    \begin{tikzpicture}
    
    \node[] (0) {$\mathbf{(a)}$};
    
    \node[block, thick, right of=0] (a) {$A^{s_1}$};
        \node[above of=a] (a1) {$s_1$};
    
    \node[block, thick, right of=a, xshift=.10cm] (b) {$A^{s_2}$};
        \node[above of=b] (b1) {$s_2$};
    
    \node[right of=b, xshift=.10cm] (d) {$\dots$};
    
    \node[block, thick, right of=d, xshift=.10cm] (e) {$A^{s_k}$};
        \node[above of=e] (e1) {$s_k$};
    
    \node[right of=e, xshift=.10cm] (f) {$\dots$};
    
    
    \node[block, thick, right of=f, xshift=.10cm] (h) {$A^{s_n}$};
        \node[above of=h] (h1) {$s_n$};
    
    \draw[black,very thick] (a) -- (b) -- (d) -- (e) -- (f) -- (h);
    \draw[black,very thick] (a) -- (a1);
    \draw[black,very thick] (b) -- (b1);
    \draw[black,very thick] (e) -- (e1);
    \draw[black,very thick] (h) -- (h1);
    
    \node[below of=0] (1) {};
    \node[below of=1] (2) {$\mathbf{(b)}$};
    \node[block, thick, right of=2,xshift=.5cm] (i) {$G$};
        \node[above of=i] (i1) {};
    \node[block, thick, below of=i, minimum size=0.685cm] (j) {$A^{s_k}$};
        \node[left of=j] (j1) {};
        \node[right of=j] (j2) {};
    \node[right of=j,xshift=.6cm] (k) {\textbf{$\Rightarrow$}};
    \node[block, thick, right of=k,xshift=.5cm, minimum size=0.685cm] (l) {$B^{s_k}$};
        \node[left of=l] (l1) {};
        \node[above of=l] (l2) {};
        \node[right of=l] (l3) {};
    \draw[black,very thick] (i) -- (i1);
    \draw[black,very thick] (i) -- (j);
    \draw[black,very thick] (j1) -- (j) -- (j2);
    \draw[black,very thick] (l1) -- (l) -- (l2);
    \draw[black,very thick] (l) -- (l3);
    

    \node[below of=2] (3) {};
    \node[below of=3] (4) {};
    \node[below of=4] (5) {$\mathbf{(c)}$};
    \node[bigblock, thick, right of=5,xshift=.51cm] (m) {$T$};
        \node[right of=5,minimum size=0.685cm] (m1) {};
        \node[right of=m1,minimum size=0.685cm] (m2) {};
        \node[above of=m1,yshift=-0.1cm] (m11) {};
        \node[above of=m2,yshift=-0.1cm] (m22) {};
    \node[block, thick, below of=m1] (n) {$A^{s_j}$};
        \node[left of=n] (n1) {};
    \node[block, thick, below of=m2] (o) {$A^{s_{k}}$};
        \node[right of=o] (o1) {};
    
    \draw[black,very thick] (m1) -- (m11);
    \draw[black,very thick] (m2) -- (m22);
    \draw[black,very thick] (m1) -- (n);
    \draw[black,very thick] (m2) -- (o);
    \draw[black,very thick] (n) -- (n1);
    \draw[black,very thick] (o) -- (o1);
    \draw[black,very thick] (o) -- (n);

    \node[right of=m,xshift=1.25cm,yshift=-.5cm] (p) {\textbf{$\Rightarrow$}};
    \node[bigblock, thick, right of=p,xshift=1.25cm] (r) {$W$};
        \node[right of=p,xshift=0.69cm,minimum size=0.685cm] (r1) {};
        \node[right of=r1,xshift=0.12cm,minimum size=0.685cm] (r2) {};
        \node[above of=r1,yshift=-0.1cm] (r1u) {};
        \node[above of=r2,yshift=-0.1cm] (r2u) {};
        \node[left of=r1,xshift=0.1cm] (r1l) {};
        \node[right of=r2,xshift=-0.1cm] (r2r) {};
        
        \node[above of=r] (r5) {};
        \node[below of=r] (r6) {};
        
        
    \draw[black,very thick] (r1) -- (r1u);
    \draw[black,very thick] (r1) -- (r1l);
    \draw[black,very thick] (r2) -- (r2u);
    \draw[black,very thick] (r2) -- (r2r);
    \draw[black,very thick,dashed,opacity=0.5] (r5) -- (r6);
    
    \node[below of=r6,yshift=0.25cm] (r66) {\textbf{$\Downarrow$}};
    \node[block, thick, below of=r66,yshift=-0.1cm,rotate=45,scale=.8] (t) {\rotatebox{-45}{$S$}};
    \node[block, thick, right of=t,xshift=.15cm] (u) {$V^{\dagger}$};
        \node[right of=u,xshift=-.1cm] (u1) {};
        \node[above of=u] (u2) {};
    \node[block, thick, left of=t,xshift=-.15cm] (s) {$U$};
        \node[left of=s,xshift=.1cm] (s1) {};
        \node[above of=s] (s2) {};
    \node[left of=s1,xshift=.525cm] (s11) {\textbf{$\Leftarrow$}};

    
    \draw[black,very thick] (s1) -- (s);
    \draw[black,very thick] (s) -- (s2);
    \draw[black,very thick] (s) -- (t);
    \draw[black,very thick] (t) -- (u);
    \draw[black,very thick] (u) -- (u1);
    \draw[black,very thick] (u) -- (u2);
    
    \node[left of=s11,xshift=.525cm] (w1) {};
        \node[block, thick, left of=w1] (w) {$B^{s_k}$};
        \node[above of=w] (w2) {};
    \node[block, thick, left of=w] (x) {$B^{s_j}$};
        \node[above of=x] (x1) {};
        \node[left of=x] (x2) {};
        
    \draw[black,very thick] (w1) -- (w);
    \draw[black,very thick] (w2) -- (w);
    \draw[black,very thick] (w) -- (x);
    \draw[black,very thick] (x1) -- (x);
    \draw[black,very thick] (x2) -- (x);

        

\end{tikzpicture}
    \caption{$\mathbf{(a)}$ Diagrammatic description of an $n$-qubit MPS, with physical indices $\{s_k\}$ and tensors $\{A^{s_k}\}$ corresponding to qubit $k$. $\mathbf{(b)}$ Single-qubit operation $G$ acting on qubit $k$ and modifying the local $k$th tensor from $A^{s_k}$ to $B^{s_k}$.  $\mathbf{(c)}$ Two-qubit operation $T$ on adjacent qubits $j$ and $k = j+1$, which may increase the bond dimension $d_j$ between $j$ and $k$. To truncate the bond dimension, we move the MPS orthogonality center to site $j$ and contract the tensors $A^{s_j}$, $A^{s_k}$, and $T$ to obtain the four-tensor $W$. The singular value decomposition $W = USV^{\dagger}$ determines the bond dimension $d_j$---i.e. the singular matrix $S$ rank. We retain the $D$ largest singular values of $S$, and rewrite the MPS in a right-canonical form with the set of tensors $\{B^{s_i}\}$.}
    \label{fig:graphicsMPS}
\end{figure}
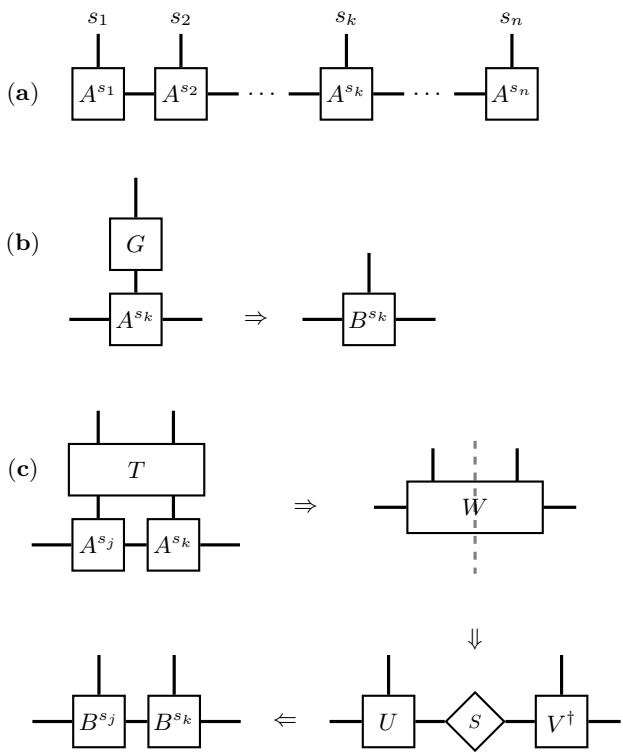

Analogously, to decompose the $n$ qubit state $\ket{\psi}$ into an MPS, we consider all $n-1$ possible bipartitions of the one-dimensional qubit chain and perform a sequence of SVDs on each one to obtain a final tensor canonical form~\cite{schollwock2011the-density-matrix}
\begin{align}
\label{eq:MPSform}
    \ket{\psi} &= \sum_{s_1, \dots s_n} \alpha_{s_1, \dots, s_n} \ket{s_1 \dots s_n } \nonumber \\
    &= \sum_{s_1, \dots s_n} \sum_{u_1, \dots u_{n-1}} A^{s_1}_{u_1} A^{s_2}_{u_1 u_2} \cdots A^{s_n}_{u_{n-1}} \ket{s_1 \dots s_n } \nonumber \\
    &= \sum_{s_1, \dots s_n} A^{s_1} A^{s_2} \cdots A^{s_n} \ket{s_1 \dots s_n},
\end{align}
with $\{\ket{s_k}\}$ the local standard basis for the $k$th qubit in the chain. The complex coefficients $\alpha_{s_1, \dots, s_n}$ are compactly represented by the matrix multiplication $A^{s_1} A^{s_2} \cdots A^{s_n}$. The matrix dimensions depend on the rank $d_k$ of the corresponding SVDs, i.e. the dimension of the $u_k$ indices, such that $A^{s_k}_{u_{k-1} u_k}$ is a $(d_{k-1}\times d_k)$ matrix. These ranks $d_k$ are also called \emph{bond dimensions} and indicates the degree of entanglement in the system, as they are the number of Schmidt weights retained after each SVD. Restricting the bond dimension allows us to study the role of entanglement in our quantum circuits. As mentioned previously, the dimension of the smallest Hilbert space in the bipartition limits the bond dimension $d_k$ such that $d_k \leq \min(2^k, 2^{n-k})$.
Therefore, the maximum possible bond dimension grows exponentially along the chain until its center $k = \lfloor n/2 \rfloor$, where both bipartitions have similar size, to then decrease again.
The MPS representation of a quantum state and the operations performed can be conveniently depicted in tensor diagram notation, as shown in Fig.~\ref{fig:graphicsMPS}.

The QAOA quantum circuits described in Sec.~\ref{sec:QAOA} comprise a sequence of single-qubit gates, two-qubit gates between nearest neighbors in the linear arrangement of qubits, and measurements. 
Single-qubit gates acting on an MPS register are easily computed, since local operations do not increase the degree of entanglement. These gates operate on the individual tensors $A^{s_i}$ of Eq.~(\ref{eq:MPSform}), which does not modify the bond dimensions $d_k$.
In contrast, two-qubit gates acting on neighboring qubits affect not only the individual tensors, but also their bond, as shown in Fig.~\ref{fig:graphicsMPS}. Also, an MPS representation is not unique, and here, we will consider a mixed-canonical form~\cite{schollwock2011the-density-matrix}, with the matrices to the left of the bond left-normalized, and the ones to the right, right-normalized. Such canonical form ensures that the basis states on the left and right sides of the chain around the bond are orthonormal, leading to a Schmidt decomposition as in Eq.~(\ref{eq:schmidtDeriv}).
Thus, the singular values between the neighboring qubits correspond to Schmidt weights, and the subsequent truncation of the bond dimension by removing the smallest values corresponds to an approximation of the degree of entanglement.

In all our QAOA circuit calculations, the entanglement truncation occurs after two-qubit gates. Since the initial state is the product state $\ket{\psi_0} =\ket{+}^{\otimes n}$ with the lowest bond dimensions possible, $d_k=1$ for all $k$, our MPS representation is always exact initially. The entanglement entropy may increase after a two-qubit gate, where we approximate the resulting MPS with a cutoff in the bond dimensions $d_k\leq D$ by keeping the $D$ largest Schmidt weights. Except for the two-qubit gates simulation described previously, we always consider an MPS in right-canonical form, with the first qubit as the orthogonality center. That is, the matrices $A^{s_i}$ in Eq.~(\ref{eq:MPSform}) are right-normalized with $\sum_{s_i} A^{s_i} \left( A^{s_i} \right)^{\dagger} = I$.
In Sec.~\ref{sec:QAOA}, we had rewritten the algorithm for one-dimensional arrays of qubits using SWAP networks~\cite{kivlichan2018quantum}. Notice that, although SWAP gates cannot generate entanglement, they can redistribute it within the circuit. Their behavior may lead to further truncations in MPS calculations, as they shift entanglement between bipartitions. Consequently, finding algorithmic descriptions minimizing the number of non-local qubit interactions could result in fewer approximations.
We use the TensorNetwork package~\cite{roberts2019tensornetwork:} to perform the tensor network operations of state creation, tensor contractions, and bond dimension truncations. 

Finally, it is possible to simulate the measurement of the resulting $n$-qubit state $\ket{\psi}$ on the standard basis. One could evaluate the overlap of each $2^n$ basis states and the final state $\braket{s_1 \dots s_n}{\psi}$ to obtain their probability amplitudes. 
Similarly, it is possible to sample bitstrings $s=s_1 \dots s_n$ from the probability distribution $|\braket{s}{\psi}|^2$ with a reduced computational cost~\cite{ferris2012perfect}.
Nevertheless, in our work, we do not simulate the execution of the measurement at the end of the algorithm. Instead, we evaluate the performance of QAOA with low entanglement based on an individual final sample. To that end, we collapse the MPS form into a single classical product state in the computational basis following a deterministic sampling method described below in Sec.~\ref{ssec:det_sampling}.

\subsection{Deterministic sequential sampling}
\label{ssec:det_sampling}

In classical simulations, the standard metric to characterize the performance of QAOA is given in terms of the expectation value of the cost in the final quantum state, $\bra{\angamma,\anbeta} H_C \ket{\angamma,\anbeta}$, as defined in Eq.~(\ref{eq:cost_p}). That is, the analysis is based on the cost averaged over the possible measurement outcomes in the computational basis, weighted by their likelihood. Then, given a suitable average cost, one could measure with high probability an approximate optimal solution.

Here, we simplify further the computational analysis by evaluating the cost of a single candidate solution $s$ sampled from the final quantum state with probability $\abs{\braket{s}{\psi}}^2 \geq 1/2^n$. Given the final $n$ qubit quantum state of QAOA, represented as a right-canonicalized MPS, we consider a deterministic sampling algorithm that projects it in the standard basis, qubit by qubit, from left to right according to the highest measurement probability of the individual elements. If the probability of the states $\ket{0}$ and $\ket{1}$ coincides, then we project the qubit onto $\ket{1}$ by default.
Thus, following Algorithm~\ref{alg:detsam_method}, we input an $n$ qubit quantum state presented in MPS form and output a single $n$-bitstring. Additionally, one could optimize the calculations by relocating the orthogonality center of the MPS on the site on which the projector acts, which simplifies expectation value estimations~\cite{stoudenmire2010minimally}. 

\begin{algorithm}[H]
    \caption{Deterministic sequential sampling}
    \label{alg:detsam_method}
    \begin{algorithmic}[1]
        \Require An $n$ qubit state $\ket{\psi}$ represented as an MPS.
        \Ensure  A single bitstring $s=s_1 \dots s_n$ with $s_k \in \{0,1 \}$ and $\abs{\braket{s}{\psi}}^2 \geq 1/2^n$.
        \For{$k = 1$ to $n$}
            \State Compute $\rho_k = \Tr_{i\in \{1,\dots,n \}\setminus \{k \}}{\ketbra{\psi}}$ and $P(s_k)$.
            \If{$P(0) > P(1)$}
                \State $s_k \gets 0$
                \State $\ket{\psi}  \gets \ket{0_k}\braket{0_k}{\psi}/P(0)$
            \Else
                \State $s_k \gets 1$
                \State $\ket{\psi} \gets \ket{1_k}\braket{1_k}{\psi}/P(1)$
            \EndIf
        \EndFor \\
        \Return $s=s_1 \dots s_n$.
    \end{algorithmic}
\end{algorithm}

\begin{figure}[!hbtp]
    \begin{tikzpicture}
    
    \node[block, very thick] (a) {$A^{\dagger s_1}$};
        \node[circle,draw,very thick,below of=a,yshift=.2cm] (a1) {};
        \node[right of=a1, xshift=-0.6cm, yshift =-0.3cm] (a1b) {$\bra{s_1}$};
        
            \node[circle,draw,very thick,below of=a1,yshift=-.4cm] (a11) {};
            \node[right of=a11, xshift=-0.6cm, yshift =0.3cm] (a1k) {$\ket{s_1}$};
        
                \node[block, very thick, below of=a11,,yshift=.2cm] (A) {$A^{s_1}$};
            \draw[black, very thick] (a1) -- (a11);
            
    \node[block, very thick, right of=a, xshift=.25cm] (b) {$A^{\dagger s_2}$};
        \node[circle,draw,very thick,below of=b,yshift=.2cm] (b1) {};
        \node[right of=b1, xshift=-0.6cm, yshift =-0.3cm] (b1d) {$\bra{s_2}$};
        
            \node[circle,draw,very thick,below of=b1,yshift=-.4cm] (b11) {};
            \node[right of=b11, xshift=-0.6cm, yshift =0.3cm] (b1d) {$\ket{s_2}$};
        
                \node[block, very thick, below of=b11,yshift=.2cm] (B) {$A^{s_2}$};
            \draw[black, very thick] (b1) -- (b11);
            
            
    
    \node[right of=b, xshift=.25cm] (d) {$\dots$};
    \node[right of=B, xshift=.25cm] (D) {$\dots$};
    
    \node[block, very thick, right of=d, xshift=.25cm] (e) {$A^{\dagger s_k}$};
        \node[below of=e] (e1) {};
            \node[below of=e1] (e11) {};
                \node[block, very thick, below of=e11] (E) {$A^{s_k}$};
            
    \node[block, very thick, right of=e, xshift=.25cm] (f) {$A^{\dagger s_{l}}$};
        \node[below of=f] (f1) {};
            \node[below of=f1] (f11) {};
                \node[block, very thick, below of=f11] (F) {$A^{s_{l}}$};
            
    \node[right of=f, xshift=.25cm] (g) {$\dots$};
    \node[right of=F, xshift=.25cm] (G) {$\dots$};
    
            
    \node[block, very thick, right of=g, xshift=.25cm] (h) {$A^{\dagger s_n}$};
        \node[below of=h] (h1) {};
            \node[below of=h1] (h11) {};
                \node[block, very thick, below of=h11] (H) {$A^{s_n}$};;
            
            
            
    
    \node[below of=E] (i) {$\Downarrow$};

    \node[block, very thick, below of=i] (j) {$\rho_{k}$};
        \node[right of=j] (jr) {};
            \node[left of=j] (jl) {};
        
    \draw[black,very thick] (j) -- (jr);
    \draw[black,very thick] (jl) -- (j);

    \draw[black,very thick] (a) -- (b) -- (d) -- (e) -- (f) -- (g) -- (h);
    \draw[black,very thick] (A) -- (B) -- (D) -- (E) -- (F) -- (G) -- (H);
    
    

    \draw[black,very thick] (a) -- (a1);
    \draw[black,very thick] (a11) -- (A);
    \draw[black,very thick] (b) -- (b1);
    \draw[black,very thick] (b11) -- (B);
    

    \draw[black,very thick] (e) -- (e1);
    \draw[black,very thick] (e11) -- (E);

    \draw[black,very thick] (f) -- (F);
    
    \draw[black,very thick] (h) -- (H);

\end{tikzpicture}   
        \caption{Calculation of the $k$th qubit conditional reduced density matrix $\rho_k$ of Algorithm~\ref{alg:detsam_method}. The MPS form of $\ket{\psi}$ ($\bra{\psi}$) contains the set of tensors $\{A^{s_k}\}$ ($\{A^{\dagger s_k}\}$) for $k = {1,\dots,n}$ qubits. The first $k-1$ qubits have been previously projected onto states $\ket{s_1} \dots \ket{s_{k-1}}$ (circles) in the sampling procedure. We contract all qubit physical indices of the MPS representation except $s_k$ to obtain $\rho_k$.}
    \label{fig:detsam_method}
\end{figure}
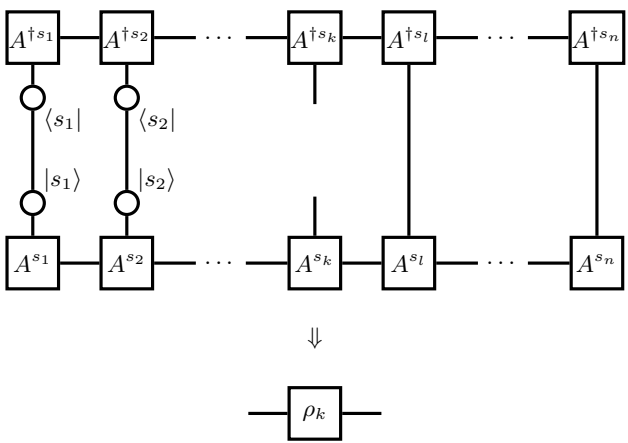

Our deterministic sequential sampling method outlined in Algorithm~\ref{alg:detsam_method} consist of computing single-qubit density matrices $\rho_k$ and single-qubit probabilities $P(s_k)$ of local computational states $\ket{s_k}$, with $s_k \in \{0,1 \}$. 
First, we calculate the reduced density matrix of qubit 1, $\rho_1 = \Tr_{i\in \{1,\dots,n \}\setminus \{1 \}}{\ketbra{\psi}}$. From $\rho_1$, we compute the probabilities $P(s_1)=\bra{\psi}\rho_1\ket{\psi}$, and \emph{deterministically project} this first qubit onto the computational state with highest probability $\ket{s_1}$. 
Therefore, we update the quantum state as $\ket{\psi} \leftarrow \ket{s_1}\braket{s_1}{\psi}/P(s_1)$. Using this updated quantum state, we repeat the process for the next qubit on the chain, obtaining the conditional reduced density matrix $\rho_2$ and conditional probabilities $P(s_2)$. Specifically, $\rho_2$ and $P(s_2)$ describe the updated system in which $s_1$ has already been measured. From here on, we follow the same protocol, graphically represented in Fig.~\ref{fig:detsam_method}, to output a final bitstring $s=s_1 \dots s_n$. We note that the associated basis state $\ket{s}$ may not correspond to that with the highest probability in the computational basis, as shown in Appendix~\ref{appx:counterex_sampling}.

Essentially, the QAOA is a heuristic technique designed to return the best candidate solution to a cost problem. It includes an outer-loop parameter classical optimization guided by the cost expected value. Increasing the circuit depth $p$, and consequently, the number $2p$ of optimal parameters $\angamma_{\text{opt}}, \anbeta_{\text{opt}}$ improves the quality of the output quantum state and the probability of sampling a successful solution rises. In this work, we circumvent the task of randomly drawing $n$ qubit configurations according to the output probability distribution of QAOA quantum circuits, that is, we avoid simulating the quantum measurement in QAOA. Instead, we select a single configuration $s$ following Algorithm~\ref{alg:detsam_method}. 
Therefore, we use MPSs to analyze the role of entanglement in a restricted representation of QAOA, where we limit the access to the full output distribution. Despite this additional limitation, in Sec.~\ref{sec:QAOA_det_sam_perform} we show that such restricted QAOA simulation still provides successful results for systems of up to 60 qubits with maximum bond dimension $D=100$.

\section{QAOA performances with restricted entanglement}
\label{sec:QAOA_det_sam_perform}

Here, we use the standard tensor network techniques described in Sec.~\ref{sec:MPS_Simulation} to analyze the performance of QAOA with restricted entanglement for the MaxCut and EC3 problems.
We consider one hundred $14$-qubit instances, one hundred $40$-qubit instances, and ten $60$-qubit instances for both MaxCut and EC3 problems, with a simplified circuit parameter choice.

The QAOA circuit depth increases its success but, in turn, the global classical optimization subroutine becomes intractable. Limitations in the optimization loop---rigorously studied~\cite{mcclean2018barren,cerezo2021cost}---demand sophisticated strategies to select adequate algorithm parameters.
Here, we consider two sets of optimized angles $\{\angamma_{\text{opt}}, \anbeta_{\text{opt}}\}$, one for MaxCut problem instances, and another one for EC3. 
To create these sets, we generate randomly ten Erd\H{o}s–R\'{e}nyi $12$-node graphs as MaxCut instances and ten $12$-qubit EC3 instances. We obtain the circuit parameters by using GlobalSearch and MultiStart algorithms as global optimization for $p=1$ and linearly extrapolating the results to $p=100$ QAOA steps~\cite{zhou2020quantum} together with the Nelder–Mead method. For each problem, we average the angles of the ten instances to create the effective set of parameters $\angamma_{\text{opt}}, \anbeta_{\text{opt}}$.
Then, we use these two sets of parameters to study QAOA for MaxCut and EC3 with up to $p\leq100$ steps, regardless of the problem size. For intermediate circuits with $p<100$ steps, we select the first $p$ angles from the full set $\angamma_{\text{opt}}, \anbeta_{\text{opt}}$.
In general, using a set of angles for all different instances of a given problem may not be a valid approximation and hinder the QAOA performance, restricting its potential capabilities. On the other hand, results concerning the concentration of optimal parameters regardless of the system size in MaxCut problems with $3$-regular graphs~\cite{brandao2018for-fixed,streif2020training} and the Sherrington-Kirkpatrick model~\cite{farhi2021the-quantum} suggest that
parameter concentrations can be leveraged to shorten the training time in QAOA~\cite{akshay2021parameter}. In our case, choosing the same set of optimal angles without studying the concentration needs to be accounted for as a possible error source when analyzing the performance. 

Once the QAOA circuits are defined, that is, after selecting the problem instances and fixing the gate parametrization, we create an MPS representation of the QAOA ansatz. In this representation, we approximate the states by setting an upper bound $D$ to the bond dimension such that in $\ket{\angamma_{\text{opt}}, \anbeta_{\text{opt}}}_{D}$ the amount of entanglement that can be retained is limited. Note that for $D = 2^{\lfloor n/2 \rfloor}$ the representation is exact. Then, given the approximated states $\ket{\angamma_{\text{opt}}, \anbeta_{\text{opt}}}_{D}$, we deterministically sample a single bitstring following Algorithm~\ref{alg:detsam_method}. Finally, we compare the quality of the sampled solutions $s$ from the approximated QAOA states to the actual solutions of MaxCut and EC3, addressing the performance of our classical simulation.
Moreover, for the $14$-qubit problems analyzed, we reach the bond dimension $D=128$ corresponding to the exact state representation. There, we study the fidelity between the exact states $\ket{\angamma_{\text{opt}},\anbeta_{\text{opt}}}$ and the approximated ones $\ket{\angamma_{\text{opt}}, \anbeta_{\text{opt}}}_{D}$ with the aim to understand the role of entanglement in the quantum algorithm.

\subsection{Performances for MaxCut}
\label{ssec:mxc_per_results}
\begin{figure*}[!hbtp]
    \includegraphics[width = 0.95\textwidth]{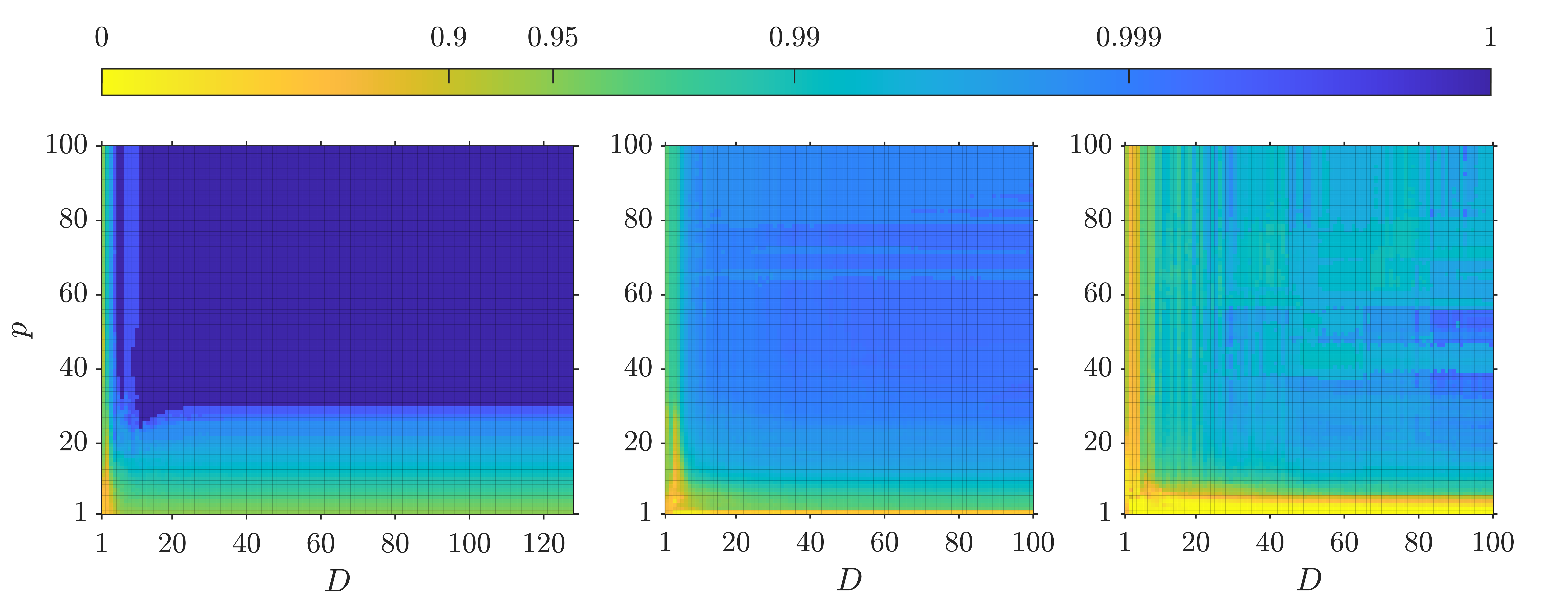}
    \caption{Average approximation ratios $\Bar{r}$ of samples obtained by a restricted QAOA simulation---in terms of the bond dimension $D$ and algorithm depth $p$---for the MaxCut problem with randomly generated Erd\H{o}s–R\'{e}nyi graphs. From left to right we show $\Bar{r}$ for one hundred $14$-qubit instances, one hundred $40$-qubit instances, and ten $60$-qubit instances. }
    \label{fig:tiled_qioa_all_MC}
\end{figure*}

We analyze the behavior of the approximated QAOA---with reduced bond dimension, restricted parameters choice, and deterministic samples---for the MaxCut problem framed as a minimization problem. We consider randomly generated Erd\H{o}s–R\'{e}nyi graphs with an edge probability of $1/2$, which can be challenging for classical solvers~\cite{coppersmith2004random} (see Appendix ~\ref{appx:rand_inst_gen_appx}). To study the performance of our classical simulation of QAOA, we use the approximation ratio
\begin{equation}
\label{eq:app_ratio}
    r(s) = \frac{\bra{s}H^{\textsc{MC}}_C\ket{s}}{C_{\text{min}}},
\end{equation}
with $\ket{s}$ the product state related to the sampled bitstring, $H^{\textsc{MC}}_C$ the cost Hamiltonian defined in Eq.~(\ref{eq:mxc_hamiltonian}), and $C_{\text{min}}$ the minimum energy connected to the exact solution. We notice that, after redefining MaxCut as a minimization problem, all the costs are negative and the approximation ratio remains positive. We compute an approximation ratio $r$ for every problem instance, for different bond dimensions $D$ and algorithm depths $p$. We then average the approximation ratios over all the instances with the same size $n$ to obtain $\Bar{r}$, represented in Fig.~\ref{fig:tiled_qioa_all_MC} for different bond dimensions $D$ and depths $p$. Therefore, Fig.~\ref{fig:tiled_qioa_all_MC} depicts the performance of the restricted QAOA simulation over all the different instances studied (see Appendix~\ref{appx:per_single_ins} for the analysis of a single instance).

For $14$ qubits, the results  range up to $D = 128$, the full bond dimension. With this size, our classical simulation finds the exact solution for all instances, $\Bar{r} = 1$, with bond dimensions beyond $D \approx 10$ and algorithm depths beyond $p \approx 30$. In fact, even for the lowest bond dimension $D =1$ corresponding to a product state, we obtain an average approximation ratio $\Bar{r} > 0.9$ for $p \geq 11$. We observe a reduction in the average approximation ratio for $7 \leq D \leq 10$, consequence of a single instance approximation ratio of $r = 0.978$. In Appendix~\ref{appx:per_single_ins}, we analyze the behavior of the single instance relating the performance decrease with the restricted choice of circuit parameters. Selecting optimized angles for this particular instance led to the optimal solution $r = 1$ with those bond dimensions.

For the cases with $40$ and $60$ qubits, the average approximation ratio reaches $\Bar{r} \approx 0.999$. Among the hundred $40$-qubit instances, for $\approx 90$ we find the optimal solution $r = 1$ within the studied range of bond dimensions and circuit depths, while for the remaining ones we reach $r \geq 0.99$. We observe analogous results for the ten $60$-qubit instances, reaching an average ratio $\Bar{r} \geq 0.95$ for $D\geq 5$ and $p\geq 15$. In contrast to the $14$-qubit instances, we do not compute the exact solution in all cases, that is, we do not reach $\Bar{r} = 1$. We attribute this fact to two possible causes: the sub-optimal parameter choice, and the low bond dimension $D \leq 100$---$4$ and $7$ orders of magnitude smaller than the highest bond dimension possible for $40$- and $60$-qubit chains, respectively.

\subsection{Performances for EC3}
\label{ssec:exc_per_results}
Here, we examine the EC3 problem tackled with our approximated QAOA. In contrast to the previous case, only exact solutions constitute an exact cover, and therefore a valid answer. That is, only the ground state of the corresponding Ising Hamiltonian $H^{\textsc{MC}}_C$ in Eq.~(\ref{eq:mxc_hamiltonian}) is a solution. Thus, to analyze the performance of the restricted QAOA for all the instances of a specific size, we define the success rate
\begin{equation}
\label{eq:ec3_success_rate}
    \bar{x}=\frac{1}{N}\sum_{i = 1}^N x_i ,
\end{equation}
with $N$ the number of instances, and $x_i = 1$ if the simulation finds a solution for the $i$th EC3 problem instance, and $x_i =0$ otherwise. As in the previous case, we obtain a statistical description of the algorithm performance for different bond dimensions $D$ and circuit depths $p$ (see Appendix~\ref{appx:per_single_ins} for a study with a single instance). In other words, the success rate $\Bar{x}$ represents the ratio of the total number of instances for which the simulation---given a certain bond dimension $D$ and depth $p$---finds a solution.

\begin{figure*}[!hbtp]
    \includegraphics[width = 0.95\textwidth]{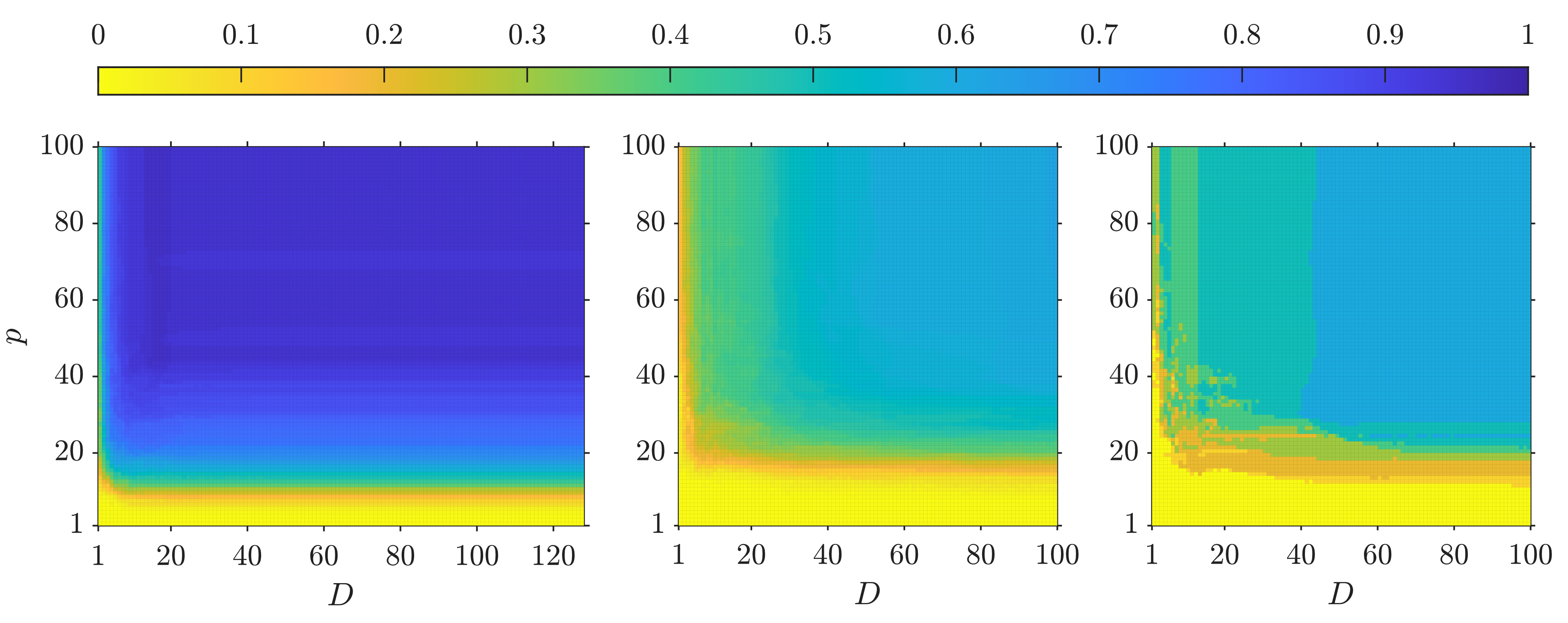}
    \caption{
    Success rates $\Bar{x}$ of the restricted QAOA simulation together with a deterministic sampling method---for different bond dimensions $D$ and algorithm depths $p$---applied to the EC3 problem instances. From left to right we show $\Bar{x}$ for one hundred $14$-qubit instances, one hundred $40$-qubit instances, and ten $60$-qubit instances.}
    \label{fig:tiled_qioa_all_EC}
\end{figure*}

Fig.~\ref{fig:tiled_qioa_all_EC} shows the success rates $\Bar{x}$ for the different EC3 problem instances involving $14$, $40$, and $60$ qubits. As previously with the MaxCut problem, we study the full range of bond dimensions $1 \leq D \leq 128$ for $14$ qubits. The simulation with the smallest bond dimension $D = 1$ outputs an exact cover for at least $50$ of the hundred instances, $\Bar{x} \approx 0.5$, when $p\geq 60$. Increasing the bond dimension up to $D=5$ improves the success rate with lower depths $p=20$ and $p=30$, reaching $\Bar{x}\approx 0.7$ and $\Bar{x} \approx 0.85$, respectively. The success rates remain similar for $D>5$, with the performance depending of the circuit depth $p$. In particular, the circuit exact simulation with $D=128$ reaches similar success rates $\Bar{x} \approx 0.85$ and $\Bar{x} \approx 0.95$ for depths $p=30$ and $p=100$, respectively.

The success rate for the hundred EC3 problem instances with $40$ qubits increases with the bond dimension, reaching a plateau for $D\approx 40$. First, the simulation with the smallest bond dimension $D=1$ succeed with $p=60$ for fifteen cases, $\Bar{x} \approx 0.15$. For $D=5$, we observe $\Bar{x} \approx 0.2$ and $\Bar{x} \approx 0.3$ for $p=20$ and $p=30$, respectively. The success rate slowly increases from $\Bar{x}\approx 0.5$, for $D\approx 40$ and $p\approx30$, to $\Bar{x}\approx 0.6$ for $D=100$ and $p=100$. Similarly, with the ten $60$-qubit problem instances we observe that even for $D = 1$, we obtain $\Bar{x}=0.3$ for $p=60$. The performance plateaus for $D\approx 50$ and $p\approx 30$ with $\Bar{x}=0.6$, the same success rate obtained with $D=100$ and $p=100$.

\subsection{Entanglement in QAOA}
\label{ssec:mps_qaoa_fidelity}

Besides the performance study of the restricted QAOA simulation for the MaxCut and EC3 problems, we address how such simulation deviates from the predicted behavior of the quantum algorithm. This second characterization adds to recent works analyzing how the entanglement is generated in different VQAs~\cite{wiersema2020exploring,diez-valle2021quantum}, and in particular QAOAs~\cite{chen2022how-much,dupont2022an-entanglement}. 

\begin{figure}
    \centering
    \includegraphics[width=0.95\columnwidth]{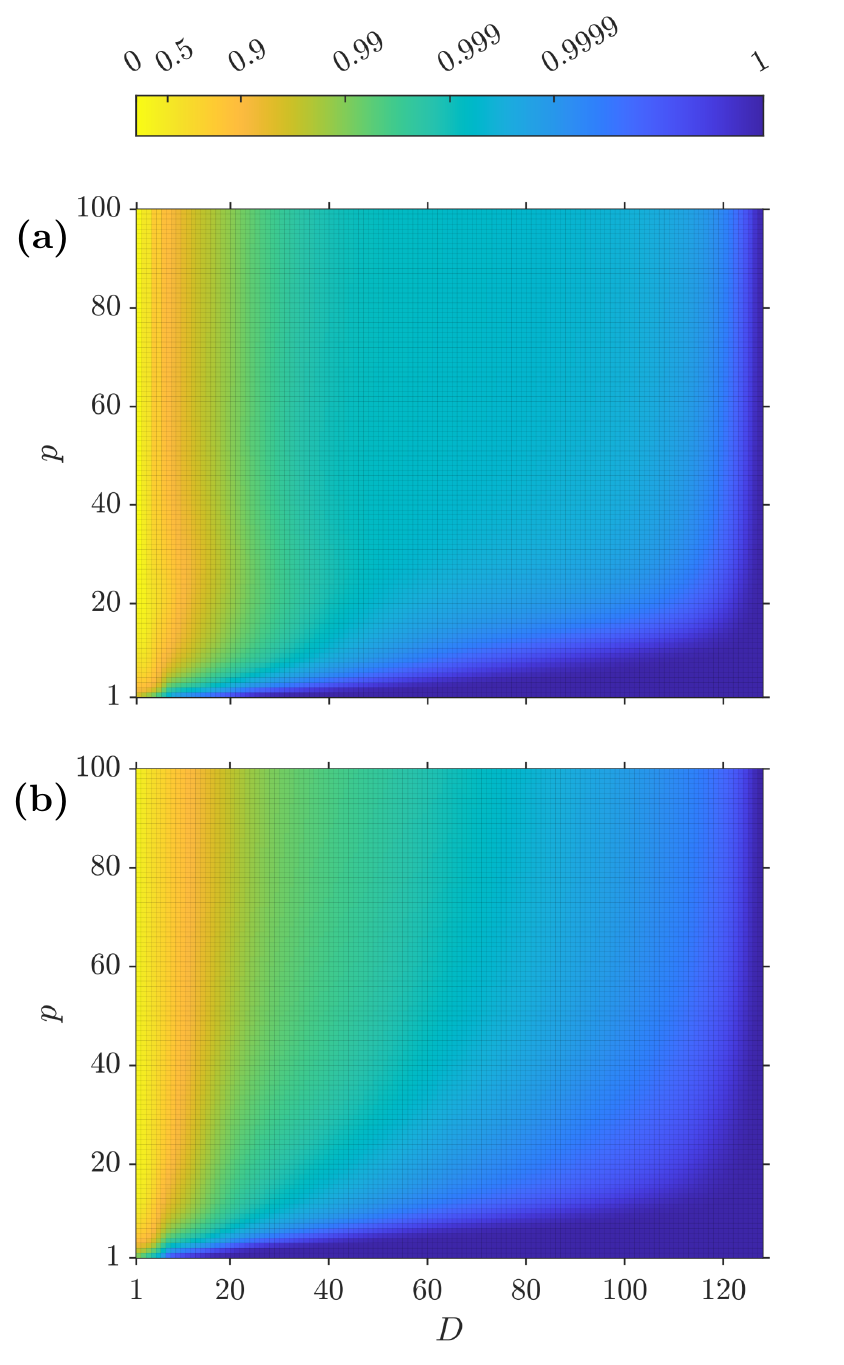}
    \caption{Averaged fidelity $\Bar{F}$ of the exact QAOA state $\ket{\angamma_{\text{opt}},\anbeta_{\text{opt}}}$ and $\ket{\angamma_{\text{opt}}, \anbeta_{\text{opt}}}_{D}$ with a reduced bond dimension $D$, calculated for one hundred $14$-qubit instances of \textbf{(a)} MaxCut and \textbf{(b)} EC3 problems. }
    \label{fig:q14_fidelity_all}
\end{figure}

On the one hand, the amount of entanglement limits the simulation of quantum systems with tensor-network techniques. On the other hand, there exist cases for which the output distribution of QAOA with the lowest depth $p=1$ cannot be efficiently simulated with classical computers~\cite{farhi2019quantum}. Here, we analyze how the entanglement---one of the resources related to classical complexity---grows throughout different depths $p$ of QAOA for MaxCut and EC3 problems. Indeed, the success of our restricted simulation with low bond dimensions may be attributed to low-entangled exact quantum states $\ket{\angamma_{\text{opt}},\anbeta_{\text{opt}}}$. To test the accuracy of the simulation with low bond dimensions $D$, we calculate the average fidelity $\Bar{F}$ between the approximated states and the exact ones for all the $14$-qubit problem instances. That is, for each instance we consider the fidelity
\begin{equation}
\label{eq:fidelity}
    F(D,p) = \abs{\braket{\angamma_{\text{opt}},\anbeta_{\text{opt}}}{\angamma_{\text{opt}}, \anbeta_{\text{opt}}}_{D}}^2,
\end{equation}
with $\ket{\angamma_{\text{opt}},\anbeta_{\text{opt}}}$ the exact ans\"{a}tze of a $p$-depth QAOA, and $\ket{\angamma_{\text{opt}},\anbeta_{\text{opt}}}_D$ the corresponding approximated state generated with the QAOA circuit limited to a bond dimension $D$. Then, we average it over all one hundred MaxCut and EC3 $14$-qubit instances for each bond dimension $D$ and depth $p$, with the results shown in Fig.~\ref{fig:q14_fidelity_all}. The fidelity decreases with increasing circuit depth $p$, particularly for smaller $D<10$, which aligns with the predicted entanglement generation with large-depth QAOA~\cite{chen2022how-much,dupont2022an-entanglement}. 

For the MaxCut problem instances, the average fidelity quickly approach $1$ from $D \approx 10$. Let us focus on the highest depth $p = 100$ and bond-dimension $D=10$, where we obtain $\Bar{F}(10,100) \approx 0.9$. This value increases progressively with the bond-dimension $D$, with $\Bar{F}(30,100) \approx 0.99$, $\Bar{F}(40,100) \approx 0.999$  and  $\Bar{F}(120,100) \approx 0.9999$. We recall that our classical simulation of QAOA together with the deterministic sampling method described in Algorithm~\ref{alg:detsam_method} outputs the exact solution for all one hundred $14$-qubit MaxCut instances if $D \geq 6$ and any $p>30$, which relates to an average fidelity  $\Bar{F}(6,100) \approx 0.88$.

Similarly, the average fidelity of the one hundred $14$-qubit EC3 instances reaches a high value for low bond dimensions. Again, for the highest depth studied $p=100$ we obtain $\Bar{F}(10,100) \approx 0.8$ for the low bond dimension $D=10$. The fidelity increases quickly to $\Bar{F}(30,100) \approx 0.99$, $\Bar{F}(40,100) \approx 0.99$, and $\Bar{F}(120,100) \approx 1$.

The high performance of the QAOA classical simulation---with reduced bond dimension and followed by our restricted sampling---of $40$- and $60$-qubit problem instances might be related to low entanglement in the quantum algorithm. However, due to the size of the systems, an analysis of the fidelity of the exact and approximated states becomes intractable.

\section{QAOA training with low entanglement in MPS representations}
\label{sec:QAOA_training}
The classical parameter optimization in variational quantum algorithms quickly becomes intractable with the circuit depth. 
In Sec.~\ref{sec:QAOA_det_sam_perform}, we have simplified the selection of parameters to analyze the performances of QAOA with depths up to $p=100$. Specifically, we used averaged parameters derived from $12$-qubit instances by linearly extrapolating $p=1$ results to different steps~\cite{zhou2020quantum}.
We explore here alternative circuit optimization strategies, since our previous highly simplified parameter choice may obstruct the algorithm's success.
Several approaches have been devised for circuit training, including tree tensor-network techniques applicable to large system sizes and lower circuit depths~\cite{streif2020training}.
Here, we investigate the circuit optimization using MPS representations with low bond dimensions.

As discussed in Sec.~\ref{sec:MPS_Simulation}, a full MPS representation of quantum states becomes intractable if the required bond dimension grows exponentially with the system size. In this work, we set upper bounds on the bond dimension $D$ and generate QAOA ans{\"a}tze with limited entanglement, $\ket{\angamma, \anbeta}_{D}$. We consider an average cost function similar to Eq.~(\ref{eq:cost_p}), related to the cost Hamiltonian $H_C$ and the approximated MPSs
\begin{equation}
\label{eq:mps_qaoa_cost}
    C_{D}(\angamma, \anbeta) = \prescript{\phantom{+}}{D} {\bra{\angamma, \anbeta}} H_C \ket{\angamma, \anbeta}_{D}.
\end{equation}
Throughout this section, we use non-normalized MPSs $\ket{\angamma, \anbeta}_{D}$ (see Appendix~\ref{appx:normalization_details} for a detailed explanation), as we observe that it improves the classical parameter training. Therefore, the costs $C_{D}(\angamma, \anbeta)$ and $C(\angamma, \anbeta)$ can only be compared if one accounts for the normalization factor in the first one. 

\begin{figure*}[!hbtp]
    \includegraphics[width= 0.95 \linewidth]{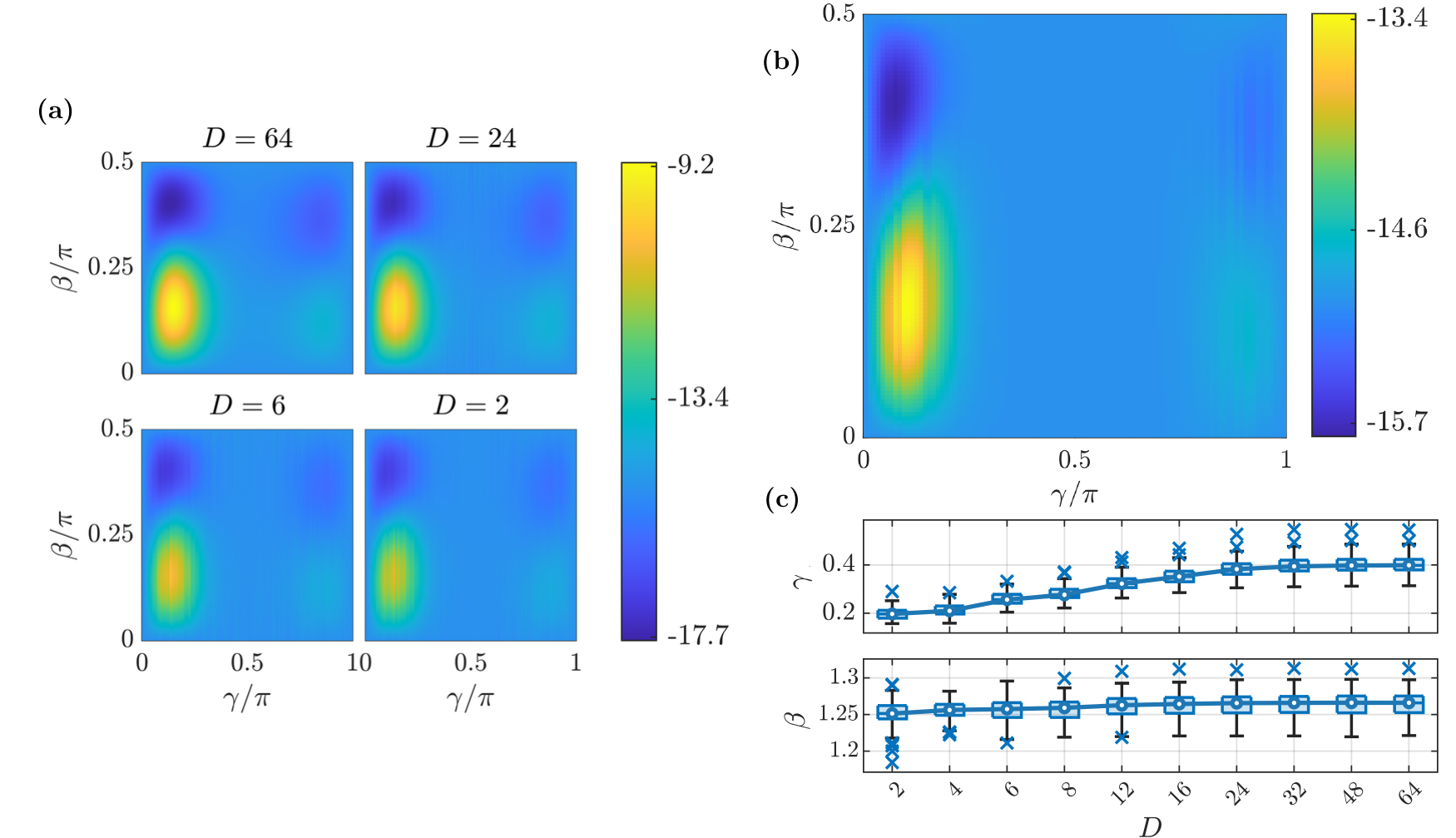}
    \caption{\textbf{(a)} QAOA cost landscapes $C_{D}(\gamma, \beta)$ of a $12$-qubit MaxCut instance for different bond dimensions, with $D=64$ the exact cost. \textbf{(b)} Magnified plot of $C_{2}(\gamma, \beta)$. \textbf{(c)} Statistical distribution of the optimal angles $\gamma_{\text{opt}}^{D}$ and $ \beta_{\text{opt}}^{D}$ for one hundred $12$-qubit MaxCut instances. The line connects the median angles for different bond dimensions, the boxes are delimited by the lower and upper quartiles, and the bars have endpoints at minimum and maximum values that are not outliers. The crosses represent outlier points, that is, values lying more than $1.5$ inter-quartile range away from the box edges.}
    \label{fig:p1_12R0cost_anglesDist}
\end{figure*}

We defined as $\angamma_{\text{opt}}$ and $\anbeta_{\text{opt}}$ the $2p$ parameters obtained from the optimization of the exact cost function $C(\angamma, \anbeta)$ of Eq.~(\ref{eq:cost_p}). Similarly, we denote as $\angamma_{\text{opt}}^{D,p}$ and $\anbeta_{\text{opt}}^{D,p}$ the $2p$ parameters obtained from the optimization of the approximated cost in Eq.~(\ref{eq:mps_qaoa_cost}). Note that for the full bond dimension $D = 2^{\lfloor n/2 \rfloor}$, these angles correspond to the exact ones.

In the following study, we consider randomly generated $12$-node Erd\H{o}s–R\'{e}nyi instances of the MaxCut problem. We consider approximated cost functions computed with reduced bond dimensions and perform a global optimization to obtain the approximated parameters (see Appendix~\ref{appx:bayesian_opt} for details on the global optimization methods). First, we compare these parameters obtained from the approximated training to those obtained from an exact simulation. 

\begin{figure*}[!htb]
    \includegraphics[width= 0.96\linewidth]{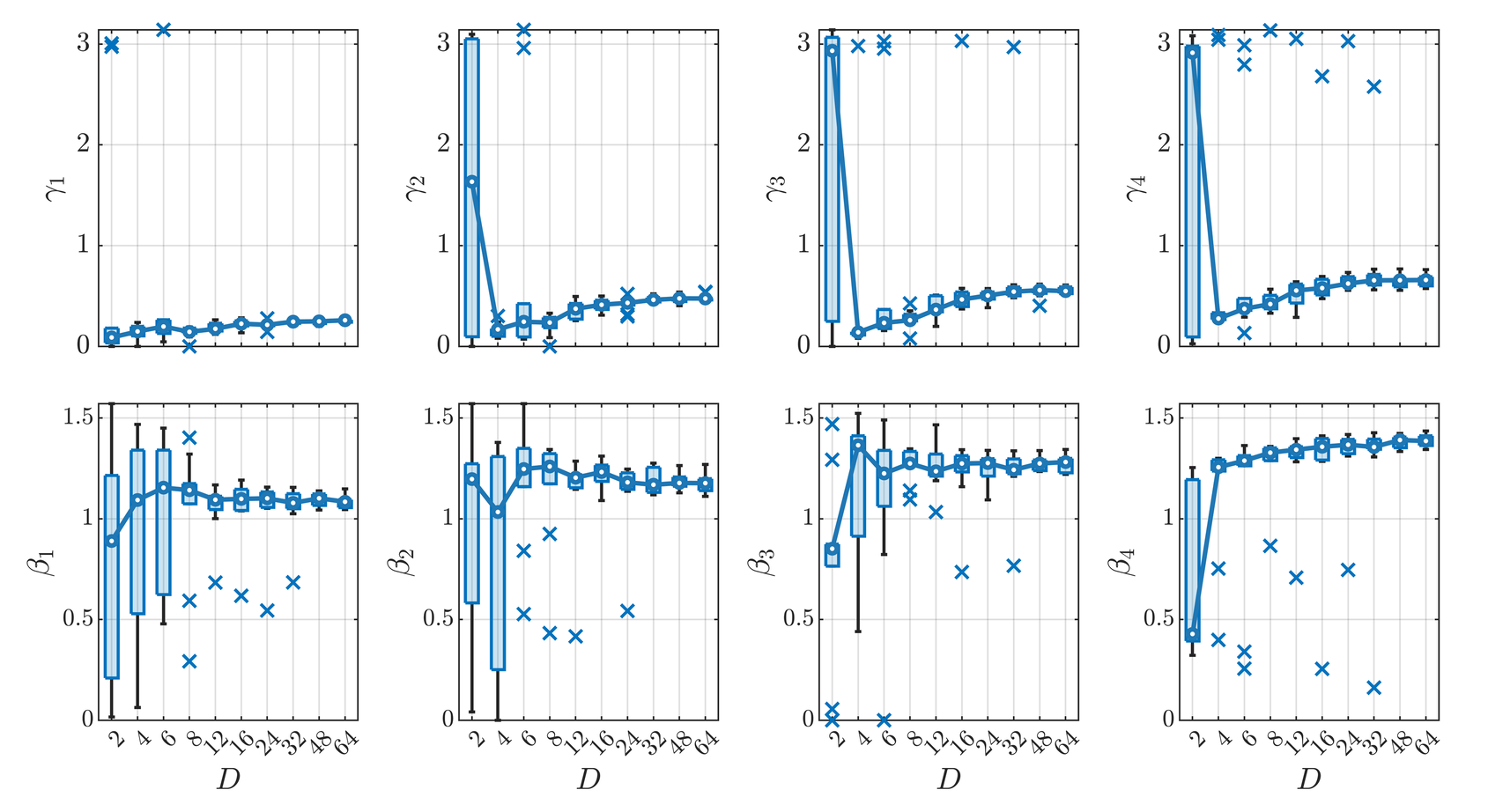}
    \caption{Statistical distribution of the optimal angles $\angamma_{\text{opt}}^{D}$ and $\anbeta_{\text{opt}}^{D}$ for ten $12$-qubit MaxCut instances and the QAOA with depth $p=4$. The line connects the median angles for different bond dimensions, the boxes are delimited by the lower and upper quartiles, and the bars have endpoints at the minimum and maximum values that are not outliers. The cross markers correspond to outlier values, farther away than $1.5$ inter-quartile range from the box edges.}
    \label{fig:Q12P4_opt_para_all}
\end{figure*}

\paragraph{Training of QAOA with algorithm depth $p = 1$.}
We can visualize the approximated cost function $C_D(\gamma,\beta)$ corresponding to the QAOA with depth $p=1$. We study the landscape change and the optimal parameters for decreasing bond dimensions in the circuit simulation. For our $12$-qubit instances, the exact cost function corresponds to bond dimension $D=64$. In Fig.~\ref{fig:p1_12R0cost_anglesDist}, we observe that the overall shape of the cost landscape---the position of the maxima and minima---is retained even on reduction of the bond dimension. The use of non-normalized MPSs flattens the landscape as we reduce the bond dimension to $D = 2$ but minimizes its distortion (see Appendix~\ref{appx:normalization_details}). 
Besides the lack of normalization, we may attribute this compressed cost landscape to the entanglement cutoff. Our observations are consistent with experimental realizations and benchmarking results~\cite{qiang2018large-scale,willsch2020benchmarking,bengtsson2020improved,pagano2020quantum,abrams2020implementation,harrigan2021quantum,dupont2022calibrating} since small bond dimensions are linked with lower fidelity and hardware noise~\cite{zhou2020what}.
In addition, we observe high-frequency noise in the $\gamma$ parameter---related to the circuit entangling gates---for lower bond dimensions $D = 2$. This feature also appears in experimental realizations with a higher number of qubits~\cite{harrigan2021quantum}. 

We study the variation of the optimal angles $\gamma_{\text{opt}}^{D}$ and $ \beta_{\text{opt}}^{D}$ for different bond dimensions in a $p = 1$ QAOA applied to MaxCut. To that end, we use one hundred $12$-qubit instances of randomly generated Erd\H{o}s–R\'{e}nyi graphs with $12$ nodes and an edge probability of $1/2$. As we observe in Fig.~\ref{fig:p1_12R0cost_anglesDist}, the optimal $\gamma_{\text{opt}}^{D}$ parameter value gradually decreases with decreasing $D$, whereas $ \beta_{\text{opt}}^{D}$ remains unchanged.

\paragraph{Training of QAOA with algorithm depth $p > 1$.}
Adding more layers in QAOA increases the number of entangling gates, which can worsen the parameter training with an MPS simulation for truncated bond dimensions. Here, we study whether the robustness of the minima observed in the cost landscapes of QAOA with depth $p = 1$ persists in $p > 1$. Again, we consider $12$-qubit instances and QAOA with depth $p=4$ for solving MaxCut and compute the distributions of the optimal angles $\angamma_{\text{opt}}^{D,p}$ and $\anbeta_{\text{opt}}^{D,p}$ (similar results for $p=2$ and $p=3$ in Appendix~\ref{appx:angles23}). Due to the numerical challenges in the landscape global optimization (see Appendix~\ref{appx:bayesian_opt}), we restrict our analysis to the first ten instances of the one hundred considered previously.

Fig.~\ref{fig:Q12P4_opt_para_all} shows the optimal parameters variations with different bond dimensions $D$ for $p = 4$. Their values change gradually from the exact simulation with $D = 64$ to a restricted bond dimension simulation with $D=6$. For lower bond dimensions, the parameters abruptly change with high dispersion.

The global landscape optimization task in variational quantum algorithms becomes intractable with the circuit depth $p$. In Sec.~\ref{sec:QAOA_det_sam_perform}, we considered layer-wise training and extrapolation strategies~\cite{zhou2020quantum} to choose suitable algorithm parameters up to $p = 100$. Moreover, we used the averaged parameters for $12$-qubit instances to larger system sizes. The success of such a crude approximation relies on the concentration of parameters for different problem sizes, and the patterns observed in low-depth circuits MaxCut with $3$-regular graphs~\cite{zhou2020quantum}. In Appendix~\ref{appx:patterns_parameters}, we analyze the patterns of the approximated parameters computed in this Section.

\begin{figure*}[!htbp]
    \includegraphics[width= 0.98\linewidth]{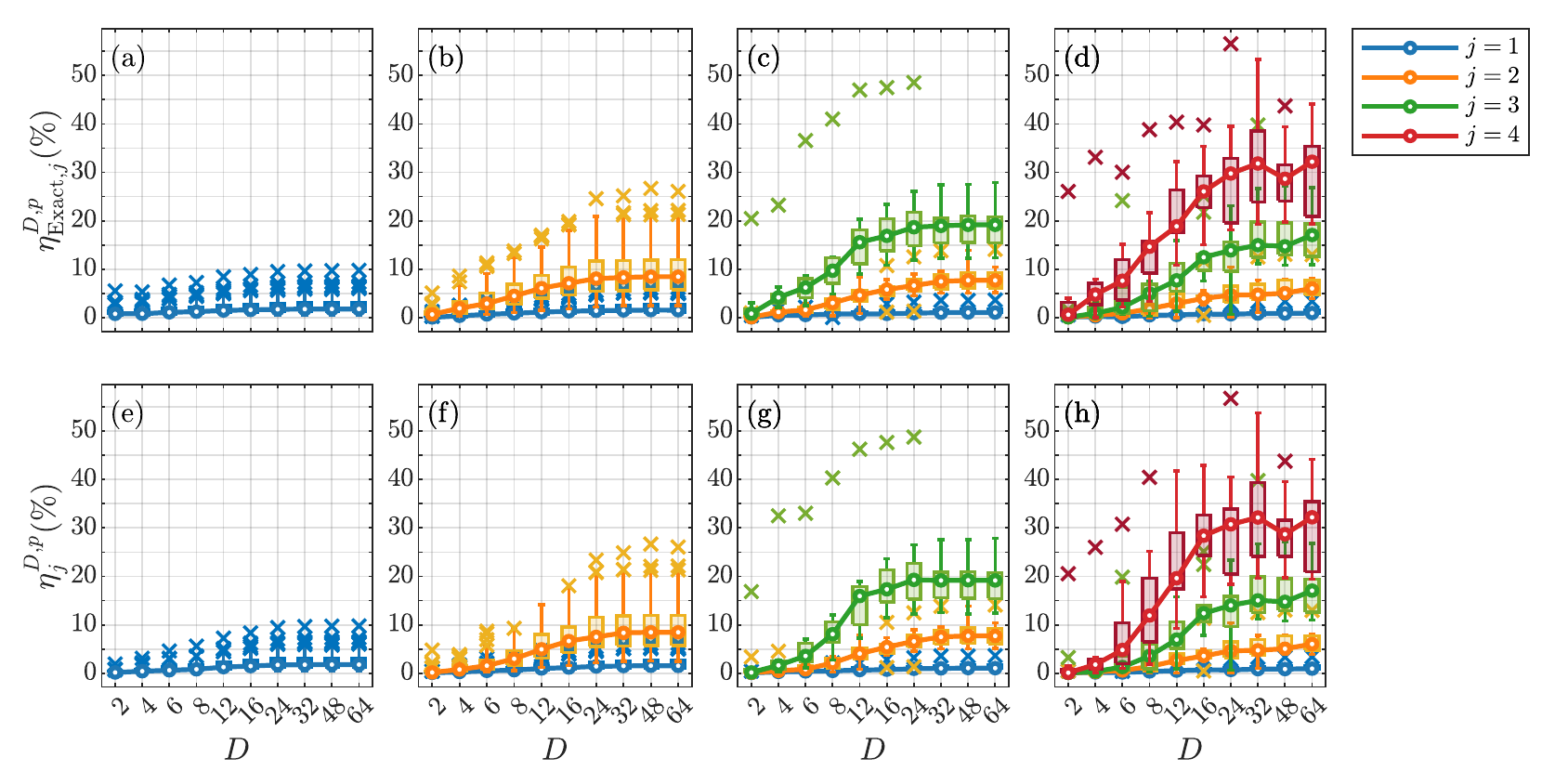}
    \caption{Statistical distribution of the algorithm success percentages using approximated parameters $(\angamma, \anbeta)^{D, p}_{\text{opt};1:j}$ in an exact QAOA simulation---i.e. Eq.~(\ref{eq:non_norm_success_exact})---for depths (a) $p=1$, (b) $p=2$, (c) $p=3$, and (d) $p=4$. Statistical distribution of the success of the approximated algorithm with bond dimension $D$ using approximated parameters from simulations with the same reduced bond dimension $D$---i.e. Eq.~(\ref{eq:non_norm_success_approx})---for depths (e) $p=1$, (f) $p=2$, (g) $p=3$, and (h) $p=4$. We analyze one hundred $12$-node MaxCut problem instances for $p=1$ and $p=2$, and ten similar instances for $p=3$ and $p=4$. For all cases, the exact simulation corresponds to $D=64$. The line connects the median values across all bond dimensions, the boxes are delimited by the lower and upper quartiles, and the bars have endpoints at the minimum and maximum values that are not outliers. The outlier points, that is, values lying more than $1.5$ inter-quartile range away from the box edges are marked as crosses.
    }
    \label{fig:nonnorm_success_percent}
\end{figure*}

\subsection{QAOA performances with approximated training}
\label{ssec:para_calc_success_prob}
Besides analyzing how the approximated optimal circuit parameters---related to a cost function computed with low bond dimensions---deviate from the exact ones with decreasing allowed entanglement, we assess their usefulness by studying the algorithm's success using them.
First, we consider finding the optimal circuit parameters for the quantum algorithm without utilizing a quantum computer~\cite{streif2020training}. Here, the training corresponds to classical circuit simulations with small bond dimensions and the consequent cost function optimization. Thus, to address this case, we analyze the performance of QAOA simulated exactly with the approximated parameters $\angamma_{\text{opt}}^{D,p}$ and $\anbeta_{\text{opt}}^{D,p}$.
Second, we examine the case of a purely classical method with the approximated optimal parameters used in a QAOA simulation with the same reduced bond dimension.

We analyze the algorithm success percentages to describe the performances of the exact and approximated QAOAs with approximated training for the Maxcut problem instances considered previously. Given a $12$-node graph instance, we consider the set of $N$ bitstrings (at least two due to the problem $\mathbb{Z}_2$ symmetry) representing optimal solutions to the MaxCut problem $\{s^k\}$ with $s^k=s^k_1 \dots s^k_{12}$, $s^k_i \in \{0,1\}$, and $1 \leq j \leq N$. Then, to assess the performance of the exact QAOA with a total depth $p$ at the $j$th step, we define the success percentage as
\begin{equation}
\label{eq:non_norm_success_exact}
    \eta^{D,p}_{\text{Exact},j} = \sum_k\abs{\bra{s^k} \ket{ (\angamma, \anbeta)^{D, p}_{\text{opt};1:j} }}^{2} \times 100 \%,
\end{equation}
with $(\angamma, \anbeta)^{D, p}_{\text{opt};1:j}$ the first $j$ components of the approximated optimal angles computed for a depth $p$ and bond dimension $D$, and $\ket{s^k}$ the product state of the optimal solution.

On the other hand, to describe the performance of a complete classical simulation of QAOA with reduced entanglement, we consider the success percentage
\begin{equation}
\label{eq:non_norm_success_approx}
    \eta^{D,p}_j = \sum_k\abs{\bra{s^k} \ket{ (\angamma, \anbeta)^{D, p}_{\text{opt};1:j} }_D}^{2} \times 100 \%,
\end{equation}
with the approximated QAOA state $\ket{ (\angamma, \anbeta)^{D, p}_{\text{opt};1:j} }_D$---properly normalized---computed using approximated angles and with a reduced bond dimension.

We consider the same sets of $12$-qubit instances representing $12$-node Erd\H{o}s–R\'{e}nyi graphs as in the previous analysis. That is, we analyze one hundred instances for $p=1$ and $p=2$, and ten instances for $p=3$ and $p=4$.
In MaxCut, we find at least two bitstrings representing an optimal solution. Hence, for $12$ qubits with at least two solutions out of the total $2^{12}$ computational basis states can be solutions, the success percentage of random sampling would be $2 \times \frac{1}{2^{12}} \times 100\% \approx 0.0488\%$. Fig.~\ref{fig:nonnorm_success_percent} shows the statistical distributions of the success percentages in Eqs.~(\ref{eq:non_norm_success_exact}) and (\ref{eq:non_norm_success_approx}) for every step in QAOAs for MaxCut with circuit depths $1 \leq p \leq 4$. Note that the values obtained for $D = 64$ correspond to a standard simulation of QAOA. We observe that the success percentage increases with the circuit depth $p$ for every bond dimension $D$, and that reducing the bond dimension for the approximated parameters and for the algorithm simulation reduces the performance. Interestingly, the success percentage for intermediate bond dimensions $D\approx 12$ approaches the standard QAOA one, and we observe similar performances considering a QAOA exact simulation with approximated angles than the corresponding QAOA approximated simulation.

\section{Conclusion}
\label{sec:conclusion}

Variational quantum algorithms are heuristic approaches developed to utilize current noisy quantum processors for optimization tasks, among others. Their development relies on understanding their limitations and identifying the fundamental quantum ingredients that may lead to an advantage over classical algorithms. 
Recent works address the role of entanglement in different VQAs~\cite{wiersema2020exploring,diez-valle2021quantum,mcclean2021low-depth,chen2022how-much,dupont2022an-entanglement}, analyzing how much entanglement is generated by them and whether more entanglement is desirable. In this work, we focus on the latter open question and investigate the need for entanglement on the performance of QAOA for MaxCut on Erd\H{o}s–R\'{e}nyi graphs and EC3 problems.
Our simulation restricts the allowed entanglement in QAOA with reduced bond dimensions in MPS representations. Moreover, we extend our analysis to high algorithm depths ($p\leq100$) by utilizing the same set of layer-wise optimized parameters for all instances. Finally, we introduce a deterministic method to sample only one final bitstring of the algorithm. 
Interestingly, we observe that, for depths $p\approx 30$, such a highly restricted simulation of QAOA provides successful results even for small bond dimensions---for instance, bond dimensions $D$ comparable to the system size---with system sizes of $14$, $40$, and $60$ qubits. Even if large-depth QAOAs have an entanglement barrier that limits their classical simulation with MPS-based techniques~\cite{dupont2022an-entanglement,chen2022how-much}, such exact simulation might not be necessary to obtain a solution to the classical problem. In particular, for the $14$-qubit systems, the average fidelity between the approximated states and the exact ones reaches $\Bar{F} \geq 0.9$ for $D \approx 14$. Nevertheless, our simulation with the lowest bond dimension $D=2$---for which the fidelity is close to zero---obtains a solution to the classical problem providing the depth of the circuit is high enough $p\approx 20$, even with a sub-optimal choice of parameters and single deterministic sampling of bitstrings.
Furthermore, we find no contradiction between our findings and a recent study of the QAOA using MPSs~\cite{dupont2022calibrating}. Their numerical analysis of QAOA applied to MaxCut problems on $3$-regular graphs and weighted complete graphs with nodes $n \leq 20$ for depths $p \leq 4$ shows that MPS techniques would require an exponentially scaling bond dimension $D \approx 2^{O(n)}$ to obtain success probabilities comparable to a standard QAOA experiment. While we cannot compare the results directly given the different kinds of graphs and the additional restricted sampling and sub-optimal parameters in our case, our algorithm also fails to solve the problem for up to $60$ qubits with reduced bond dimensions and $p \leq 4$. However, increasing the QAOA layers to $p \geq 30$---with a simple parameter choice---allows our restricted simulation to solve the optimization problem exactly or approximately even with small bond dimensions.
These results motivate further studies of VQAs with tensor network techniques for different kinds of problems to determine whether entanglement provides any advantage for optimization.

Additionally, we consider another possibility of a completely classical training of the QAOA~\cite{streif2020training}. We compute the optimal circuit parameters of QAOA applied to $12$-qubit MaxCut on Erd\H{o}s–R\'{e}nyi graphs with cost landscapes calculated using non-normalized MPSs having reduced bond dimensions for $1 \leq p \leq 4$. We observe that the approximated parameters gradually approach the exact ones ($D=64$) from bond dimensions $D \approx 6$. 
Patterns in these angles with circuit depth $p$ for different instances~\cite{zhou2020quantum} persist for smaller $D$. To assess the viability of using this MPS-based optimization, we analyze the performance of QAOA---simulated exactly---with these parameters. Moreover, we also examine how a fully approximated QAOA simulation with reduced bond dimensions behaves with these corresponding approximated parameters. We observe that, in both cases, for $D \approx 12$ and $1 \leq p \leq 4$, the median success percentage using approximated parameters approaches the success of a standard QAOA with optimal angles, surpassing it for outlier instances.

In conclusion, we observe that entanglement plays a minor role in finding the solution to the classical problems studied here for large-depth QAOA. Since we only explore system sizes of $60$ qubits, more work is needed to determine whether low-entanglement simulations of QAOA can solve optimization problems with larger graphs. Moreover, one could test whether the deterministic sampling method used in this work remains successful in other contexts. Finally, we believe future studies analyzing quantum resources in variational quantum algorithms are crucial for their development and understanding.

The data and source code necessary to reproduce this work are publicly available~\cite{sreedhar2022}.

\begin{acknowledgments}
We acknowledge support from the Knut and Alice Wallenberg Foundation through the Wallenberg Center for Quantum Technology (WACQT).
\end{acknowledgments}

\appendix

\section{Choice of MaxCut and Exact Cover 3 instances}
\label{appx:rand_inst_gen_appx}
In a MaxCut optimization task, given a graph $G$, we aim to find two graph partitions with the most shared edges between them. In this work, we generate random Erd\H{o}s–R\'{e}nyi graphs using the NetworkX package~\cite{hagberg2008exploring} for our MaxCut problem instances. 
The graphs are constructed following the $G(n,w)$ model, with $n$ the number of vertices and $w$ the individual edge probability, so that the expected number of edges is $\tbinom{n}{2}w$. 
We produce $14$, $40$, and $60$ node instances with $w=1/2$, thus exceeding the $1/2$ edge-to-node ratio, below which one finds efficient classical algorithms for solving the problem~\cite{coppersmith2004random}. Each graph can be represented with an adjacency matrix $A$, with elements $A_{ij}=1$ if there is an edge between vertices $i$ and $j$, and $0$ otherwise.
To benchmark QAOA, we need the solutions to these problem instances. For $14$ vertices, we find the solutions by exact diagonalization of the cost Hamiltonian in Eq.~(\ref{eq:mxc_hamiltonian}). For the $40$ and $60$ node instances, we used LocalSolver~\cite{localsolver}.

Another hard problem in classical computation is the EC3 problem, studied in the context of adiabatic quantum computation. In a general Exact Cover problem, we consider a set $A = \{a_1, a_2, \dots a_m \}$ of $m$ elements and a set of $n$ subsets of $A$, $B = \{B_1, B_2, \dots, B_n\}$. The task of finding a cover of $A$ consists in choosing a subset $Y \subseteq \{1, 2, \dots, n \}$ such that $\bigcup_{k \in Y}B_k = A$. Moreover, we require the additional constraint of $B_k \cap B_l = \emptyset$ for $k \neq l \in Y$ for it to be an \emph{exact cover} of $A$. The EC3 problem is a special case of Exact Cover in which each element $a_k \in A$ is only present in three subsets of $B$. 
To create the EC3 problem instances, we define an $n$-bit binary string $x = x_1 x_2 \dots x_n$ such that a subset $B_k$ is included in the exact cover when $x_k=1$, and excluded when $x_k=0$. Let $B_{k_1}, B_{k_2}, B_{k_3}$ be the three subsets containing element $a_k \in A$, and $x_{k_1}, x_{k_2}, x_{k_3}$ their corresponding binary variables. Each element $a_k$ must only be covered once in the solution, and therefore exactly one subset $B_{k_1}, B_{k_2}, B_{k_3}$ can be included in the exact cover---that is, $x_{k_1} + x_{k_2}+ x_{k_3} =1$. Similarly, we find analogous constraints for each of the $m$ elements $a_k \in A$, leading to $m$ clauses that need to be satisfied to find an exact cover. 
In other words, EC3 is a particular case of the 3SAT satisfiability problem that can be formulated with the cost function
\begin{equation}
\label{eq:exc_bincost_appx}
    C(x) = \sum_{k=1}^m \qty[\qty(\sum^{3}_{i = 1} x_{k_i}) -1]^2.
\end{equation}
An exact cover exists if a bitstring $x=x_1 x_2 \dots x_n$ minimizes the cost such that $C(x)=0$. 

We generate EC3 problem instances by considering an incidence matrix $K$ of size $m \times n$, such that $\sum_{j=1}^{n} K_{ij} x_{j} = 1$. The $m$ previous clauses in EC3 are thus represented with a matrix $K$ with exactly three ones per row, which we select randomly per row. To build a satisfiable problem, we add clauses one at a time and use a mixed-integer linear programming solver to find a satisfying assignment to the problem. We keep creating clauses until we cannot find a solution. We retrieve then the satisfiable clauses from the previous step, generating a solvable EC3 problem instance. Using this procedure the number of clauses is not fixed but varies from instance to instance. In Sec.~\ref{ssec:exc_per_results}, we consider EC3 instances with a bitstring solution size $x$ of $14$, $40$, and $60$, corresponding to the number of subsets considered for creating an exact cover.

Encoding the problem into qubits requires replacing the binary variables $x_j$ in Eq.~(\ref{eq:exc_bincost_appx}) by Pauli operators as $x_j\leftarrow (\pauli{z}{j}+1)/2$. This substitution leads to the Ising cost Hamiltonian of Eq.~(\ref{eq:exc_hamiltonian}) in the main text. 
The single-qubit and two-qubit terms can be calculated from the incidence matrix, following an analogous derivation to the one used in a general exact cover problem~\cite{vikstaal2020applying}.

\begin{figure}[htbp]
    \includegraphics[width=0.9\columnwidth]{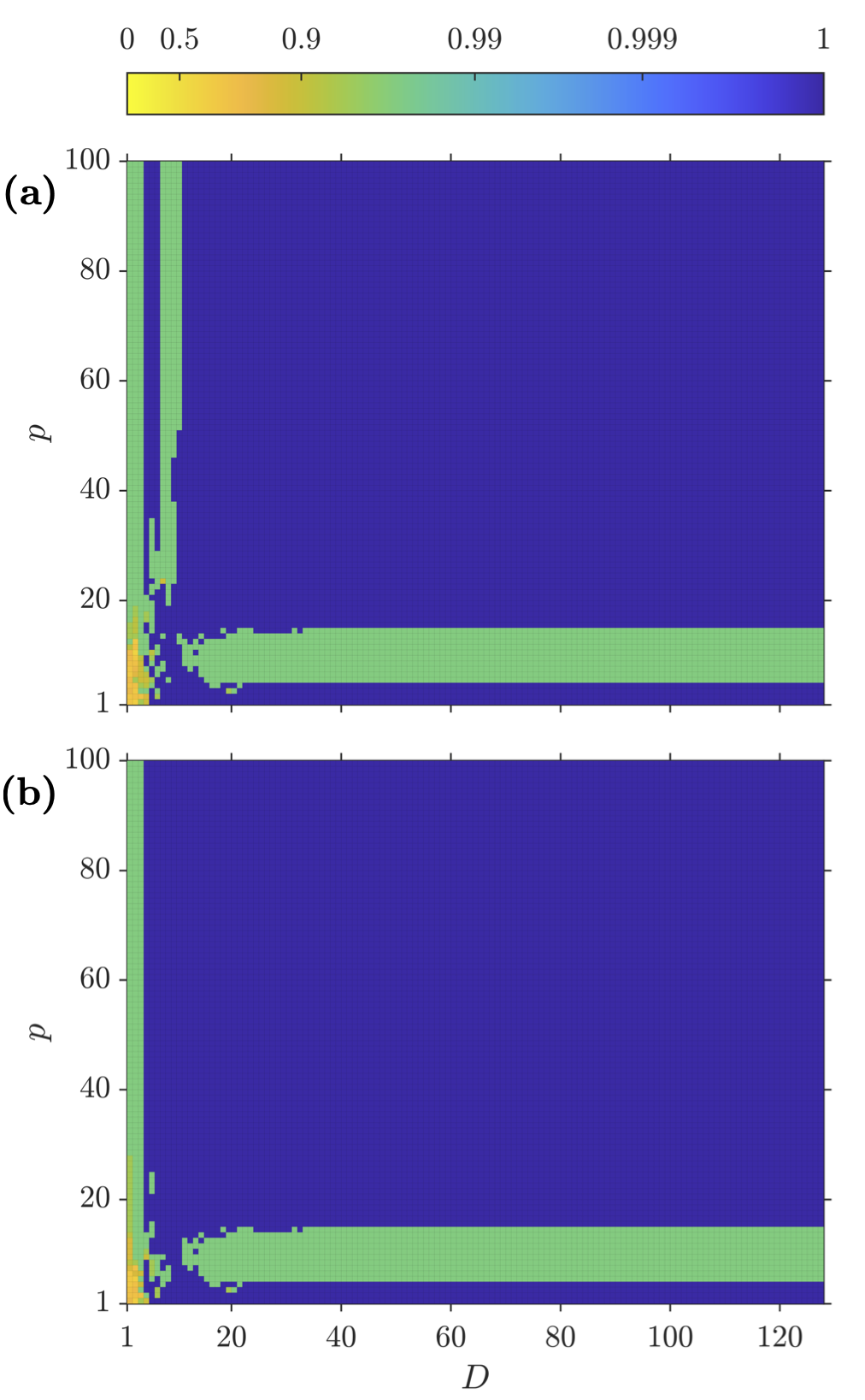}
    \caption{Approximation ratio $r$ of samples obtained by a restricted QAOA simulation---in terms of the bond dimension $D$ and algorithm depth $p$---for a single $14$-qubit MaxCut problem case (instance Q14R90 in repository~\cite{sreedhar2022}) with \textbf{(a)} sub-optimal angles used in the general analysis, and \textbf{(b)} optimized angles for this particular case.}
    \label{fig:ratio_single_mxc}
\end{figure}

\section{Probability of deterministic sample}
\label{appx:counterex_sampling}
The deterministic sequential sampling method outlined in Algorithm~\ref{alg:detsam_method} outputs a single bitstring $s=s_1 \dots s_n$ from an $n$ qubit state $\ket{\psi}$ given in an MPS form. The probability of measuring such configuration $s$ in the computational basis is given by $\abs{\braket{s}{\psi}}^2 \geq 1/2^n$ and may not correspond necessarily to the bitstring with highest probability, that is, $s^* =\arg \max_{x} \abs{\braket{x}{\psi}}^2$, with $x\in\{ 0,1\}^n$. To illustrate this situation, we consider a two-qubit example in which the probability associated with the sample provided by our method is larger than $1/4$, but not the highest. We begin with the two-qubit state decomposition in the computational basis
\begin{equation}
    \ket{\psi} = \sum_{s_{1},s_{2}=0}^{1} c_{s_{1} s_{2}} \ket{s_{1} s_{2}},
\end{equation}
with $\abs{c_{00}}^2=0.32$, $\abs{c_{01}}^2=0.28$, $\abs{c_{10}}^2=0.05$, and $\abs{c_{11}}^2=0.35$. For these specific values, the bitstring most likely to be measured is $s^*=11$. Nevertheless, following Algorithm~\ref{alg:detsam_method}, the probabilities of measuring qubit 1 states $\ket{0_1}$ and $\ket{1_1}$ are $P(0_1)=\abs{c_{00}}^2+\abs{c_{01}^2}$ and $P(1_1)=\abs{c_{10}}^2+\abs{c_{11}}^2$, respectively. Since $P(0_1)=0.60>P(1_1)=0.40$, we update the quantum state $\ket{\psi} \leftarrow \ket{0_1}\braket{0_1}{\psi}/P(0_1)$. Now, the conditional probability of measuring the second qubit states $\ket{0_2}$ and $\ket{1_2}$ given that the first qubit state is $\ket{0_1}$ are $P(0_2)=\abs{c_{00}}^2/P(0_1)$  and $P(1_2)=\abs{c_{01}}^2/P(0_1)$, respectively. Likewise, we compare $P(0_2)\approx 0.53>P(1_2)\approx 0.47$, and the sample retrieved is $s=00 \neq s^*$.

\begin{figure}[htbp]
    \includegraphics[width=0.9\columnwidth]{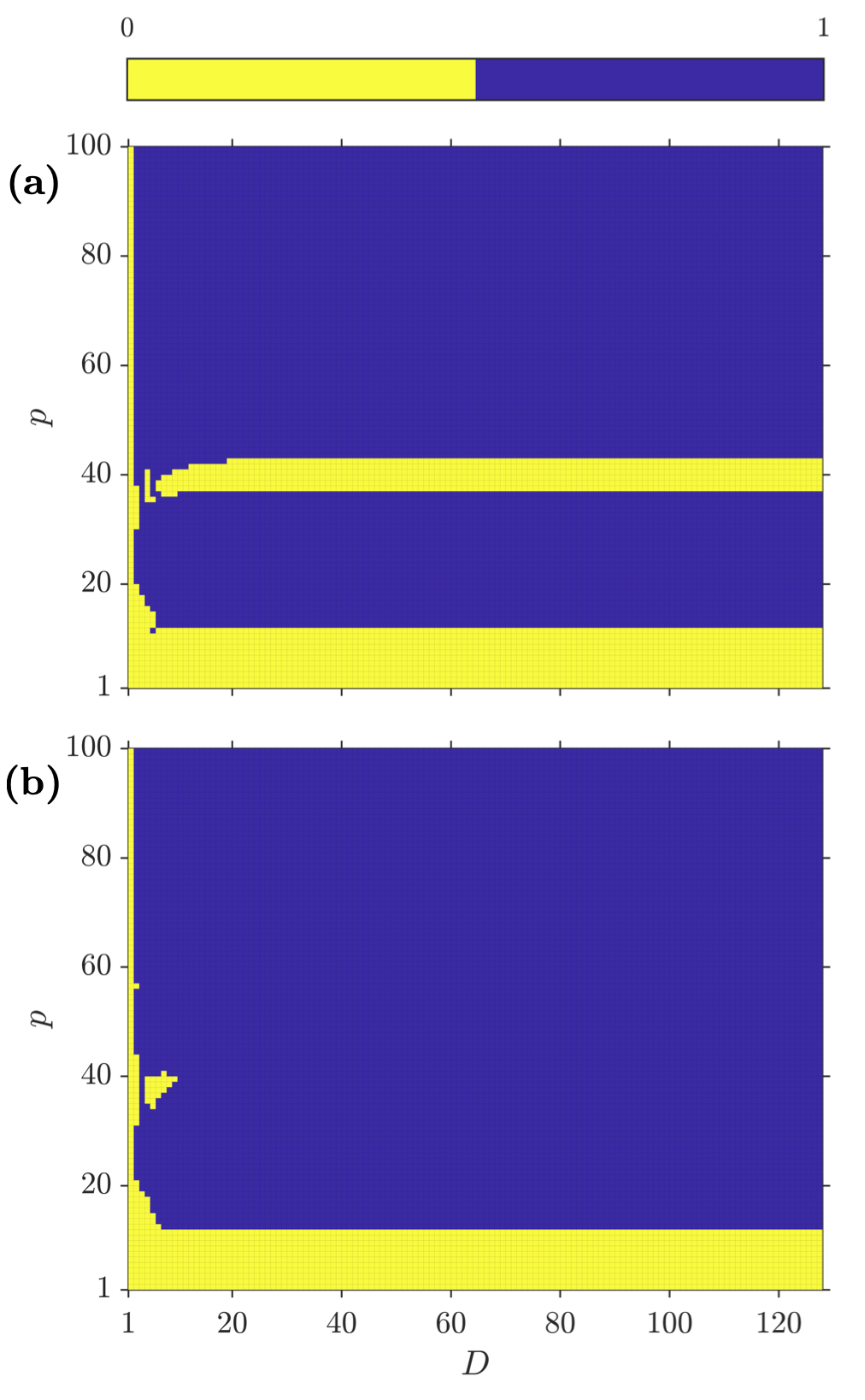}
    \caption{Success $x$ of samples obtained by a restricted QAOA simulation---in terms of the bond dimension $D$ and algorithm depth $p$---for a single $14$-qubit EC3 problem case (instance Q14R71 in repository~\cite{sreedhar2022}) with \textbf{(a)} sub-optimal angles used in the general analysis, and \textbf{(b)} optimized angles for this particular case.
    The algorithm either succeeds in finding an exact cover, $x=1$, or fails $x=0$.}
    \label{fig:success_single_ec3}
\end{figure}

\section{QAOA performances with restricted entanglement for single MaxCut and EC3 instances}
\label{appx:per_single_ins}
For the MaxCut problem, we show in Sec.~\ref{ssec:mxc_per_results} an average approximation ratio $\Bar{r}$ in terms of different bond dimensions $D$ and algorithm depths $p$ considered. Such statistical analysis provides a general view of the approximated QAOA with reduced bond dimension followed by deterministic sampling. However, choosing the same circuit parameters for all instances may affect the success of the simulation. In fact, we observe that the average approximation ratio $\Bar{r}$ decreases for bond dimensions $7 \leq D \leq 10$ in the $14$-qubit problems. We identify the cause of this behavior in the sub-optimal choice of circuit parameters for one single instance, evaluated in Fig.~\ref{fig:ratio_single_mxc}. There we show that the approximation ratio for an optimized choice of angles increases, which in turn solves the performance decay.

To illustrate the analysis for a single EC3 problem instance, we choose a $14$-qubit example. In Fig.~\ref{fig:success_single_ec3}, we observe how the sample obtained from the approximated QAOA simulation with a given bond dimension $D$ and depth $p$ is either an exact cover, $x=1$, or not, $x=0$. In Sec.~\ref{ssec:mxc_per_results}, we show a statistical behavior with the $\Bar{x}$, defined in Eq.~(\ref{eq:ec3_success_rate}). Fig.~\ref{fig:success_single_ec3} presents a narrow band around $p\approx 40$ where the simulated algorithm does not find an exact cover, regardless of the bond dimension, including the exact case $D=128$. We relate this transitory diabatic behavior of the simulated QAOA to the sub-optimal choice of the circuit parameters for this particular instance, disappearing when using better angles.

In both Figs.~\ref{fig:ratio_single_mxc} and \ref{fig:success_single_ec3}, the bond dimension $D = 128$ corresponds to an exact simulation of QAOA in terms of entanglement. In line with the statistical results, these two examples show that reducing the bond to $D \approx 5$ does not affect significantly the simulated algorithm performance.

\begin{figure}[tbp]
    \includegraphics[width=0.9\columnwidth]{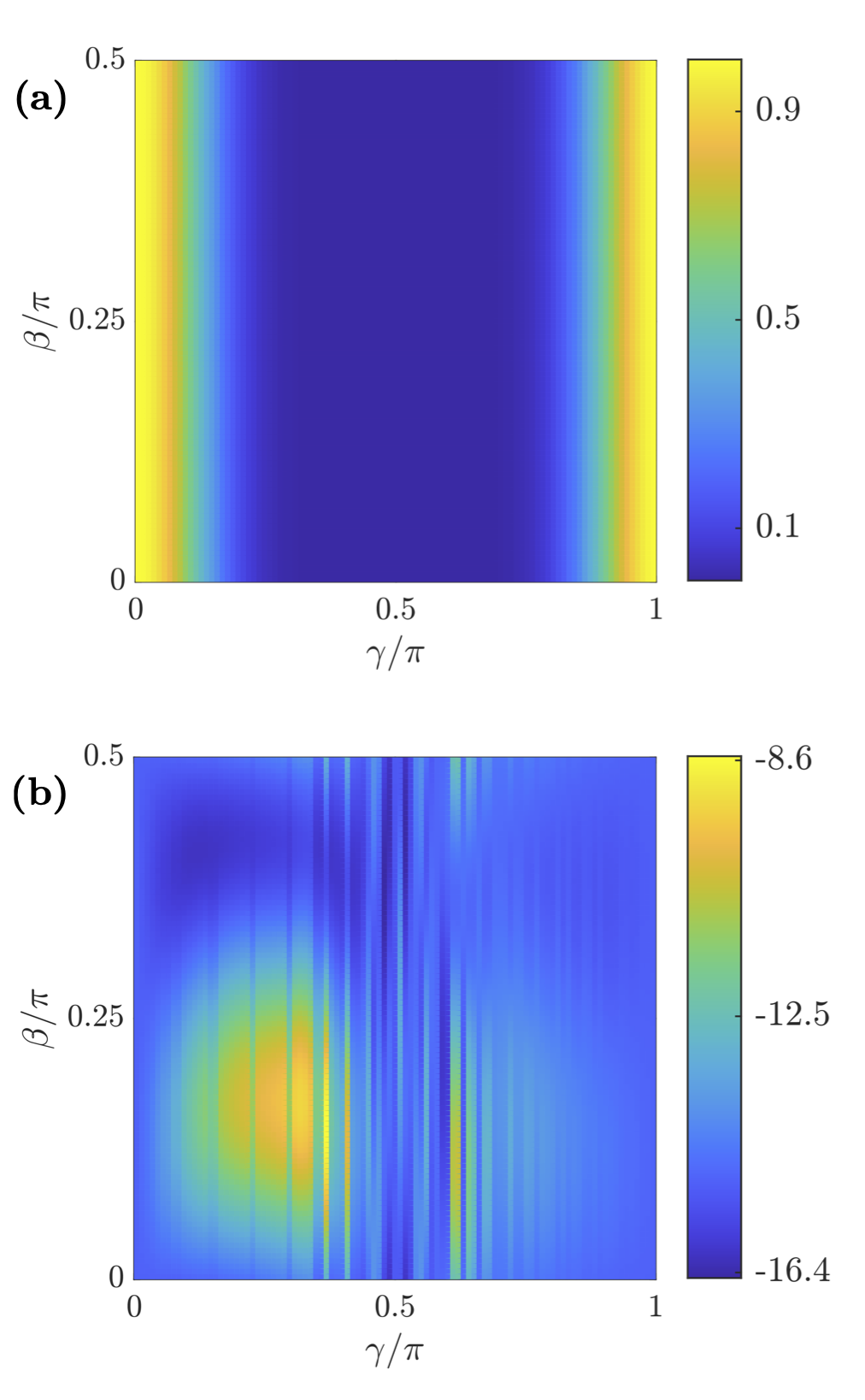}
    \caption{ Norm of the non-normalized MPS $\ket{\gamma,\beta}_D$ with $D=2$, computed with $p=1$ QAOA for a MaxCut $12$-qubit problem, and \textbf{(b)} energy cost landscape $C_{D}(\gamma, \beta)$ for the same $12$-qubit MaxCut problem using normalized MPSs (instance Q12R0D2 in repository~\cite{sreedhar2022}). The energy cost landscape for non-normalized states appears in Fig.~\ref{fig:p1_12R0cost_anglesDist} of the main text.}
    \label{fig:why_non_normalized_training}
\end{figure}

\section{Non-normalized MPS representation of QAOA states for classical training} 
\label{appx:normalization_details}
In Sec.~\ref{sec:QAOA_training}, we explore the possibility of training the QAOA via its classical simulation with MPSs and tractable bond dimensions $D$. We observe that the optimal parameters $\angamma_{\text{opt}}^{D,p}$ and $\anbeta_{\text{opt}}^{D,p}$ obtained from a $p$-depth QAOA simulation with bond dimension $D$ and non-normalized MPSs outperform the ones corresponding to the same simulation with normalized MPSs.

As described in Sec.~\ref{sec:MPS_Simulation}, we perform our simulations right-canonicalized MPS, i.e. the matrices $A^{s_k}$ in Eq.~(\ref{eq:MPSform}) obey $\sum_{s_i} A^{s_i} \left( A^{s_i} \right)^{\dagger} = I$. During the canonicalization of an MPS, we iteratively reshape the matrices $A^{s_k}$ from right to left, such that the right-normalization holds. At the last site, the normalization condition may not hold, as we find a scalar $A^{s_1}$ on the first qubit---our orthogonality center---corresponding to the norm of the state $\ket{\psi}$. During the optimization loop, we keep this scalar to perform calculations with non-normalized states. In general, considering a reduced bond dimension $D$ and hence truncating the smallest Schmidt weights leads to a state norm $\abs{\braket{\psi}}\leq 1$. In our case, the norm of the final states depends on the QAOA parameters related to two-qubit gates and entanglement, $\angamma$, as we show in Fig.~\ref{fig:why_non_normalized_training} for QAOA with $p=1$. We observe that the state norm is $1$ for $\gamma = 0,\pi$ corresponding to unentangled states and decreases towards a minimum for $\gamma = \pi/2$, the most entangled state. Moreover, the norm is independent of the parameter $\beta$, related to single-qubit gates.

To understand the convenience of working with non-normalized states for the optimization loop, one can compare the parameter landscapes of a single MaxCut instance of $12$ nodes for $D = 2$ with non-normalized MPSs and normalized ones, shown in Figs.~\ref{fig:p1_12R0cost_anglesDist} and \ref{fig:why_non_normalized_training}, respectively. The corresponding exact calculation of the cost landscape with bond dimension $D=64$---also in Fig.~\ref{fig:why_non_normalized_training} of the main text---shows that using normalized states with reduced bond dimensions alters the minima positions and leads to noisier surfaces.

\section{Global optimization of approximated cost functions}
\label{appx:bayesian_opt}

In Sec.~\ref{sec:QAOA_training}, we consider global optimization techniques to obtain the optimized circuit parameters, in contrast to the extrapolation method used previously.
For circuit depths $p = 1$, we use grid search to determine the optimal angles $\angamma_{\text{opt}}^{D,p}, \anbeta_{\text{opt}}^{D,p}$ for different bond dimensions $D$. For $p \leq 2$, we use the Bayesian optimization tool, GpyOpt~\cite{gpyopt2016}. Reducing the bond dimension of the MPS representation leads to high-frequency noise in the $\angamma$ parameters, as observed for $p=1$ depths in Fig.~\ref{fig:p1_12R0cost_anglesDist}. We observed worse performances with local gradient-based optimizers for lower bond dimensions and thus higher noise landscapes.

For a successful landscape optimization, we have considered different options in the Bayesian optimization tool. We have obtained the results presented in this work using Gaussian process (GP) as our probabilistic model, the expected improvement (EI) as the acquisition function, and Limited-memory Broyden–Fletcher–Goldfarb–Shanno algorithm (LBFGS) as the optimizer. 
We utilize the Latin Hypercube sampling method to randomly initialize the parameters. Our numerical bottleneck with Bayesian optimization is the number of function evaluations it is allowed to use before finding the minima.

\begin{figure}[tb]
    \includegraphics[width=\linewidth]{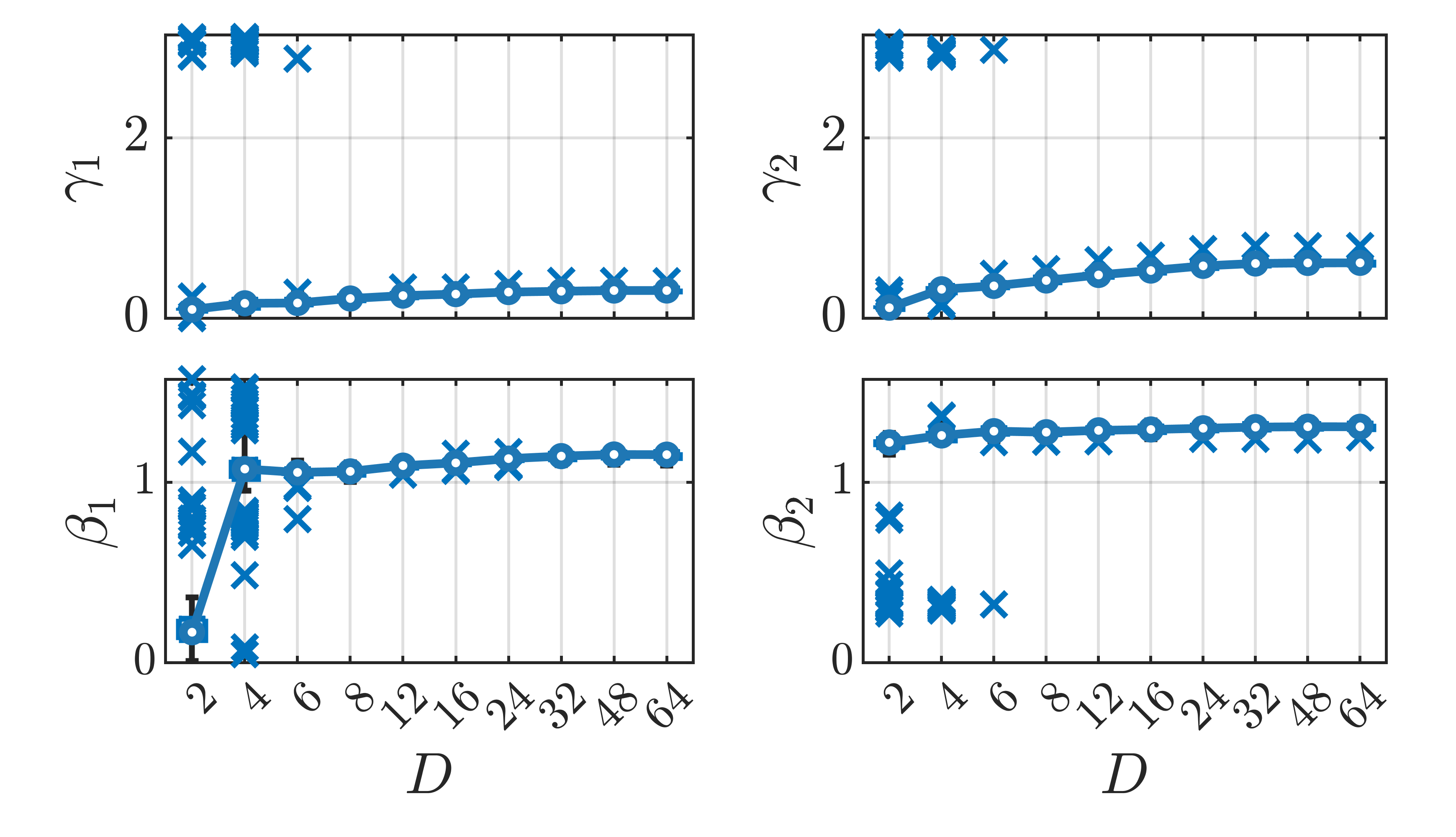}
    \caption{Statistical distribution of the four optimal angles $\angamma_{\text{opt}}^{D}$ and $ \anbeta_{\text{opt}}^{D}$ for one hundred $12$-qubit MaxCut instances and QAOA depth $p=2$. The line connects the median angles for different bond dimensions, the boxes are delimited by the lower and upper quartiles, and the bars have endpoints at the minimum and maximum values that are not outliers. The crosses indicate outlier points lying more than $1.5$ inter-quartile range away from the box edges.}
    \label{fig:Q12P2_opt_par}
\end{figure}

\begin{figure*}[htb]
    \includegraphics[width=0.86\linewidth]{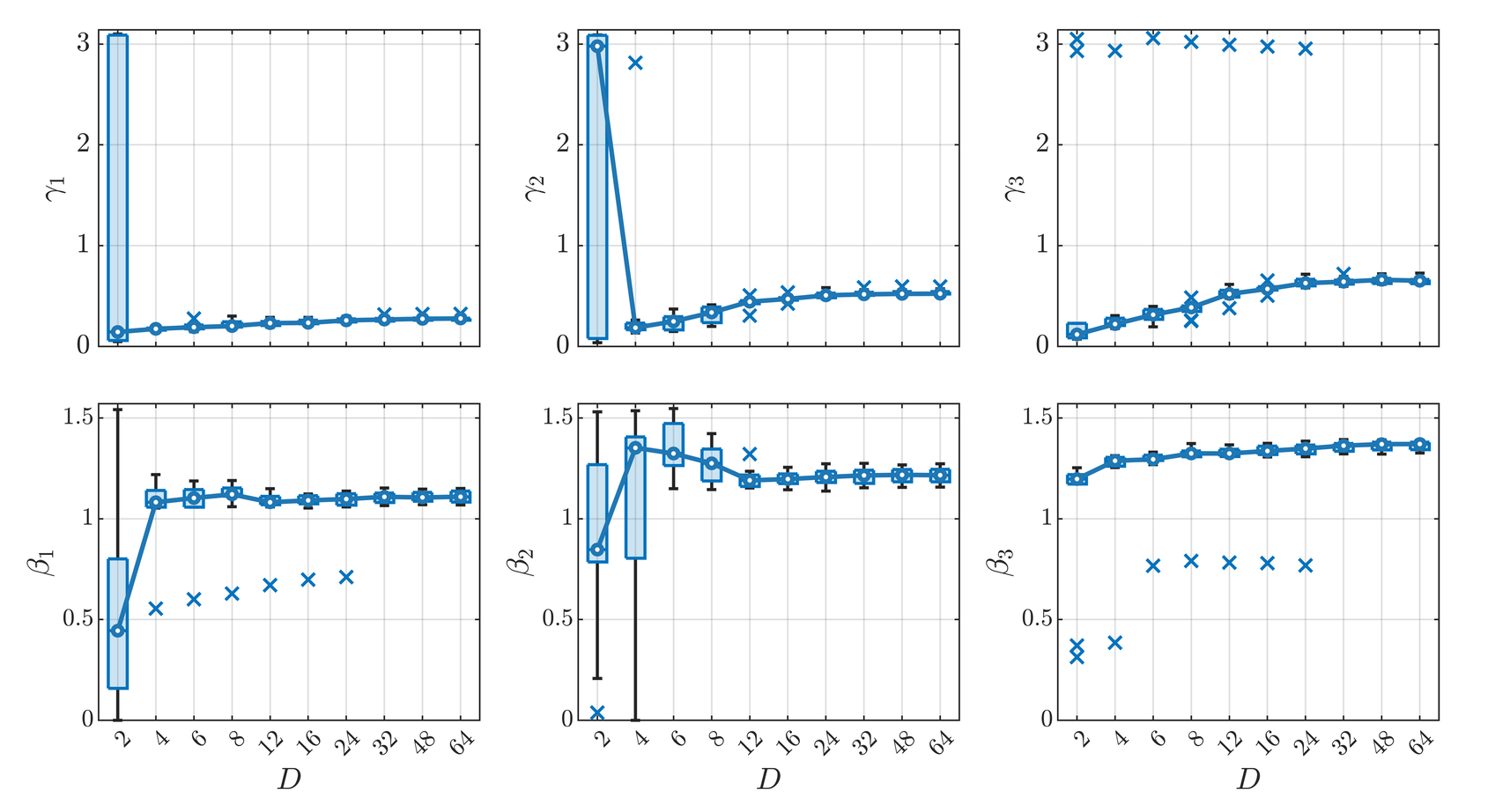}
    \caption{Statistical distribution of the four optimal angles $\angamma_{\text{opt}}^{D}$ and $ \anbeta_{\text{opt}}^{D}$ for ten $12$-qubit MaxCut instances and QAOA depth $p=3$. The line connects the median angles for different bond dimensions, the boxes are delimited by the lower and upper quartiles, and the bars have endpoints at the minimum and maximum values that are not outliers. The crosses indicate outlier points lying more than $1.5$ inter-quartile range away from the box edges.}
    \label{fig:Q12P3_opt_par}
\end{figure*}

For $p = 2$, we used $200$ initialization points and a total of of $500$ functional evaluations (including initialization). Due to the stochastic nature of Bayesian optimization, we repeated the optimization ten times for all the ten $12$-qubit instances and ten bond dimensions, always converging to the same solutions.

We limit the number of function evaluations for $p = 3$ and $p = 4$ depths. The results presented correspond to the parameters obtained with the following pairs of initialization points and total function evaluations in the Bayesian optimization: $(125,300)$, $(150,400)$, $(200,500)$, $(250,600)$, and $(300,700)$. We select the optimal parameters after repeating the calculation $300$ times for each pair.

\section{Training of QAOA with algorithm depths $p=2$ and $p=3$}
\label{appx:angles23}

In Sec.~\ref{sec:QAOA_training}, we consider a purely classical training of the QAOA using MPSs with reduced bond dimensions $D$. Here, we show the results for algorithm depths $p = 2$ and $p = 3$ for MaxCut $12$-qubit problems.

Fig.~\ref{fig:Q12P2_opt_par} shows the optimal angles $\angamma_{\text{opt}}^{D,p = 2}$ and $\anbeta_{\text{opt}}^{D,p = 2}$ variation with different bond dimensions, with $D=64$ the exact calculation. We use one hundred $12$-qubit instances corresponding to $12$-node Erd\H{o}s–R\'{e}nyi graphs with $1/2$ edge probability. We observe a low deviation on the optimal angles for bond dimensions $D \geq 6$, and abrupt changes for smaller bond dimensions.

For $p = 3$, we restrict our study to the first ten $12$-qubit instances of the hundred considered previously due to the numerical bottlenecks in the landscape optimization (see Appendix~\ref{appx:bayesian_opt}).
Fig.~\ref{fig:Q12P3_opt_par} shows the distribution of the optimal angles $\angamma_{\text{opt}}^{D,p=3}$ and $\beta_{\text{opt},2}^{D,p=3}$. Similarly, the optimal values vary abruptly for bond dimensions $D\leq 6$, and remain close to the exact one for higher bond dimensions.

\section{Patterns in approximated parameters}
\label{appx:patterns_parameters}
\begin{figure*}[htbp]
    \centering
    \includegraphics[height=0.74\textheight]{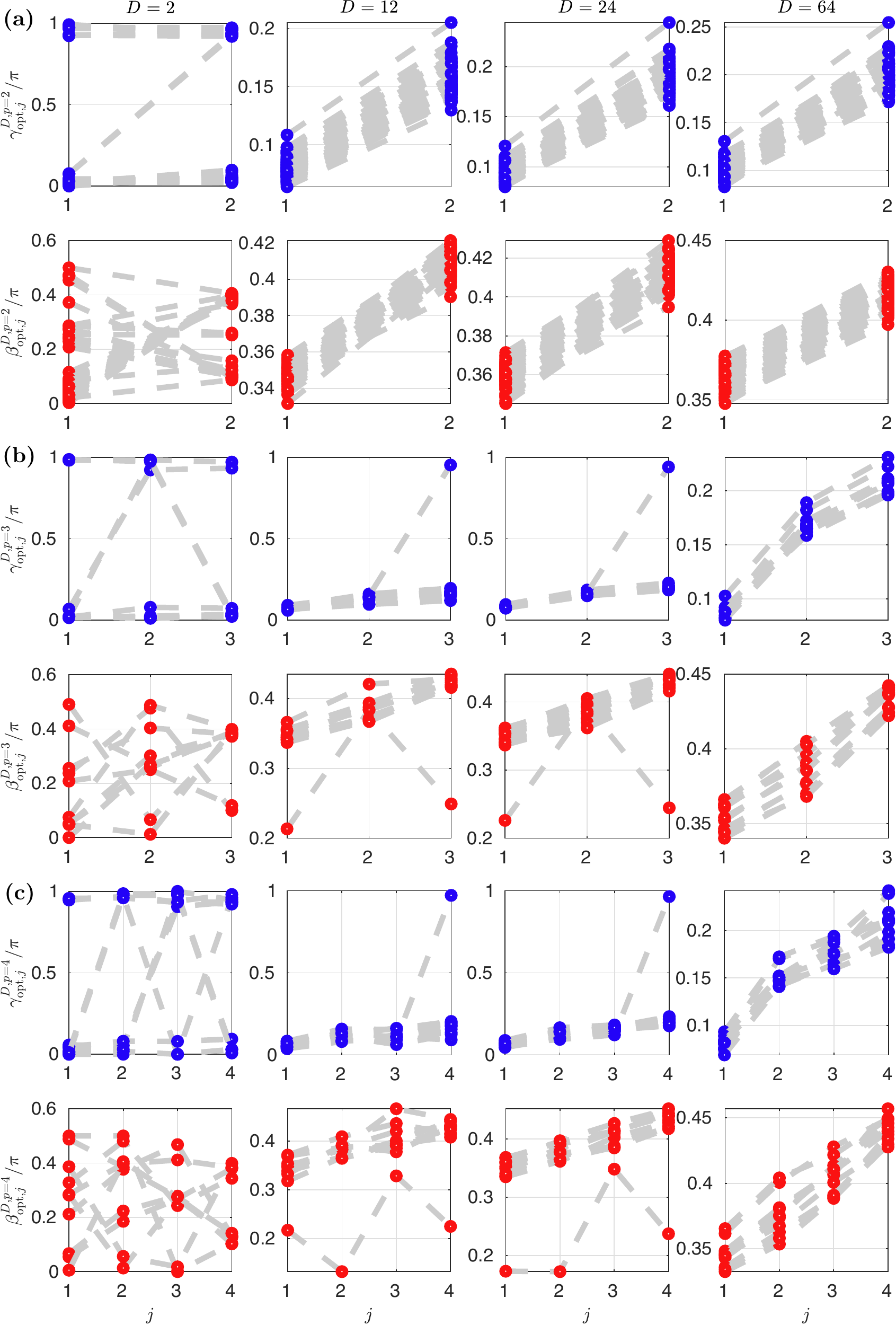}
    \caption{Optimal parameters $\angamma_{\text{opt}}^{D}$ (blue) and $\anbeta_{\text{opt}}^{D}$ (red) for approximated MaxCut cost landscapes computed with reduced bond dimensions. We consider one hundred $12$-node Erd\H{o}s–R\'{e}nyi instances encoded in $12$ qubits for depths $p=2$ and ten instances for $p=3$ and $p=4$. 
    We show the optimal parameter values for each instance connected by lines for bond dimensions $D=2$, $D=12$, $D=24$, and $D=64$ (exact simulation) circuit depths \textbf{(a)} $p=2$, \textbf{(b)} $p=3$, and \textbf{(c)} $p=4$.}
    \label{fig:angles_interp_wrt_dmax}
\end{figure*}

Here, we analyze whether the optimal parameters for an approximated landscape computed with reduced bond dimensions $D$ in Sec.~\ref{sec:QAOA_training} exhibit patterns. Finding trends among the approximated parameters can lead to methods to prepare good initial ans\"{a}tze in QAOA~\cite{zhou2020quantum}. For circuit depths up $2\leq p \leq 4$, Fig.~\ref{fig:angles_interp_wrt_dmax} shows a high dispersion for the approximated optimal parameters with bond dimension $D=2$.
For an intermediate bond dimension $D = 12$---and a $12$-qubit system---the patterns emerge. For $p=2$ and $D \geq 12$, the approximated optimal angles of the hundred instances are close to the exact ones corresponding to $D=64$. For $p = 3$, $p=4$, and $D \geq 12$, the parameters of nine of the ten instances follow a pattern similar to the one corresponding to the exact parameters. The survival of the optimal parameters patterns with different algorithm depths $p$ for small bond dimensions suggests that one may address the quantum circuit training challenge by classically simulating the cost landscape with low entanglement.

\section{Comparison of exact and approximate QAOAs with approximated training with the standard QAOA}
\label{appx:q12_norm_success_prob_appx}
In Eqs.~(\ref{eq:non_norm_success_exact}) and (\ref{eq:non_norm_success_approx}) of Sec.~\ref{ssec:para_calc_success_prob}, we defined the success percentages $\eta^{D,p}_{\text{Exact},j}$ and $\eta^{D,p}_{j}$ of an exact simulation of QAOA using approximated parameters and an approximated QAOA simulation with bond dimension $D$, respectively. In both cases, the approximated parameters $(\angamma, \anbeta)^{D, p}_{\text{opt};1:j}$ correspond to the first $2j$ values optimized with a cost landscape from a circuit simulation of depth $p$ and bond dimension $D$. In in the main text, Fig.~\ref{fig:nonnorm_success_percent} shows that both success percentages approach the standard QAOA one ($D=64$) from intermediate bond dimensions $D\approx 12$.

\begin{figure*}[!hbtp]
    \includegraphics[width=0.95\linewidth]{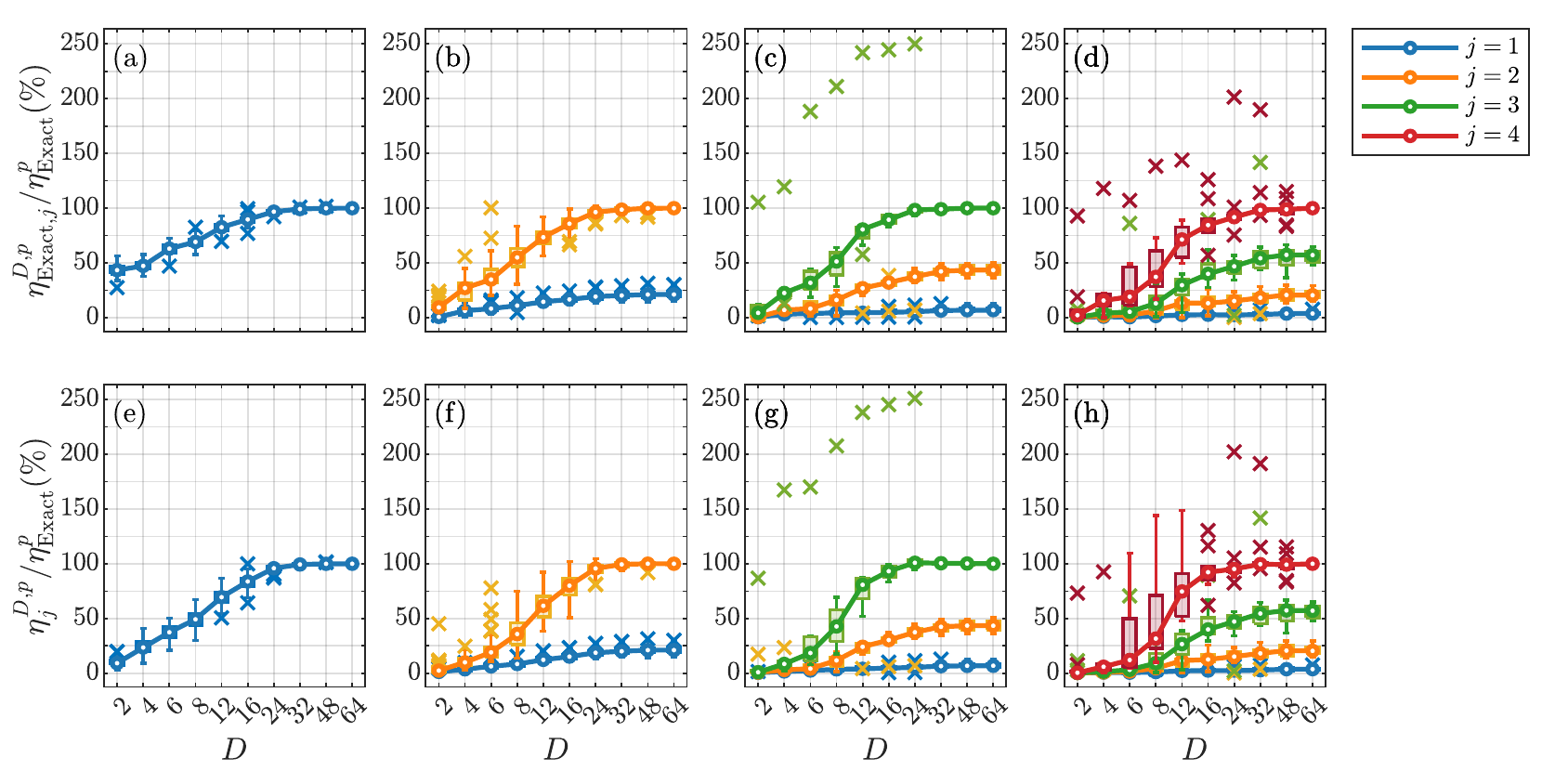}
    \caption{Statistical distribution of the algorithm normalized success percentages using approximated parameters $(\angamma, \anbeta)^{D, p}_{\text{opt};1:j}$ in an exact QAOA simulation---i.e. $\eta^{D,p}_{\text{Exact},j}/\eta^{p}_{\text{Exact}} \times 100 \%$---for depths (a) $p=1$, (b) $p=2$, (c) $p=3$, and (d) $p=4$. Statistical distribution of the normalized success of the approximated algorithm with bond dimension $D$ using approximated parameters from simulations with the same reduced bond dimension $D$---i.e. $\eta^{D,p}_{j}/\eta^{p}_{\text{Exact}} \times 100 \%$---for depths (e) $p=1$, (f) $p=2$, (g) $p=3$, and (h) $p=4$. We analyze one hundred $12$-node MaxCut problem instances for $p=1$ and $p=2$, and ten similar instances for $p=3$ and $p=4$. For all cases, the exact simulation corresponds to $D=64$.
    The line connects the median values across all bond dimensions, the boxes are delimited by the lower and upper quartiles, and the bars have endpoints at the minimum and maximum values that are not outliers. The outlier points, that is, values lying more than $1.5$ inter-quartile range away from the box edges are marked as crosses.}
    \label{fig:norm_success_percent}
\end{figure*}

Here, to easily compare the performance of the approximated cases with the standard algorithm, we consider the normalized success percentages $\eta^{D,p}_{\text{Exact},j}/\eta^{p}_{\text{Exact}} \times 100 \%$  and $\eta^{D,p}_{j}/\eta^{p}_{\text{Exact}} \times 100 \%$, with $\eta^{p}_{\text{Exact}}$ the success for the standard version of QAOA with depth $p$ and $D=64$, with its corresponding optimal parameters. 
Fig.~\ref{fig:norm_success_percent} shows that even with low bond dimensions $D\approx 8$, the approximated cases reach a median performance $\approx 50\%$ as good as the standard QAOA, even improving its results for certain instances with $>100\%$.

\newpage
\thispagestyle{empty}
\newpage

\bibliographystyle{apsrev4-2}
\bibliography{mpsQAOA}

\begin{thebibliography}{61}%
\makeatletter
\providecommand \@ifxundefined [1]{%
 \@ifx{#1\undefined}
}%
\providecommand \@ifnum [1]{%
 \ifnum #1\expandafter \@firstoftwo
 \else \expandafter \@secondoftwo
 \fi
}%
\providecommand \@ifx [1]{%
 \ifx #1\expandafter \@firstoftwo
 \else \expandafter \@secondoftwo
 \fi
}%
\providecommand \natexlab [1]{#1}%
\providecommand \enquote  [1]{``#1''}%
\providecommand \bibnamefont  [1]{#1}%
\providecommand \bibfnamefont [1]{#1}%
\providecommand \citenamefont [1]{#1}%
\providecommand \href@noop [0]{\@secondoftwo}%
\providecommand \href [0]{\begingroup \@sanitize@url \@href}%
\providecommand \@href[1]{\@@startlink{#1}\@@href}%
\providecommand \@@href[1]{\endgroup#1\@@endlink}%
\providecommand \@sanitize@url [0]{\catcode `\\12\catcode `\$12\catcode
  `\&12\catcode `\#12\catcode `\^12\catcode `\_12\catcode `\%12\relax}%
\providecommand \@@startlink[1]{}%
\providecommand \@@endlink[0]{}%
\providecommand \url  [0]{\begingroup\@sanitize@url \@url }%
\providecommand \@url [1]{\endgroup\@href {#1}{\urlprefix }}%
\providecommand \urlprefix  [0]{URL }%
\providecommand \Eprint [0]{\href }%
\providecommand \doibase [0]{https://doi.org/}%
\providecommand \selectlanguage [0]{\@gobble}%
\providecommand \bibinfo  [0]{\@secondoftwo}%
\providecommand \bibfield  [0]{\@secondoftwo}%
\providecommand \translation [1]{[#1]}%
\providecommand \BibitemOpen [0]{}%
\providecommand \bibitemStop [0]{}%
\providecommand \bibitemNoStop [0]{.\EOS\space}%
\providecommand \EOS [0]{\spacefactor3000\relax}%
\providecommand \BibitemShut  [1]{\csname bibitem#1\endcsname}%
\let\auto@bib@innerbib\@empty
\bibitem [{\citenamefont {Preskill}(2018)}]{preskill2018quantum}%
  \BibitemOpen
  \bibfield  {author} {\bibinfo {author} {\bibfnamefont {J.}~\bibnamefont
  {Preskill}},\ }\href {https://doi.org/10.22331/q-2018-08-06-79} {\bibfield
  {journal} {\bibinfo  {journal} {{Quantum}}\ }\textbf {\bibinfo {volume}
  {2}},\ \bibinfo {pages} {79} (\bibinfo {year} {2018})}\BibitemShut {NoStop}%
\bibitem [{\citenamefont {Arute}\ \emph {et~al.}(2019)\citenamefont {Arute},
  \citenamefont {Arya}, \citenamefont {Babbush}, \citenamefont {Bacon},
  \citenamefont {Bardin}, \citenamefont {Barends}, \citenamefont {Biswas},
  \citenamefont {Boixo}, \citenamefont {Brandao}, \citenamefont {Buell},
  \citenamefont {Burkett}, \citenamefont {Chen}, \citenamefont {Chen},
  \citenamefont {Chiaro}, \citenamefont {Collins}, \citenamefont {Courtney},
  \citenamefont {Dunsworth}, \citenamefont {Farhi}, \citenamefont {Foxen},
  \citenamefont {Fowler}, \citenamefont {Gidney}, \citenamefont {Giustina},
  \citenamefont {Graff}, \citenamefont {Guerin}, \citenamefont {Habegger},
  \citenamefont {Harrigan}, \citenamefont {Hartmann}, \citenamefont {Ho},
  \citenamefont {Hoffmann}, \citenamefont {Huang}, \citenamefont {Humble},
  \citenamefont {Isakov}, \citenamefont {Jeffrey}, \citenamefont {Jiang},
  \citenamefont {Kafri}, \citenamefont {Kechedzhi}, \citenamefont {Kelly},
  \citenamefont {Klimov}, \citenamefont {Knysh}, \citenamefont {Korotkov},
  \citenamefont {Kostritsa}, \citenamefont {Landhuis}, \citenamefont
  {Lindmark}, \citenamefont {Lucero}, \citenamefont {Lyakh}, \citenamefont
  {Mandr{\`a}}, \citenamefont {McClean}, \citenamefont {McEwen}, \citenamefont
  {Megrant}, \citenamefont {Mi}, \citenamefont {Michielsen}, \citenamefont
  {Mohseni}, \citenamefont {Mutus}, \citenamefont {Naaman}, \citenamefont
  {Neeley}, \citenamefont {Neill}, \citenamefont {Niu}, \citenamefont {Ostby},
  \citenamefont {Petukhov}, \citenamefont {Platt}, \citenamefont {Quintana},
  \citenamefont {Rieffel}, \citenamefont {Roushan}, \citenamefont {Rubin},
  \citenamefont {Sank}, \citenamefont {Satzinger}, \citenamefont {Smelyanskiy},
  \citenamefont {Sung}, \citenamefont {Trevithick}, \citenamefont
  {Vainsencher}, \citenamefont {Villalonga}, \citenamefont {White},
  \citenamefont {Yao}, \citenamefont {Yeh}, \citenamefont {Zalcman},
  \citenamefont {Neven},\ and\ \citenamefont {Martinis}}]{arute2019quantum}%
  \BibitemOpen
  \bibfield  {author} {\bibinfo {author} {\bibfnamefont {F.}~\bibnamefont
  {Arute}}, \bibinfo {author} {\bibfnamefont {K.}~\bibnamefont {Arya}},
  \bibinfo {author} {\bibfnamefont {R.}~\bibnamefont {Babbush}}, \bibinfo
  {author} {\bibfnamefont {D.}~\bibnamefont {Bacon}}, \bibinfo {author}
  {\bibfnamefont {J.~C.}\ \bibnamefont {Bardin}}, \bibinfo {author}
  {\bibfnamefont {R.}~\bibnamefont {Barends}}, \bibinfo {author} {\bibfnamefont
  {R.}~\bibnamefont {Biswas}}, \bibinfo {author} {\bibfnamefont
  {S.}~\bibnamefont {Boixo}}, \bibinfo {author} {\bibfnamefont {F.~G. S.~L.}\
  \bibnamefont {Brandao}}, \bibinfo {author} {\bibfnamefont {D.~A.}\
  \bibnamefont {Buell}}, \bibinfo {author} {\bibfnamefont {B.}~\bibnamefont
  {Burkett}}, \bibinfo {author} {\bibfnamefont {Y.}~\bibnamefont {Chen}},
  \bibinfo {author} {\bibfnamefont {Z.}~\bibnamefont {Chen}}, \bibinfo {author}
  {\bibfnamefont {B.}~\bibnamefont {Chiaro}}, \bibinfo {author} {\bibfnamefont
  {R.}~\bibnamefont {Collins}}, \bibinfo {author} {\bibfnamefont
  {W.}~\bibnamefont {Courtney}}, \bibinfo {author} {\bibfnamefont
  {A.}~\bibnamefont {Dunsworth}}, \bibinfo {author} {\bibfnamefont
  {E.}~\bibnamefont {Farhi}}, \bibinfo {author} {\bibfnamefont
  {B.}~\bibnamefont {Foxen}}, \bibinfo {author} {\bibfnamefont
  {A.}~\bibnamefont {Fowler}}, \bibinfo {author} {\bibfnamefont
  {C.}~\bibnamefont {Gidney}}, \bibinfo {author} {\bibfnamefont
  {M.}~\bibnamefont {Giustina}}, \bibinfo {author} {\bibfnamefont
  {R.}~\bibnamefont {Graff}}, \bibinfo {author} {\bibfnamefont
  {K.}~\bibnamefont {Guerin}}, \bibinfo {author} {\bibfnamefont
  {S.}~\bibnamefont {Habegger}}, \bibinfo {author} {\bibfnamefont {M.~P.}\
  \bibnamefont {Harrigan}}, \bibinfo {author} {\bibfnamefont {M.~J.}\
  \bibnamefont {Hartmann}}, \bibinfo {author} {\bibfnamefont {A.}~\bibnamefont
  {Ho}}, \bibinfo {author} {\bibfnamefont {M.}~\bibnamefont {Hoffmann}},
  \bibinfo {author} {\bibfnamefont {T.}~\bibnamefont {Huang}}, \bibinfo
  {author} {\bibfnamefont {T.~S.}\ \bibnamefont {Humble}}, \bibinfo {author}
  {\bibfnamefont {S.~V.}\ \bibnamefont {Isakov}}, \bibinfo {author}
  {\bibfnamefont {E.}~\bibnamefont {Jeffrey}}, \bibinfo {author} {\bibfnamefont
  {Z.}~\bibnamefont {Jiang}}, \bibinfo {author} {\bibfnamefont
  {D.}~\bibnamefont {Kafri}}, \bibinfo {author} {\bibfnamefont
  {K.}~\bibnamefont {Kechedzhi}}, \bibinfo {author} {\bibfnamefont
  {J.}~\bibnamefont {Kelly}}, \bibinfo {author} {\bibfnamefont {P.~V.}\
  \bibnamefont {Klimov}}, \bibinfo {author} {\bibfnamefont {S.}~\bibnamefont
  {Knysh}}, \bibinfo {author} {\bibfnamefont {A.}~\bibnamefont {Korotkov}},
  \bibinfo {author} {\bibfnamefont {F.}~\bibnamefont {Kostritsa}}, \bibinfo
  {author} {\bibfnamefont {D.}~\bibnamefont {Landhuis}}, \bibinfo {author}
  {\bibfnamefont {M.}~\bibnamefont {Lindmark}}, \bibinfo {author}
  {\bibfnamefont {E.}~\bibnamefont {Lucero}}, \bibinfo {author} {\bibfnamefont
  {D.}~\bibnamefont {Lyakh}}, \bibinfo {author} {\bibfnamefont
  {S.}~\bibnamefont {Mandr{\`a}}}, \bibinfo {author} {\bibfnamefont {J.~R.}\
  \bibnamefont {McClean}}, \bibinfo {author} {\bibfnamefont {M.}~\bibnamefont
  {McEwen}}, \bibinfo {author} {\bibfnamefont {A.}~\bibnamefont {Megrant}},
  \bibinfo {author} {\bibfnamefont {X.}~\bibnamefont {Mi}}, \bibinfo {author}
  {\bibfnamefont {K.}~\bibnamefont {Michielsen}}, \bibinfo {author}
  {\bibfnamefont {M.}~\bibnamefont {Mohseni}}, \bibinfo {author} {\bibfnamefont
  {J.}~\bibnamefont {Mutus}}, \bibinfo {author} {\bibfnamefont
  {O.}~\bibnamefont {Naaman}}, \bibinfo {author} {\bibfnamefont
  {M.}~\bibnamefont {Neeley}}, \bibinfo {author} {\bibfnamefont
  {C.}~\bibnamefont {Neill}}, \bibinfo {author} {\bibfnamefont {M.~Y.}\
  \bibnamefont {Niu}}, \bibinfo {author} {\bibfnamefont {E.}~\bibnamefont
  {Ostby}}, \bibinfo {author} {\bibfnamefont {A.}~\bibnamefont {Petukhov}},
  \bibinfo {author} {\bibfnamefont {J.~C.}\ \bibnamefont {Platt}}, \bibinfo
  {author} {\bibfnamefont {C.}~\bibnamefont {Quintana}}, \bibinfo {author}
  {\bibfnamefont {E.~G.}\ \bibnamefont {Rieffel}}, \bibinfo {author}
  {\bibfnamefont {P.}~\bibnamefont {Roushan}}, \bibinfo {author} {\bibfnamefont
  {N.~C.}\ \bibnamefont {Rubin}}, \bibinfo {author} {\bibfnamefont
  {D.}~\bibnamefont {Sank}}, \bibinfo {author} {\bibfnamefont {K.~J.}\
  \bibnamefont {Satzinger}}, \bibinfo {author} {\bibfnamefont {V.}~\bibnamefont
  {Smelyanskiy}}, \bibinfo {author} {\bibfnamefont {K.~J.}\ \bibnamefont
  {Sung}}, \bibinfo {author} {\bibfnamefont {M.~D.}\ \bibnamefont
  {Trevithick}}, \bibinfo {author} {\bibfnamefont {A.}~\bibnamefont
  {Vainsencher}}, \bibinfo {author} {\bibfnamefont {B.}~\bibnamefont
  {Villalonga}}, \bibinfo {author} {\bibfnamefont {T.}~\bibnamefont {White}},
  \bibinfo {author} {\bibfnamefont {Z.~J.}\ \bibnamefont {Yao}}, \bibinfo
  {author} {\bibfnamefont {P.}~\bibnamefont {Yeh}}, \bibinfo {author}
  {\bibfnamefont {A.}~\bibnamefont {Zalcman}}, \bibinfo {author} {\bibfnamefont
  {H.}~\bibnamefont {Neven}},\ and\ \bibinfo {author} {\bibfnamefont {J.~M.}\
  \bibnamefont {Martinis}},\ }\href {https://doi.org/10.1038/s41586-019-1666-5}
  {\bibfield  {journal} {\bibinfo  {journal} {Nature}\ }\textbf {\bibinfo
  {volume} {574}},\ \bibinfo {pages} {505} (\bibinfo {year}
  {2019})}\BibitemShut {NoStop}%
\bibitem [{\citenamefont {Zhong}\ \emph {et~al.}(2020)\citenamefont {Zhong},
  \citenamefont {Wang}, \citenamefont {Deng}, \citenamefont {Chen},
  \citenamefont {Peng}, \citenamefont {Luo}, \citenamefont {Qin}, \citenamefont
  {Wu}, \citenamefont {Ding}, \citenamefont {Hu}, \citenamefont {Hu},
  \citenamefont {Yang}, \citenamefont {Zhang}, \citenamefont {Li},
  \citenamefont {Li}, \citenamefont {Jiang}, \citenamefont {Gan}, \citenamefont
  {Yang}, \citenamefont {You}, \citenamefont {Wang}, \citenamefont {Li},
  \citenamefont {Liu}, \citenamefont {Lu},\ and\ \citenamefont
  {Pan}}]{zhong2020quantum}%
  \BibitemOpen
  \bibfield  {author} {\bibinfo {author} {\bibfnamefont {H.-S.}\ \bibnamefont
  {Zhong}}, \bibinfo {author} {\bibfnamefont {H.}~\bibnamefont {Wang}},
  \bibinfo {author} {\bibfnamefont {Y.-H.}\ \bibnamefont {Deng}}, \bibinfo
  {author} {\bibfnamefont {M.-C.}\ \bibnamefont {Chen}}, \bibinfo {author}
  {\bibfnamefont {L.-C.}\ \bibnamefont {Peng}}, \bibinfo {author}
  {\bibfnamefont {Y.-H.}\ \bibnamefont {Luo}}, \bibinfo {author} {\bibfnamefont
  {J.}~\bibnamefont {Qin}}, \bibinfo {author} {\bibfnamefont {D.}~\bibnamefont
  {Wu}}, \bibinfo {author} {\bibfnamefont {X.}~\bibnamefont {Ding}}, \bibinfo
  {author} {\bibfnamefont {Y.}~\bibnamefont {Hu}}, \bibinfo {author}
  {\bibfnamefont {P.}~\bibnamefont {Hu}}, \bibinfo {author} {\bibfnamefont
  {X.-Y.}\ \bibnamefont {Yang}}, \bibinfo {author} {\bibfnamefont {W.-J.}\
  \bibnamefont {Zhang}}, \bibinfo {author} {\bibfnamefont {H.}~\bibnamefont
  {Li}}, \bibinfo {author} {\bibfnamefont {Y.}~\bibnamefont {Li}}, \bibinfo
  {author} {\bibfnamefont {X.}~\bibnamefont {Jiang}}, \bibinfo {author}
  {\bibfnamefont {L.}~\bibnamefont {Gan}}, \bibinfo {author} {\bibfnamefont
  {G.}~\bibnamefont {Yang}}, \bibinfo {author} {\bibfnamefont {L.}~\bibnamefont
  {You}}, \bibinfo {author} {\bibfnamefont {Z.}~\bibnamefont {Wang}}, \bibinfo
  {author} {\bibfnamefont {L.}~\bibnamefont {Li}}, \bibinfo {author}
  {\bibfnamefont {N.-L.}\ \bibnamefont {Liu}}, \bibinfo {author} {\bibfnamefont
  {C.-Y.}\ \bibnamefont {Lu}},\ and\ \bibinfo {author} {\bibfnamefont {J.-W.}\
  \bibnamefont {Pan}},\ }\href {https://doi.org/10.1126/science.abe8770}
  {\bibfield  {journal} {\bibinfo  {journal} {Science}\ }\textbf {\bibinfo
  {volume} {370}},\ \bibinfo {pages} {1460} (\bibinfo {year}
  {2020})}\BibitemShut {NoStop}%
\bibitem [{\citenamefont {Zhong}\ \emph {et~al.}(2021)\citenamefont {Zhong},
  \citenamefont {Deng}, \citenamefont {Qin}, \citenamefont {Wang},
  \citenamefont {Chen}, \citenamefont {Peng}, \citenamefont {Luo},
  \citenamefont {Wu}, \citenamefont {Gong}, \citenamefont {Su}, \citenamefont
  {Hu}, \citenamefont {Hu}, \citenamefont {Yang}, \citenamefont {Zhang},
  \citenamefont {Li}, \citenamefont {Li}, \citenamefont {Jiang}, \citenamefont
  {Gan}, \citenamefont {Yang}, \citenamefont {You}, \citenamefont {Wang},
  \citenamefont {Li}, \citenamefont {Liu}, \citenamefont {Renema},
  \citenamefont {Lu},\ and\ \citenamefont {Pan}}]{zhong2021phase-programmable}%
  \BibitemOpen
  \bibfield  {author} {\bibinfo {author} {\bibfnamefont {H.-S.}\ \bibnamefont
  {Zhong}}, \bibinfo {author} {\bibfnamefont {Y.-H.}\ \bibnamefont {Deng}},
  \bibinfo {author} {\bibfnamefont {J.}~\bibnamefont {Qin}}, \bibinfo {author}
  {\bibfnamefont {H.}~\bibnamefont {Wang}}, \bibinfo {author} {\bibfnamefont
  {M.-C.}\ \bibnamefont {Chen}}, \bibinfo {author} {\bibfnamefont {L.-C.}\
  \bibnamefont {Peng}}, \bibinfo {author} {\bibfnamefont {Y.-H.}\ \bibnamefont
  {Luo}}, \bibinfo {author} {\bibfnamefont {D.}~\bibnamefont {Wu}}, \bibinfo
  {author} {\bibfnamefont {S.-Q.}\ \bibnamefont {Gong}}, \bibinfo {author}
  {\bibfnamefont {H.}~\bibnamefont {Su}}, \bibinfo {author} {\bibfnamefont
  {Y.}~\bibnamefont {Hu}}, \bibinfo {author} {\bibfnamefont {P.}~\bibnamefont
  {Hu}}, \bibinfo {author} {\bibfnamefont {X.-Y.}\ \bibnamefont {Yang}},
  \bibinfo {author} {\bibfnamefont {W.-J.}\ \bibnamefont {Zhang}}, \bibinfo
  {author} {\bibfnamefont {H.}~\bibnamefont {Li}}, \bibinfo {author}
  {\bibfnamefont {Y.}~\bibnamefont {Li}}, \bibinfo {author} {\bibfnamefont
  {X.}~\bibnamefont {Jiang}}, \bibinfo {author} {\bibfnamefont
  {L.}~\bibnamefont {Gan}}, \bibinfo {author} {\bibfnamefont {G.}~\bibnamefont
  {Yang}}, \bibinfo {author} {\bibfnamefont {L.}~\bibnamefont {You}}, \bibinfo
  {author} {\bibfnamefont {Z.}~\bibnamefont {Wang}}, \bibinfo {author}
  {\bibfnamefont {L.}~\bibnamefont {Li}}, \bibinfo {author} {\bibfnamefont
  {N.-L.}\ \bibnamefont {Liu}}, \bibinfo {author} {\bibfnamefont {J.~J.}\
  \bibnamefont {Renema}}, \bibinfo {author} {\bibfnamefont {C.-Y.}\
  \bibnamefont {Lu}},\ and\ \bibinfo {author} {\bibfnamefont {J.-W.}\
  \bibnamefont {Pan}},\ }\href {https://doi.org/10.1103/PhysRevLett.127.180502}
  {\bibfield  {journal} {\bibinfo  {journal} {Phys. Rev. Lett.}\ }\textbf
  {\bibinfo {volume} {127}},\ \bibinfo {pages} {180502} (\bibinfo {year}
  {2021})}\BibitemShut {NoStop}%
\bibitem [{\citenamefont {Wu}\ \emph {et~al.}(2021)\citenamefont {Wu},
  \citenamefont {Bao}, \citenamefont {Cao}, \citenamefont {Chen}, \citenamefont
  {Chen}, \citenamefont {Chen}, \citenamefont {Chung}, \citenamefont {Deng},
  \citenamefont {Du}, \citenamefont {Fan}, \citenamefont {Gong}, \citenamefont
  {Guo}, \citenamefont {Guo}, \citenamefont {Guo}, \citenamefont {Han},
  \citenamefont {Hong}, \citenamefont {Huang}, \citenamefont {Huo},
  \citenamefont {Li}, \citenamefont {Li}, \citenamefont {Li}, \citenamefont
  {Li}, \citenamefont {Liang}, \citenamefont {Lin}, \citenamefont {Lin},
  \citenamefont {Qian}, \citenamefont {Qiao}, \citenamefont {Rong},
  \citenamefont {Su}, \citenamefont {Sun}, \citenamefont {Wang}, \citenamefont
  {Wang}, \citenamefont {Wu}, \citenamefont {Xu}, \citenamefont {Yan},
  \citenamefont {Yang}, \citenamefont {Yang}, \citenamefont {Ye}, \citenamefont
  {Yin}, \citenamefont {Ying}, \citenamefont {Yu}, \citenamefont {Zha},
  \citenamefont {Zhang}, \citenamefont {Zhang}, \citenamefont {Zhang},
  \citenamefont {Zhang}, \citenamefont {Zhao}, \citenamefont {Zhao},
  \citenamefont {Zhou}, \citenamefont {Zhu}, \citenamefont {Lu}, \citenamefont
  {Peng}, \citenamefont {Zhu},\ and\ \citenamefont {Pan}}]{wu2021strong}%
  \BibitemOpen
  \bibfield  {author} {\bibinfo {author} {\bibfnamefont {Y.}~\bibnamefont
  {Wu}}, \bibinfo {author} {\bibfnamefont {W.-S.}\ \bibnamefont {Bao}},
  \bibinfo {author} {\bibfnamefont {S.}~\bibnamefont {Cao}}, \bibinfo {author}
  {\bibfnamefont {F.}~\bibnamefont {Chen}}, \bibinfo {author} {\bibfnamefont
  {M.-C.}\ \bibnamefont {Chen}}, \bibinfo {author} {\bibfnamefont
  {X.}~\bibnamefont {Chen}}, \bibinfo {author} {\bibfnamefont {T.-H.}\
  \bibnamefont {Chung}}, \bibinfo {author} {\bibfnamefont {H.}~\bibnamefont
  {Deng}}, \bibinfo {author} {\bibfnamefont {Y.}~\bibnamefont {Du}}, \bibinfo
  {author} {\bibfnamefont {D.}~\bibnamefont {Fan}}, \bibinfo {author}
  {\bibfnamefont {M.}~\bibnamefont {Gong}}, \bibinfo {author} {\bibfnamefont
  {C.}~\bibnamefont {Guo}}, \bibinfo {author} {\bibfnamefont {C.}~\bibnamefont
  {Guo}}, \bibinfo {author} {\bibfnamefont {S.}~\bibnamefont {Guo}}, \bibinfo
  {author} {\bibfnamefont {L.}~\bibnamefont {Han}}, \bibinfo {author}
  {\bibfnamefont {L.}~\bibnamefont {Hong}}, \bibinfo {author} {\bibfnamefont
  {H.-L.}\ \bibnamefont {Huang}}, \bibinfo {author} {\bibfnamefont {Y.-H.}\
  \bibnamefont {Huo}}, \bibinfo {author} {\bibfnamefont {L.}~\bibnamefont
  {Li}}, \bibinfo {author} {\bibfnamefont {N.}~\bibnamefont {Li}}, \bibinfo
  {author} {\bibfnamefont {S.}~\bibnamefont {Li}}, \bibinfo {author}
  {\bibfnamefont {Y.}~\bibnamefont {Li}}, \bibinfo {author} {\bibfnamefont
  {F.}~\bibnamefont {Liang}}, \bibinfo {author} {\bibfnamefont
  {C.}~\bibnamefont {Lin}}, \bibinfo {author} {\bibfnamefont {J.}~\bibnamefont
  {Lin}}, \bibinfo {author} {\bibfnamefont {H.}~\bibnamefont {Qian}}, \bibinfo
  {author} {\bibfnamefont {D.}~\bibnamefont {Qiao}}, \bibinfo {author}
  {\bibfnamefont {H.}~\bibnamefont {Rong}}, \bibinfo {author} {\bibfnamefont
  {H.}~\bibnamefont {Su}}, \bibinfo {author} {\bibfnamefont {L.}~\bibnamefont
  {Sun}}, \bibinfo {author} {\bibfnamefont {L.}~\bibnamefont {Wang}}, \bibinfo
  {author} {\bibfnamefont {S.}~\bibnamefont {Wang}}, \bibinfo {author}
  {\bibfnamefont {D.}~\bibnamefont {Wu}}, \bibinfo {author} {\bibfnamefont
  {Y.}~\bibnamefont {Xu}}, \bibinfo {author} {\bibfnamefont {K.}~\bibnamefont
  {Yan}}, \bibinfo {author} {\bibfnamefont {W.}~\bibnamefont {Yang}}, \bibinfo
  {author} {\bibfnamefont {Y.}~\bibnamefont {Yang}}, \bibinfo {author}
  {\bibfnamefont {Y.}~\bibnamefont {Ye}}, \bibinfo {author} {\bibfnamefont
  {J.}~\bibnamefont {Yin}}, \bibinfo {author} {\bibfnamefont {C.}~\bibnamefont
  {Ying}}, \bibinfo {author} {\bibfnamefont {J.}~\bibnamefont {Yu}}, \bibinfo
  {author} {\bibfnamefont {C.}~\bibnamefont {Zha}}, \bibinfo {author}
  {\bibfnamefont {C.}~\bibnamefont {Zhang}}, \bibinfo {author} {\bibfnamefont
  {H.}~\bibnamefont {Zhang}}, \bibinfo {author} {\bibfnamefont
  {K.}~\bibnamefont {Zhang}}, \bibinfo {author} {\bibfnamefont
  {Y.}~\bibnamefont {Zhang}}, \bibinfo {author} {\bibfnamefont
  {H.}~\bibnamefont {Zhao}}, \bibinfo {author} {\bibfnamefont {Y.}~\bibnamefont
  {Zhao}}, \bibinfo {author} {\bibfnamefont {L.}~\bibnamefont {Zhou}}, \bibinfo
  {author} {\bibfnamefont {Q.}~\bibnamefont {Zhu}}, \bibinfo {author}
  {\bibfnamefont {C.-Y.}\ \bibnamefont {Lu}}, \bibinfo {author} {\bibfnamefont
  {C.-Z.}\ \bibnamefont {Peng}}, \bibinfo {author} {\bibfnamefont
  {X.}~\bibnamefont {Zhu}},\ and\ \bibinfo {author} {\bibfnamefont {J.-W.}\
  \bibnamefont {Pan}},\ }\href {https://doi.org/10.1103/PhysRevLett.127.180501}
  {\bibfield  {journal} {\bibinfo  {journal} {Phys. Rev. Lett.}\ }\textbf
  {\bibinfo {volume} {127}},\ \bibinfo {pages} {180501} (\bibinfo {year}
  {2021})}\BibitemShut {NoStop}%
\bibitem [{\citenamefont {Cerezo}\ \emph
  {et~al.}(2021{\natexlab{a}})\citenamefont {Cerezo}, \citenamefont
  {Arrasmith}, \citenamefont {Babbush}, \citenamefont {Benjamin}, \citenamefont
  {Endo}, \citenamefont {Fujii}, \citenamefont {McClean}, \citenamefont
  {Mitarai}, \citenamefont {Yuan}, \citenamefont {Cincio},\ and\ \citenamefont
  {Coles}}]{cerezo2021variational}%
  \BibitemOpen
  \bibfield  {author} {\bibinfo {author} {\bibfnamefont {M.}~\bibnamefont
  {Cerezo}}, \bibinfo {author} {\bibfnamefont {A.}~\bibnamefont {Arrasmith}},
  \bibinfo {author} {\bibfnamefont {R.}~\bibnamefont {Babbush}}, \bibinfo
  {author} {\bibfnamefont {S.~C.}\ \bibnamefont {Benjamin}}, \bibinfo {author}
  {\bibfnamefont {S.}~\bibnamefont {Endo}}, \bibinfo {author} {\bibfnamefont
  {K.}~\bibnamefont {Fujii}}, \bibinfo {author} {\bibfnamefont {J.~R.}\
  \bibnamefont {McClean}}, \bibinfo {author} {\bibfnamefont {K.}~\bibnamefont
  {Mitarai}}, \bibinfo {author} {\bibfnamefont {X.}~\bibnamefont {Yuan}},
  \bibinfo {author} {\bibfnamefont {L.}~\bibnamefont {Cincio}},\ and\ \bibinfo
  {author} {\bibfnamefont {P.~J.}\ \bibnamefont {Coles}},\ }\href
  {https://doi.org/10.1038/s42254-021-00348-9} {\bibfield  {journal} {\bibinfo
  {journal} {Nature Reviews Physics}\ }\textbf {\bibinfo {volume} {3}},\
  \bibinfo {pages} {625} (\bibinfo {year} {2021}{\natexlab{a}})}\BibitemShut
  {NoStop}%
\bibitem [{\citenamefont {Peruzzo}\ \emph {et~al.}(2014)\citenamefont
  {Peruzzo}, \citenamefont {McClean}, \citenamefont {Shadbolt}, \citenamefont
  {Yung}, \citenamefont {Zhou}, \citenamefont {Love}, \citenamefont
  {Aspuru-Guzik},\ and\ \citenamefont {O'Brien}}]{peruzzo2014a-variational}%
  \BibitemOpen
  \bibfield  {author} {\bibinfo {author} {\bibfnamefont {A.}~\bibnamefont
  {Peruzzo}}, \bibinfo {author} {\bibfnamefont {J.}~\bibnamefont {McClean}},
  \bibinfo {author} {\bibfnamefont {P.}~\bibnamefont {Shadbolt}}, \bibinfo
  {author} {\bibfnamefont {M.-H.}\ \bibnamefont {Yung}}, \bibinfo {author}
  {\bibfnamefont {X.-Q.}\ \bibnamefont {Zhou}}, \bibinfo {author}
  {\bibfnamefont {P.~J.}\ \bibnamefont {Love}}, \bibinfo {author}
  {\bibfnamefont {A.}~\bibnamefont {Aspuru-Guzik}},\ and\ \bibinfo {author}
  {\bibfnamefont {J.~L.}\ \bibnamefont {O'Brien}},\ }\href
  {https://doi.org/10.1038/ncomms5213} {\bibfield  {journal} {\bibinfo
  {journal} {Nature Communications}\ }\textbf {\bibinfo {volume} {5}},\
  \bibinfo {pages} {4213} (\bibinfo {year} {2014})}\BibitemShut {NoStop}%
\bibitem [{\citenamefont {Farhi}\ \emph {et~al.}(2001)\citenamefont {Farhi},
  \citenamefont {Goldstone}, \citenamefont {Gutmann}, \citenamefont {Lapan},
  \citenamefont {Lundgren},\ and\ \citenamefont {Preda}}]{farhi2001a-quantum}%
  \BibitemOpen
  \bibfield  {author} {\bibinfo {author} {\bibfnamefont {E.}~\bibnamefont
  {Farhi}}, \bibinfo {author} {\bibfnamefont {J.}~\bibnamefont {Goldstone}},
  \bibinfo {author} {\bibfnamefont {S.}~\bibnamefont {Gutmann}}, \bibinfo
  {author} {\bibfnamefont {J.}~\bibnamefont {Lapan}}, \bibinfo {author}
  {\bibfnamefont {A.}~\bibnamefont {Lundgren}},\ and\ \bibinfo {author}
  {\bibfnamefont {D.}~\bibnamefont {Preda}},\ }\href
  {https://doi.org/10.1126/science.1057726} {\bibfield  {journal} {\bibinfo
  {journal} {Science}\ }\textbf {\bibinfo {volume} {292}},\ \bibinfo {pages}
  {472} (\bibinfo {year} {2001})}\BibitemShut {NoStop}%
\bibitem [{\citenamefont {Hadfield}\ \emph {et~al.}(2019)\citenamefont
  {Hadfield}, \citenamefont {Wang}, \citenamefont {O'Gorman}, \citenamefont
  {Rieffel}, \citenamefont {Venturelli},\ and\ \citenamefont
  {Biswas}}]{hadfield2019from}%
  \BibitemOpen
  \bibfield  {author} {\bibinfo {author} {\bibfnamefont {S.}~\bibnamefont
  {Hadfield}}, \bibinfo {author} {\bibfnamefont {Z.}~\bibnamefont {Wang}},
  \bibinfo {author} {\bibfnamefont {B.}~\bibnamefont {O'Gorman}}, \bibinfo
  {author} {\bibfnamefont {E.~G.}\ \bibnamefont {Rieffel}}, \bibinfo {author}
  {\bibfnamefont {D.}~\bibnamefont {Venturelli}},\ and\ \bibinfo {author}
  {\bibfnamefont {R.}~\bibnamefont {Biswas}},\ }\bibfield  {journal} {\bibinfo
  {journal} {Algorithms}\ }\textbf {\bibinfo {volume} {12}},\ \href
  {https://doi.org/10.3390/a12020034} {10.3390/a12020034} (\bibinfo {year}
  {2019})\BibitemShut {NoStop}%
\bibitem [{\citenamefont {Pednault}\ \emph {et~al.}(2019)\citenamefont
  {Pednault}, \citenamefont {Gunnels}, \citenamefont {Nannicini}, \citenamefont
  {Horesh},\ and\ \citenamefont {Wisnieff}}]{pednault2019leveraging}%
  \BibitemOpen
  \bibfield  {author} {\bibinfo {author} {\bibfnamefont {E.}~\bibnamefont
  {Pednault}}, \bibinfo {author} {\bibfnamefont {J.~A.}\ \bibnamefont
  {Gunnels}}, \bibinfo {author} {\bibfnamefont {G.}~\bibnamefont {Nannicini}},
  \bibinfo {author} {\bibfnamefont {L.}~\bibnamefont {Horesh}},\ and\ \bibinfo
  {author} {\bibfnamefont {R.}~\bibnamefont {Wisnieff}},\ }\href
  {http://arxiv.org/abs/1910.09534} {\bibfield  {journal} {\bibinfo  {journal}
  {arXiv:1910.09534}\ } (\bibinfo {year} {2019})}\BibitemShut {NoStop}%
\bibitem [{\citenamefont {Huang}\ \emph {et~al.}(2020)\citenamefont {Huang},
  \citenamefont {Zhang}, \citenamefont {Newman}, \citenamefont {Cai},
  \citenamefont {Gao}, \citenamefont {Tian}, \citenamefont {Wu}, \citenamefont
  {Xu}, \citenamefont {Yu}, \citenamefont {Yuan}, \citenamefont {Szegedy},
  \citenamefont {Shi},\ and\ \citenamefont {Chen}}]{huang2020classical}%
  \BibitemOpen
  \bibfield  {author} {\bibinfo {author} {\bibfnamefont {C.}~\bibnamefont
  {Huang}}, \bibinfo {author} {\bibfnamefont {F.}~\bibnamefont {Zhang}},
  \bibinfo {author} {\bibfnamefont {M.}~\bibnamefont {Newman}}, \bibinfo
  {author} {\bibfnamefont {J.}~\bibnamefont {Cai}}, \bibinfo {author}
  {\bibfnamefont {X.}~\bibnamefont {Gao}}, \bibinfo {author} {\bibfnamefont
  {Z.}~\bibnamefont {Tian}}, \bibinfo {author} {\bibfnamefont {J.}~\bibnamefont
  {Wu}}, \bibinfo {author} {\bibfnamefont {H.}~\bibnamefont {Xu}}, \bibinfo
  {author} {\bibfnamefont {H.}~\bibnamefont {Yu}}, \bibinfo {author}
  {\bibfnamefont {B.}~\bibnamefont {Yuan}}, \bibinfo {author} {\bibfnamefont
  {M.}~\bibnamefont {Szegedy}}, \bibinfo {author} {\bibfnamefont
  {Y.}~\bibnamefont {Shi}},\ and\ \bibinfo {author} {\bibfnamefont
  {J.}~\bibnamefont {Chen}},\ }\href {http://arxiv.org/abs/2005.06787}
  {\bibfield  {journal} {\bibinfo  {journal} {arXiv:2005.06787}\ } (\bibinfo
  {year} {2020})}\BibitemShut {NoStop}%
\bibitem [{\citenamefont {Pan}\ \emph {et~al.}(2020)\citenamefont {Pan},
  \citenamefont {Zhou}, \citenamefont {Li},\ and\ \citenamefont
  {Zhang}}]{pan2020contracting}%
  \BibitemOpen
  \bibfield  {author} {\bibinfo {author} {\bibfnamefont {F.}~\bibnamefont
  {Pan}}, \bibinfo {author} {\bibfnamefont {P.}~\bibnamefont {Zhou}}, \bibinfo
  {author} {\bibfnamefont {S.}~\bibnamefont {Li}},\ and\ \bibinfo {author}
  {\bibfnamefont {P.}~\bibnamefont {Zhang}},\ }\href
  {https://doi.org/10.1103/PhysRevLett.125.060503} {\bibfield  {journal}
  {\bibinfo  {journal} {Phys. Rev. Lett.}\ }\textbf {\bibinfo {volume} {125}},\
  \bibinfo {pages} {060503} (\bibinfo {year} {2020})}\BibitemShut {NoStop}%
\bibitem [{\citenamefont {Gray}\ and\ \citenamefont
  {Kourtis}(2021)}]{gray2021hyper-optimized}%
  \BibitemOpen
  \bibfield  {author} {\bibinfo {author} {\bibfnamefont {J.}~\bibnamefont
  {Gray}}\ and\ \bibinfo {author} {\bibfnamefont {S.}~\bibnamefont {Kourtis}},\
  }\href {https://doi.org/10.22331/q-2021-03-15-410} {\bibfield  {journal}
  {\bibinfo  {journal} {{Quantum}}\ }\textbf {\bibinfo {volume} {5}},\ \bibinfo
  {pages} {410} (\bibinfo {year} {2021})}\BibitemShut {NoStop}%
\bibitem [{\citenamefont {Pan}\ and\ \citenamefont
  {Zhang}(2021)}]{pan2021simulating}%
  \BibitemOpen
  \bibfield  {author} {\bibinfo {author} {\bibfnamefont {F.}~\bibnamefont
  {Pan}}\ and\ \bibinfo {author} {\bibfnamefont {P.}~\bibnamefont {Zhang}},\
  }\href {http://arxiv.org/abs/2103.03074} {\bibfield  {journal} {\bibinfo
  {journal} {arXiv:2103.03074}\ } (\bibinfo {year} {2021})}\BibitemShut
  {NoStop}%
\bibitem [{\citenamefont {Medvidovi{\'c}}\ and\ \citenamefont
  {Carleo}(2021)}]{medvidovic2021classical}%
  \BibitemOpen
  \bibfield  {author} {\bibinfo {author} {\bibfnamefont {M.}~\bibnamefont
  {Medvidovi{\'c}}}\ and\ \bibinfo {author} {\bibfnamefont {G.}~\bibnamefont
  {Carleo}},\ }\href {https://doi.org/10.1038/s41534-021-00440-z} {\bibfield
  {journal} {\bibinfo  {journal} {npj Quantum Information}\ }\textbf {\bibinfo
  {volume} {7}},\ \bibinfo {pages} {101} (\bibinfo {year} {2021})}\BibitemShut
  {NoStop}%
\bibitem [{\citenamefont {Wiersema}\ \emph {et~al.}(2020)\citenamefont
  {Wiersema}, \citenamefont {Zhou}, \citenamefont {de~Sereville}, \citenamefont
  {Carrasquilla}, \citenamefont {Kim},\ and\ \citenamefont
  {Yuen}}]{wiersema2020exploring}%
  \BibitemOpen
  \bibfield  {author} {\bibinfo {author} {\bibfnamefont {R.}~\bibnamefont
  {Wiersema}}, \bibinfo {author} {\bibfnamefont {C.}~\bibnamefont {Zhou}},
  \bibinfo {author} {\bibfnamefont {Y.}~\bibnamefont {de~Sereville}}, \bibinfo
  {author} {\bibfnamefont {J.~F.}\ \bibnamefont {Carrasquilla}}, \bibinfo
  {author} {\bibfnamefont {Y.~B.}\ \bibnamefont {Kim}},\ and\ \bibinfo {author}
  {\bibfnamefont {H.}~\bibnamefont {Yuen}},\ }\href
  {https://doi.org/10.1103/PRXQuantum.1.020319} {\bibfield  {journal} {\bibinfo
   {journal} {PRX Quantum}\ }\textbf {\bibinfo {volume} {1}},\ \bibinfo {pages}
  {020319} (\bibinfo {year} {2020})}\BibitemShut {NoStop}%
\bibitem [{\citenamefont {D{\'\i}ez-Valle}\ \emph {et~al.}(2021)\citenamefont
  {D{\'\i}ez-Valle}, \citenamefont {Porras},\ and\ \citenamefont
  {Garc{\'\i}a-Ripoll}}]{diez-valle2021quantum}%
  \BibitemOpen
  \bibfield  {author} {\bibinfo {author} {\bibfnamefont {P.}~\bibnamefont
  {D{\'\i}ez-Valle}}, \bibinfo {author} {\bibfnamefont {D.}~\bibnamefont
  {Porras}},\ and\ \bibinfo {author} {\bibfnamefont {J.~J.}\ \bibnamefont
  {Garc{\'\i}a-Ripoll}},\ }\href {https://doi.org/10.1103/PhysRevA.104.062426}
  {\bibfield  {journal} {\bibinfo  {journal} {Physical Review A}\ }\textbf
  {\bibinfo {volume} {104}},\ \bibinfo {pages} {062426} (\bibinfo {year}
  {2021})}\BibitemShut {NoStop}%
\bibitem [{\citenamefont {McClean}\ \emph {et~al.}(2021)\citenamefont
  {McClean}, \citenamefont {Harrigan}, \citenamefont {Mohseni}, \citenamefont
  {Rubin}, \citenamefont {Jiang}, \citenamefont {Boixo}, \citenamefont
  {Smelyanskiy}, \citenamefont {Babbush},\ and\ \citenamefont
  {Neven}}]{mcclean2021low-depth}%
  \BibitemOpen
  \bibfield  {author} {\bibinfo {author} {\bibfnamefont {J.~R.}\ \bibnamefont
  {McClean}}, \bibinfo {author} {\bibfnamefont {M.~P.}\ \bibnamefont
  {Harrigan}}, \bibinfo {author} {\bibfnamefont {M.}~\bibnamefont {Mohseni}},
  \bibinfo {author} {\bibfnamefont {N.~C.}\ \bibnamefont {Rubin}}, \bibinfo
  {author} {\bibfnamefont {Z.}~\bibnamefont {Jiang}}, \bibinfo {author}
  {\bibfnamefont {S.}~\bibnamefont {Boixo}}, \bibinfo {author} {\bibfnamefont
  {V.~N.}\ \bibnamefont {Smelyanskiy}}, \bibinfo {author} {\bibfnamefont
  {R.}~\bibnamefont {Babbush}},\ and\ \bibinfo {author} {\bibfnamefont
  {H.}~\bibnamefont {Neven}},\ }\href
  {https://doi.org/10.1103/PRXQuantum.2.030312} {\bibfield  {journal} {\bibinfo
   {journal} {PRX Quantum}\ }\textbf {\bibinfo {volume} {2}},\ \bibinfo {pages}
  {030312} (\bibinfo {year} {2021})}\BibitemShut {NoStop}%
\bibitem [{\citenamefont {Chen}\ \emph {et~al.}(2022)\citenamefont {Chen},
  \citenamefont {Zhu}, \citenamefont {Liu}, \citenamefont {Mayhall},
  \citenamefont {Barnes},\ and\ \citenamefont {Economou}}]{chen2022how-much}%
  \BibitemOpen
  \bibfield  {author} {\bibinfo {author} {\bibfnamefont {Y.}~\bibnamefont
  {Chen}}, \bibinfo {author} {\bibfnamefont {L.}~\bibnamefont {Zhu}}, \bibinfo
  {author} {\bibfnamefont {C.}~\bibnamefont {Liu}}, \bibinfo {author}
  {\bibfnamefont {N.~J.}\ \bibnamefont {Mayhall}}, \bibinfo {author}
  {\bibfnamefont {E.}~\bibnamefont {Barnes}},\ and\ \bibinfo {author}
  {\bibfnamefont {S.~E.}\ \bibnamefont {Economou}},\ }\href
  {https://arxiv.org/abs/2205.12283} {\bibfield  {journal} {\bibinfo  {journal}
  {arXiv:2205.12283}\ } (\bibinfo {year} {2022})}\BibitemShut {NoStop}%
\bibitem [{\citenamefont {Dupont}\ \emph
  {et~al.}(2022{\natexlab{a}})\citenamefont {Dupont}, \citenamefont {Didier},
  \citenamefont {Hodson}, \citenamefont {Moore},\ and\ \citenamefont
  {Reagor}}]{dupont2022an-entanglement}%
  \BibitemOpen
  \bibfield  {author} {\bibinfo {author} {\bibfnamefont {M.}~\bibnamefont
  {Dupont}}, \bibinfo {author} {\bibfnamefont {N.}~\bibnamefont {Didier}},
  \bibinfo {author} {\bibfnamefont {M.~J.}\ \bibnamefont {Hodson}}, \bibinfo
  {author} {\bibfnamefont {J.~E.}\ \bibnamefont {Moore}},\ and\ \bibinfo
  {author} {\bibfnamefont {M.~J.}\ \bibnamefont {Reagor}},\ }\href
  {http://arxiv.org/abs/2206.07024} {\bibfield  {journal} {\bibinfo  {journal}
  {arXiv:2206.07024}\ } (\bibinfo {year} {2022}{\natexlab{a}})}\BibitemShut
  {NoStop}%
\bibitem [{\citenamefont {Hastings}(2019)}]{hastings2019classical}%
  \BibitemOpen
  \bibfield  {author} {\bibinfo {author} {\bibfnamefont {M.~B.}\ \bibnamefont
  {Hastings}},\ }\href {http://arxiv.org/abs/1905.07047} {\bibfield  {journal}
  {\bibinfo  {journal} {arXiv:1905.07047}\ } (\bibinfo {year}
  {2019})}\BibitemShut {NoStop}%
\bibitem [{\citenamefont {Farhi}\ \emph {et~al.}(2020)\citenamefont {Farhi},
  \citenamefont {Gamarnik},\ and\ \citenamefont
  {Gutmann}}]{farhi2020the-quantum}%
  \BibitemOpen
  \bibfield  {author} {\bibinfo {author} {\bibfnamefont {E.}~\bibnamefont
  {Farhi}}, \bibinfo {author} {\bibfnamefont {D.}~\bibnamefont {Gamarnik}},\
  and\ \bibinfo {author} {\bibfnamefont {S.}~\bibnamefont {Gutmann}},\ }\href
  {http://arxiv.org/abs/2004.09002} {\bibfield  {journal} {\bibinfo  {journal}
  {arXiv:2004.09002}\ } (\bibinfo {year} {2020})}\BibitemShut {NoStop}%
\bibitem [{\citenamefont {Bravyi}\ \emph {et~al.}(2021)\citenamefont {Bravyi},
  \citenamefont {Gosset},\ and\ \citenamefont
  {Movassagh}}]{bravyi2021classical}%
  \BibitemOpen
  \bibfield  {author} {\bibinfo {author} {\bibfnamefont {S.}~\bibnamefont
  {Bravyi}}, \bibinfo {author} {\bibfnamefont {D.}~\bibnamefont {Gosset}},\
  and\ \bibinfo {author} {\bibfnamefont {R.}~\bibnamefont {Movassagh}},\ }\href
  {https://doi.org/10.1038/s41567-020-01109-8} {\bibfield  {journal} {\bibinfo
  {journal} {Nature Physics}\ }\textbf {\bibinfo {volume} {17}},\ \bibinfo
  {pages} {337} (\bibinfo {year} {2021})}\BibitemShut {NoStop}%
\bibitem [{\citenamefont {Basso}\ \emph {et~al.}(2022)\citenamefont {Basso},
  \citenamefont {Farhi}, \citenamefont {Marwaha}, \citenamefont {Villalonga},\
  and\ \citenamefont {Zhou}}]{basso2022the-quantum}%
  \BibitemOpen
  \bibfield  {author} {\bibinfo {author} {\bibfnamefont {J.}~\bibnamefont
  {Basso}}, \bibinfo {author} {\bibfnamefont {E.}~\bibnamefont {Farhi}},
  \bibinfo {author} {\bibfnamefont {K.}~\bibnamefont {Marwaha}}, \bibinfo
  {author} {\bibfnamefont {B.}~\bibnamefont {Villalonga}},\ and\ \bibinfo
  {author} {\bibfnamefont {L.}~\bibnamefont {Zhou}},\ }\href
  {http://arxiv.org/abs/2110.14206} {\bibfield  {journal} {\bibinfo  {journal}
  {arXiv:2110.14206}\ } (\bibinfo {year} {2022})}\BibitemShut {NoStop}%
\bibitem [{\citenamefont {Harrigan}\ \emph {et~al.}(2021)\citenamefont
  {Harrigan}, \citenamefont {Sung}, \citenamefont {Neeley}, \citenamefont
  {Satzinger}, \citenamefont {Arute}, \citenamefont {Arya}, \citenamefont
  {Atalaya}, \citenamefont {Bardin}, \citenamefont {Barends}, \citenamefont
  {Boixo}, \citenamefont {Broughton}, \citenamefont {Buckley}, \citenamefont
  {Buell}, \citenamefont {Burkett}, \citenamefont {Bushnell}, \citenamefont
  {Chen}, \citenamefont {Chen}, \citenamefont {{Ben Chiaro}}, \citenamefont
  {Collins}, \citenamefont {Courtney}, \citenamefont {Demura}, \citenamefont
  {Dunsworth}, \citenamefont {Eppens}, \citenamefont {Fowler}, \citenamefont
  {Foxen}, \citenamefont {Gidney}, \citenamefont {Giustina}, \citenamefont
  {Graff}, \citenamefont {Habegger}, \citenamefont {Ho}, \citenamefont {Hong},
  \citenamefont {Huang}, \citenamefont {Ioffe}, \citenamefont {Isakov},
  \citenamefont {Jeffrey}, \citenamefont {Jiang}, \citenamefont {Jones},
  \citenamefont {Kafri}, \citenamefont {Kechedzhi}, \citenamefont {Kelly},
  \citenamefont {Kim}, \citenamefont {Klimov}, \citenamefont {Korotkov},
  \citenamefont {Kostritsa}, \citenamefont {Landhuis}, \citenamefont {Laptev},
  \citenamefont {Lindmark}, \citenamefont {Leib}, \citenamefont {Martin},
  \citenamefont {Martinis}, \citenamefont {McClean}, \citenamefont {McEwen},
  \citenamefont {Megrant}, \citenamefont {Mi}, \citenamefont {Mohseni},
  \citenamefont {Mruczkiewicz}, \citenamefont {Mutus}, \citenamefont {Naaman},
  \citenamefont {Neill}, \citenamefont {Neukart}, \citenamefont {Niu},
  \citenamefont {O'Brien}, \citenamefont {O'Gorman}, \citenamefont {Ostby},
  \citenamefont {Petukhov}, \citenamefont {Putterman}, \citenamefont
  {Quintana}, \citenamefont {Roushan}, \citenamefont {Rubin}, \citenamefont
  {Sank}, \citenamefont {Skolik}, \citenamefont {Smelyanskiy}, \citenamefont
  {Strain}, \citenamefont {Streif}, \citenamefont {Szalay}, \citenamefont
  {Vainsencher}, \citenamefont {White}, \citenamefont {Yao}, \citenamefont
  {Yeh}, \citenamefont {Zalcman}, \citenamefont {Zhou}, \citenamefont {Neven},
  \citenamefont {Bacon}, \citenamefont {Lucero}, \citenamefont {Farhi},\ and\
  \citenamefont {Babbush}}]{harrigan2021quantum}%
  \BibitemOpen
  \bibfield  {author} {\bibinfo {author} {\bibfnamefont {M.~P.}\ \bibnamefont
  {Harrigan}}, \bibinfo {author} {\bibfnamefont {K.~J.}\ \bibnamefont {Sung}},
  \bibinfo {author} {\bibfnamefont {M.}~\bibnamefont {Neeley}}, \bibinfo
  {author} {\bibfnamefont {K.~J.}\ \bibnamefont {Satzinger}}, \bibinfo {author}
  {\bibfnamefont {F.}~\bibnamefont {Arute}}, \bibinfo {author} {\bibfnamefont
  {K.}~\bibnamefont {Arya}}, \bibinfo {author} {\bibfnamefont {J.}~\bibnamefont
  {Atalaya}}, \bibinfo {author} {\bibfnamefont {J.~C.}\ \bibnamefont {Bardin}},
  \bibinfo {author} {\bibfnamefont {R.}~\bibnamefont {Barends}}, \bibinfo
  {author} {\bibfnamefont {S.}~\bibnamefont {Boixo}}, \bibinfo {author}
  {\bibfnamefont {M.}~\bibnamefont {Broughton}}, \bibinfo {author}
  {\bibfnamefont {B.~B.}\ \bibnamefont {Buckley}}, \bibinfo {author}
  {\bibfnamefont {D.~A.}\ \bibnamefont {Buell}}, \bibinfo {author}
  {\bibfnamefont {B.}~\bibnamefont {Burkett}}, \bibinfo {author} {\bibfnamefont
  {N.}~\bibnamefont {Bushnell}}, \bibinfo {author} {\bibfnamefont
  {Y.}~\bibnamefont {Chen}}, \bibinfo {author} {\bibfnamefont {Z.}~\bibnamefont
  {Chen}}, \bibinfo {author} {\bibnamefont {{Ben Chiaro}}}, \bibinfo {author}
  {\bibfnamefont {R.}~\bibnamefont {Collins}}, \bibinfo {author} {\bibfnamefont
  {W.}~\bibnamefont {Courtney}}, \bibinfo {author} {\bibfnamefont
  {S.}~\bibnamefont {Demura}}, \bibinfo {author} {\bibfnamefont
  {A.}~\bibnamefont {Dunsworth}}, \bibinfo {author} {\bibfnamefont
  {D.}~\bibnamefont {Eppens}}, \bibinfo {author} {\bibfnamefont
  {A.}~\bibnamefont {Fowler}}, \bibinfo {author} {\bibfnamefont
  {B.}~\bibnamefont {Foxen}}, \bibinfo {author} {\bibfnamefont
  {C.}~\bibnamefont {Gidney}}, \bibinfo {author} {\bibfnamefont
  {M.}~\bibnamefont {Giustina}}, \bibinfo {author} {\bibfnamefont
  {R.}~\bibnamefont {Graff}}, \bibinfo {author} {\bibfnamefont
  {S.}~\bibnamefont {Habegger}}, \bibinfo {author} {\bibfnamefont
  {A.}~\bibnamefont {Ho}}, \bibinfo {author} {\bibfnamefont {S.}~\bibnamefont
  {Hong}}, \bibinfo {author} {\bibfnamefont {T.}~\bibnamefont {Huang}},
  \bibinfo {author} {\bibfnamefont {L.~B.}\ \bibnamefont {Ioffe}}, \bibinfo
  {author} {\bibfnamefont {S.~V.}\ \bibnamefont {Isakov}}, \bibinfo {author}
  {\bibfnamefont {E.}~\bibnamefont {Jeffrey}}, \bibinfo {author} {\bibfnamefont
  {Z.}~\bibnamefont {Jiang}}, \bibinfo {author} {\bibfnamefont
  {C.}~\bibnamefont {Jones}}, \bibinfo {author} {\bibfnamefont
  {D.}~\bibnamefont {Kafri}}, \bibinfo {author} {\bibfnamefont
  {K.}~\bibnamefont {Kechedzhi}}, \bibinfo {author} {\bibfnamefont
  {J.}~\bibnamefont {Kelly}}, \bibinfo {author} {\bibfnamefont
  {S.}~\bibnamefont {Kim}}, \bibinfo {author} {\bibfnamefont {P.~V.}\
  \bibnamefont {Klimov}}, \bibinfo {author} {\bibfnamefont {A.~N.}\
  \bibnamefont {Korotkov}}, \bibinfo {author} {\bibfnamefont {F.}~\bibnamefont
  {Kostritsa}}, \bibinfo {author} {\bibfnamefont {D.}~\bibnamefont {Landhuis}},
  \bibinfo {author} {\bibfnamefont {P.}~\bibnamefont {Laptev}}, \bibinfo
  {author} {\bibfnamefont {M.}~\bibnamefont {Lindmark}}, \bibinfo {author}
  {\bibfnamefont {M.}~\bibnamefont {Leib}}, \bibinfo {author} {\bibfnamefont
  {O.}~\bibnamefont {Martin}}, \bibinfo {author} {\bibfnamefont {J.~M.}\
  \bibnamefont {Martinis}}, \bibinfo {author} {\bibfnamefont {J.~R.}\
  \bibnamefont {McClean}}, \bibinfo {author} {\bibfnamefont {M.}~\bibnamefont
  {McEwen}}, \bibinfo {author} {\bibfnamefont {A.}~\bibnamefont {Megrant}},
  \bibinfo {author} {\bibfnamefont {X.}~\bibnamefont {Mi}}, \bibinfo {author}
  {\bibfnamefont {M.}~\bibnamefont {Mohseni}}, \bibinfo {author} {\bibfnamefont
  {W.}~\bibnamefont {Mruczkiewicz}}, \bibinfo {author} {\bibfnamefont
  {J.}~\bibnamefont {Mutus}}, \bibinfo {author} {\bibfnamefont
  {O.}~\bibnamefont {Naaman}}, \bibinfo {author} {\bibfnamefont
  {C.}~\bibnamefont {Neill}}, \bibinfo {author} {\bibfnamefont
  {F.}~\bibnamefont {Neukart}}, \bibinfo {author} {\bibfnamefont {M.~Y.}\
  \bibnamefont {Niu}}, \bibinfo {author} {\bibfnamefont {T.~E.}\ \bibnamefont
  {O'Brien}}, \bibinfo {author} {\bibfnamefont {B.}~\bibnamefont {O'Gorman}},
  \bibinfo {author} {\bibfnamefont {E.}~\bibnamefont {Ostby}}, \bibinfo
  {author} {\bibfnamefont {A.}~\bibnamefont {Petukhov}}, \bibinfo {author}
  {\bibfnamefont {H.}~\bibnamefont {Putterman}}, \bibinfo {author}
  {\bibfnamefont {C.}~\bibnamefont {Quintana}}, \bibinfo {author}
  {\bibfnamefont {P.}~\bibnamefont {Roushan}}, \bibinfo {author} {\bibfnamefont
  {N.~C.}\ \bibnamefont {Rubin}}, \bibinfo {author} {\bibfnamefont
  {D.}~\bibnamefont {Sank}}, \bibinfo {author} {\bibfnamefont {A.}~\bibnamefont
  {Skolik}}, \bibinfo {author} {\bibfnamefont {V.}~\bibnamefont {Smelyanskiy}},
  \bibinfo {author} {\bibfnamefont {D.}~\bibnamefont {Strain}}, \bibinfo
  {author} {\bibfnamefont {M.}~\bibnamefont {Streif}}, \bibinfo {author}
  {\bibfnamefont {M.}~\bibnamefont {Szalay}}, \bibinfo {author} {\bibfnamefont
  {A.}~\bibnamefont {Vainsencher}}, \bibinfo {author} {\bibfnamefont
  {T.}~\bibnamefont {White}}, \bibinfo {author} {\bibfnamefont {Z.~J.}\
  \bibnamefont {Yao}}, \bibinfo {author} {\bibfnamefont {P.}~\bibnamefont
  {Yeh}}, \bibinfo {author} {\bibfnamefont {A.}~\bibnamefont {Zalcman}},
  \bibinfo {author} {\bibfnamefont {L.}~\bibnamefont {Zhou}}, \bibinfo {author}
  {\bibfnamefont {H.}~\bibnamefont {Neven}}, \bibinfo {author} {\bibfnamefont
  {D.}~\bibnamefont {Bacon}}, \bibinfo {author} {\bibfnamefont
  {E.}~\bibnamefont {Lucero}}, \bibinfo {author} {\bibfnamefont
  {E.}~\bibnamefont {Farhi}},\ and\ \bibinfo {author} {\bibfnamefont
  {R.}~\bibnamefont {Babbush}},\ }\href
  {https://doi.org/10.1038/s41567-020-01105-y} {\bibfield  {journal} {\bibinfo
  {journal} {Nature Physics}\ }\textbf {\bibinfo {volume} {17}},\ \bibinfo
  {pages} {332} (\bibinfo {year} {2021})}\BibitemShut {NoStop}%
\bibitem [{\citenamefont {Yu}\ and\ \citenamefont
  {Eberly}(2004)}]{yu2004finite-time}%
  \BibitemOpen
  \bibfield  {author} {\bibinfo {author} {\bibfnamefont {T.}~\bibnamefont
  {Yu}}\ and\ \bibinfo {author} {\bibfnamefont {J.~H.}\ \bibnamefont
  {Eberly}},\ }\href {https://doi.org/10.1103/PhysRevLett.93.140404} {\bibfield
   {journal} {\bibinfo  {journal} {Phys. Rev. Lett.}\ }\textbf {\bibinfo
  {volume} {93}},\ \bibinfo {pages} {140404} (\bibinfo {year}
  {2004})}\BibitemShut {NoStop}%
\bibitem [{\citenamefont {Almeida}\ \emph {et~al.}(2007)\citenamefont
  {Almeida}, \citenamefont {de~Melo}, \citenamefont {Hor-Meyll}, \citenamefont
  {Salles}, \citenamefont {Walborn}, \citenamefont {Ribeiro},\ and\
  \citenamefont {Davidovich}}]{almeida2007environment-induced}%
  \BibitemOpen
  \bibfield  {author} {\bibinfo {author} {\bibfnamefont {M.~P.}\ \bibnamefont
  {Almeida}}, \bibinfo {author} {\bibfnamefont {F.}~\bibnamefont {de~Melo}},
  \bibinfo {author} {\bibfnamefont {M.}~\bibnamefont {Hor-Meyll}}, \bibinfo
  {author} {\bibfnamefont {A.}~\bibnamefont {Salles}}, \bibinfo {author}
  {\bibfnamefont {S.~P.}\ \bibnamefont {Walborn}}, \bibinfo {author}
  {\bibfnamefont {P.~H.~S.}\ \bibnamefont {Ribeiro}},\ and\ \bibinfo {author}
  {\bibfnamefont {L.}~\bibnamefont {Davidovich}},\ }\href
  {https://doi.org/10.1126/science.1139892} {\bibfield  {journal} {\bibinfo
  {journal} {Science}\ }\textbf {\bibinfo {volume} {316}},\ \bibinfo {pages}
  {579} (\bibinfo {year} {2007})}\BibitemShut {NoStop}%
\bibitem [{\citenamefont {Farhi}\ \emph {et~al.}(2002)\citenamefont {Farhi},
  \citenamefont {Goldstone},\ and\ \citenamefont {Gutmann}}]{farhi2002quantum}%
  \BibitemOpen
  \bibfield  {author} {\bibinfo {author} {\bibfnamefont {E.}~\bibnamefont
  {Farhi}}, \bibinfo {author} {\bibfnamefont {J.}~\bibnamefont {Goldstone}},\
  and\ \bibinfo {author} {\bibfnamefont {S.}~\bibnamefont {Gutmann}},\ }\href
  {http://arxiv.org/abs/quant-ph/0201031} {\bibfield  {journal} {\bibinfo
  {journal} {arXiv:quant-ph/0201031}\ } (\bibinfo {year} {2002})}\BibitemShut
  {NoStop}%
\bibitem [{\citenamefont {Dupont}\ \emph
  {et~al.}(2022{\natexlab{b}})\citenamefont {Dupont}, \citenamefont {Didier},
  \citenamefont {Hodson}, \citenamefont {Moore},\ and\ \citenamefont
  {Reagor}}]{dupont2022calibrating}%
  \BibitemOpen
  \bibfield  {author} {\bibinfo {author} {\bibfnamefont {M.}~\bibnamefont
  {Dupont}}, \bibinfo {author} {\bibfnamefont {N.}~\bibnamefont {Didier}},
  \bibinfo {author} {\bibfnamefont {M.~J.}\ \bibnamefont {Hodson}}, \bibinfo
  {author} {\bibfnamefont {J.~E.}\ \bibnamefont {Moore}},\ and\ \bibinfo
  {author} {\bibfnamefont {M.~J.}\ \bibnamefont {Reagor}},\ }\href
  {http://arxiv.org/abs/2206.06348} {\bibfield  {journal} {\bibinfo  {journal}
  {arXiv.2206.06348}\ } (\bibinfo {year} {2022}{\natexlab{b}})}\BibitemShut
  {NoStop}%
\bibitem [{\citenamefont
  {Schollw{\"o}ck}(2011)}]{schollwock2011the-density-matrix}%
  \BibitemOpen
  \bibfield  {author} {\bibinfo {author} {\bibfnamefont {U.}~\bibnamefont
  {Schollw{\"o}ck}},\ }\href {https://doi.org/10.1016/j.aop.2010.09.012}
  {\bibfield  {journal} {\bibinfo  {journal} {Annals of Physics}\ }\bibinfo
  {series} {January 2011 {Special} {Issue}},\ \textbf {\bibinfo {volume}
  {326}},\ \bibinfo {pages} {96} (\bibinfo {year} {2011})}\BibitemShut
  {NoStop}%
\bibitem [{\citenamefont {White}(1992)}]{white1992density}%
  \BibitemOpen
  \bibfield  {author} {\bibinfo {author} {\bibfnamefont {S.~R.}\ \bibnamefont
  {White}},\ }\href {https://doi.org/10.1103/PhysRevLett.69.2863} {\bibfield
  {journal} {\bibinfo  {journal} {Phys. Rev. Lett.}\ }\textbf {\bibinfo
  {volume} {69}},\ \bibinfo {pages} {2863} (\bibinfo {year}
  {1992})}\BibitemShut {NoStop}%
\bibitem [{\citenamefont {White}(1993)}]{white1993density-matrix}%
  \BibitemOpen
  \bibfield  {author} {\bibinfo {author} {\bibfnamefont {S.~R.}\ \bibnamefont
  {White}},\ }\href {https://doi.org/10.1103/PhysRevB.48.10345} {\bibfield
  {journal} {\bibinfo  {journal} {Phys. Rev. B}\ }\textbf {\bibinfo {volume}
  {48}},\ \bibinfo {pages} {10345} (\bibinfo {year} {1993})}\BibitemShut
  {NoStop}%
\bibitem [{\citenamefont {Hastings}(2007)}]{hastings2007an-area}%
  \BibitemOpen
  \bibfield  {author} {\bibinfo {author} {\bibfnamefont {M.~B.}\ \bibnamefont
  {Hastings}},\ }\href {https://doi.org/10.1088/1742-5468/2007/08/P08024}
  {\bibfield  {journal} {\bibinfo  {journal} {Journal of Statistical Mechanics:
  Theory and Experiment}\ }\textbf {\bibinfo {volume} {2007}},\ \bibinfo
  {pages} {P08024} (\bibinfo {year} {2007})}\BibitemShut {NoStop}%
\bibitem [{\citenamefont {Vidal}(2003)}]{vidal2003efficient}%
  \BibitemOpen
  \bibfield  {author} {\bibinfo {author} {\bibfnamefont {G.}~\bibnamefont
  {Vidal}},\ }\href {https://doi.org/10.1103/PhysRevLett.91.147902} {\bibfield
  {journal} {\bibinfo  {journal} {Phys. Rev. Lett.}\ }\textbf {\bibinfo
  {volume} {91}},\ \bibinfo {pages} {147902} (\bibinfo {year}
  {2003})}\BibitemShut {NoStop}%
\bibitem [{\citenamefont {Markov}\ and\ \citenamefont
  {Shi}(2008)}]{markov2008simulating}%
  \BibitemOpen
  \bibfield  {author} {\bibinfo {author} {\bibfnamefont {I.~L.}\ \bibnamefont
  {Markov}}\ and\ \bibinfo {author} {\bibfnamefont {Y.}~\bibnamefont {Shi}},\
  }\href {https://doi.org/10.1137/050644756} {\bibfield  {journal} {\bibinfo
  {journal} {SIAM Journal on Computing}\ }\textbf {\bibinfo {volume} {38}},\
  \bibinfo {pages} {963} (\bibinfo {year} {2008})}\BibitemShut {NoStop}%
\bibitem [{\citenamefont {Zhou}\ \emph
  {et~al.}(2020{\natexlab{a}})\citenamefont {Zhou}, \citenamefont
  {Stoudenmire},\ and\ \citenamefont {Waintal}}]{zhou2020what}%
  \BibitemOpen
  \bibfield  {author} {\bibinfo {author} {\bibfnamefont {Y.}~\bibnamefont
  {Zhou}}, \bibinfo {author} {\bibfnamefont {E.~M.}\ \bibnamefont
  {Stoudenmire}},\ and\ \bibinfo {author} {\bibfnamefont {X.}~\bibnamefont
  {Waintal}},\ }\href {https://doi.org/10.1103/PhysRevX.10.041038} {\bibfield
  {journal} {\bibinfo  {journal} {Physical Review X}\ }\textbf {\bibinfo
  {volume} {10}},\ \bibinfo {pages} {041038} (\bibinfo {year}
  {2020}{\natexlab{a}})}\BibitemShut {NoStop}%
\bibitem [{\citenamefont {Farhi}\ \emph {et~al.}(2014)\citenamefont {Farhi},
  \citenamefont {Goldstone},\ and\ \citenamefont
  {Gutmann}}]{farhi2014a-quantum}%
  \BibitemOpen
  \bibfield  {author} {\bibinfo {author} {\bibfnamefont {E.}~\bibnamefont
  {Farhi}}, \bibinfo {author} {\bibfnamefont {J.}~\bibnamefont {Goldstone}},\
  and\ \bibinfo {author} {\bibfnamefont {S.}~\bibnamefont {Gutmann}},\ }\href
  {http://arxiv.org/abs/1411.4028} {\bibfield  {journal} {\bibinfo  {journal}
  {arXiv:1411.4028}\ } (\bibinfo {year} {2014})}\BibitemShut {NoStop}%
\bibitem [{\citenamefont {Zhou}\ \emph
  {et~al.}(2020{\natexlab{b}})\citenamefont {Zhou}, \citenamefont {Wang},
  \citenamefont {Choi}, \citenamefont {Pichler},\ and\ \citenamefont
  {Lukin}}]{zhou2020quantum}%
  \BibitemOpen
  \bibfield  {author} {\bibinfo {author} {\bibfnamefont {L.}~\bibnamefont
  {Zhou}}, \bibinfo {author} {\bibfnamefont {S.-T.}\ \bibnamefont {Wang}},
  \bibinfo {author} {\bibfnamefont {S.}~\bibnamefont {Choi}}, \bibinfo {author}
  {\bibfnamefont {H.}~\bibnamefont {Pichler}},\ and\ \bibinfo {author}
  {\bibfnamefont {M.~D.}\ \bibnamefont {Lukin}},\ }\href
  {https://doi.org/10.1103/PhysRevX.10.021067} {\bibfield  {journal} {\bibinfo
  {journal} {Physical Review X}\ }\textbf {\bibinfo {volume} {10}},\ \bibinfo
  {pages} {021067} (\bibinfo {year} {2020}{\natexlab{b}})}\BibitemShut
  {NoStop}%
\bibitem [{\citenamefont {Choi}(2011)}]{choi2011different}%
  \BibitemOpen
  \bibfield  {author} {\bibinfo {author} {\bibfnamefont {V.}~\bibnamefont
  {Choi}},\ }\href {https://doi.org/10.26421/QIC11.7-8-7} {\bibfield  {journal}
  {\bibinfo  {journal} {Quantum Information and Computation}\ }\textbf
  {\bibinfo {volume} {11}},\ \bibinfo {pages} {638} (\bibinfo {year}
  {2011})}\BibitemShut {NoStop}%
\bibitem [{\citenamefont {Kivlichan}\ \emph {et~al.}(2018)\citenamefont
  {Kivlichan}, \citenamefont {McClean}, \citenamefont {Wiebe}, \citenamefont
  {Gidney}, \citenamefont {Aspuru-Guzik}, \citenamefont {Chan},\ and\
  \citenamefont {Babbush}}]{kivlichan2018quantum}%
  \BibitemOpen
  \bibfield  {author} {\bibinfo {author} {\bibfnamefont {I.~D.}\ \bibnamefont
  {Kivlichan}}, \bibinfo {author} {\bibfnamefont {J.}~\bibnamefont {McClean}},
  \bibinfo {author} {\bibfnamefont {N.}~\bibnamefont {Wiebe}}, \bibinfo
  {author} {\bibfnamefont {C.}~\bibnamefont {Gidney}}, \bibinfo {author}
  {\bibfnamefont {A.}~\bibnamefont {Aspuru-Guzik}}, \bibinfo {author}
  {\bibfnamefont {G.~K.-L.}\ \bibnamefont {Chan}},\ and\ \bibinfo {author}
  {\bibfnamefont {R.}~\bibnamefont {Babbush}},\ }\href
  {https://doi.org/10.1103/PhysRevLett.120.110501} {\bibfield  {journal}
  {\bibinfo  {journal} {Phys. Rev. Lett.}\ }\textbf {\bibinfo {volume} {120}},\
  \bibinfo {pages} {110501} (\bibinfo {year} {2018})}\BibitemShut {NoStop}%
\bibitem [{\citenamefont {Roberts}\ \emph {et~al.}(2019)\citenamefont
  {Roberts}, \citenamefont {Milsted}, \citenamefont {Ganahl}, \citenamefont
  {Zalcman}, \citenamefont {Fontaine}, \citenamefont {Zou}, \citenamefont
  {Hidary}, \citenamefont {Vidal},\ and\ \citenamefont
  {Leichenauer}}]{roberts2019tensornetwork:}%
  \BibitemOpen
  \bibfield  {author} {\bibinfo {author} {\bibfnamefont {C.}~\bibnamefont
  {Roberts}}, \bibinfo {author} {\bibfnamefont {A.}~\bibnamefont {Milsted}},
  \bibinfo {author} {\bibfnamefont {M.}~\bibnamefont {Ganahl}}, \bibinfo
  {author} {\bibfnamefont {A.}~\bibnamefont {Zalcman}}, \bibinfo {author}
  {\bibfnamefont {B.}~\bibnamefont {Fontaine}}, \bibinfo {author}
  {\bibfnamefont {Y.}~\bibnamefont {Zou}}, \bibinfo {author} {\bibfnamefont
  {J.}~\bibnamefont {Hidary}}, \bibinfo {author} {\bibfnamefont
  {G.}~\bibnamefont {Vidal}},\ and\ \bibinfo {author} {\bibfnamefont
  {S.}~\bibnamefont {Leichenauer}},\ }\href {https://arxiv.org/abs/1905.01330}
  {\bibfield  {journal} {\bibinfo  {journal} {arXiv:1905.01330}\ } (\bibinfo
  {year} {2019})}\BibitemShut {NoStop}%
\bibitem [{\citenamefont {Ferris}\ and\ \citenamefont
  {Vidal}(2012)}]{ferris2012perfect}%
  \BibitemOpen
  \bibfield  {author} {\bibinfo {author} {\bibfnamefont {A.~J.}\ \bibnamefont
  {Ferris}}\ and\ \bibinfo {author} {\bibfnamefont {G.}~\bibnamefont {Vidal}},\
  }\href {https://doi.org/10.1103/PhysRevB.85.165146} {\bibfield  {journal}
  {\bibinfo  {journal} {Phys. Rev. B}\ }\textbf {\bibinfo {volume} {85}},\
  \bibinfo {pages} {165146} (\bibinfo {year} {2012})}\BibitemShut {NoStop}%
\bibitem [{\citenamefont {Stoudenmire}\ and\ \citenamefont
  {White}(2010)}]{stoudenmire2010minimally}%
  \BibitemOpen
  \bibfield  {author} {\bibinfo {author} {\bibfnamefont {E.~M.}\ \bibnamefont
  {Stoudenmire}}\ and\ \bibinfo {author} {\bibfnamefont {S.~R.}\ \bibnamefont
  {White}},\ }\href {https://doi.org/10.1088/1367-2630/12/5/055026} {\bibfield
  {journal} {\bibinfo  {journal} {New Journal of Physics}\ }\textbf {\bibinfo
  {volume} {12}},\ \bibinfo {pages} {055026} (\bibinfo {year}
  {2010})}\BibitemShut {NoStop}%
\bibitem [{\citenamefont {McClean}\ \emph {et~al.}(2018)\citenamefont
  {McClean}, \citenamefont {Boixo}, \citenamefont {Smelyanskiy}, \citenamefont
  {Babbush},\ and\ \citenamefont {Neven}}]{mcclean2018barren}%
  \BibitemOpen
  \bibfield  {author} {\bibinfo {author} {\bibfnamefont {J.~R.}\ \bibnamefont
  {McClean}}, \bibinfo {author} {\bibfnamefont {S.}~\bibnamefont {Boixo}},
  \bibinfo {author} {\bibfnamefont {V.~N.}\ \bibnamefont {Smelyanskiy}},
  \bibinfo {author} {\bibfnamefont {R.}~\bibnamefont {Babbush}},\ and\ \bibinfo
  {author} {\bibfnamefont {H.}~\bibnamefont {Neven}},\ }\href
  {https://doi.org/10.1038/s41467-018-07090-4} {\bibfield  {journal} {\bibinfo
  {journal} {Nature Communications}\ }\textbf {\bibinfo {volume} {9}},\
  \bibinfo {pages} {4812} (\bibinfo {year} {2018})}\BibitemShut {NoStop}%
\bibitem [{\citenamefont {Cerezo}\ \emph
  {et~al.}(2021{\natexlab{b}})\citenamefont {Cerezo}, \citenamefont {Sone},
  \citenamefont {Volkoff}, \citenamefont {Cincio},\ and\ \citenamefont
  {Coles}}]{cerezo2021cost}%
  \BibitemOpen
  \bibfield  {author} {\bibinfo {author} {\bibfnamefont {M.}~\bibnamefont
  {Cerezo}}, \bibinfo {author} {\bibfnamefont {A.}~\bibnamefont {Sone}},
  \bibinfo {author} {\bibfnamefont {T.}~\bibnamefont {Volkoff}}, \bibinfo
  {author} {\bibfnamefont {L.}~\bibnamefont {Cincio}},\ and\ \bibinfo {author}
  {\bibfnamefont {P.~J.}\ \bibnamefont {Coles}},\ }\href
  {https://doi.org/10.1038/s41467-021-21728-w} {\bibfield  {journal} {\bibinfo
  {journal} {Nature Communications}\ }\textbf {\bibinfo {volume} {12}},\
  \bibinfo {pages} {1791} (\bibinfo {year} {2021}{\natexlab{b}})}\BibitemShut
  {NoStop}%
\bibitem [{\citenamefont {Brandao}\ \emph {et~al.}(2018)\citenamefont
  {Brandao}, \citenamefont {Broughton}, \citenamefont {Farhi}, \citenamefont
  {Gutmann},\ and\ \citenamefont {Neven}}]{brandao2018for-fixed}%
  \BibitemOpen
  \bibfield  {author} {\bibinfo {author} {\bibfnamefont {F.~G. S.~L.}\
  \bibnamefont {Brandao}}, \bibinfo {author} {\bibfnamefont {M.}~\bibnamefont
  {Broughton}}, \bibinfo {author} {\bibfnamefont {E.}~\bibnamefont {Farhi}},
  \bibinfo {author} {\bibfnamefont {S.}~\bibnamefont {Gutmann}},\ and\ \bibinfo
  {author} {\bibfnamefont {H.}~\bibnamefont {Neven}},\ }\href
  {http://arxiv.org/abs/1812.04170} {\bibfield  {journal} {\bibinfo  {journal}
  {arXiv:1812.04170}\ } (\bibinfo {year} {2018})}\BibitemShut {NoStop}%
\bibitem [{\citenamefont {Streif}\ and\ \citenamefont
  {Leib}(2020)}]{streif2020training}%
  \BibitemOpen
  \bibfield  {author} {\bibinfo {author} {\bibfnamefont {M.}~\bibnamefont
  {Streif}}\ and\ \bibinfo {author} {\bibfnamefont {M.}~\bibnamefont {Leib}},\
  }\href {https://doi.org/10.1088/2058-9565/ab8c2b} {\bibfield  {journal}
  {\bibinfo  {journal} {Quantum Science and Technology}\ }\textbf {\bibinfo
  {volume} {5}},\ \bibinfo {pages} {034008} (\bibinfo {year}
  {2020})}\BibitemShut {NoStop}%
\bibitem [{\citenamefont {Farhi}\ \emph {et~al.}(2021)\citenamefont {Farhi},
  \citenamefont {Goldstone}, \citenamefont {Gutmann},\ and\ \citenamefont
  {Zhou}}]{farhi2021the-quantum}%
  \BibitemOpen
  \bibfield  {author} {\bibinfo {author} {\bibfnamefont {E.}~\bibnamefont
  {Farhi}}, \bibinfo {author} {\bibfnamefont {J.}~\bibnamefont {Goldstone}},
  \bibinfo {author} {\bibfnamefont {S.}~\bibnamefont {Gutmann}},\ and\ \bibinfo
  {author} {\bibfnamefont {L.}~\bibnamefont {Zhou}},\ }\href
  {http://arxiv.org/abs/1910.08187} {\bibfield  {journal} {\bibinfo  {journal}
  {arXiv.1910.08187}\ } (\bibinfo {year} {2021})}\BibitemShut {NoStop}%
\bibitem [{\citenamefont {Akshay}\ \emph {et~al.}(2021)\citenamefont {Akshay},
  \citenamefont {Rabinovich}, \citenamefont {Campos},\ and\ \citenamefont
  {Biamonte}}]{akshay2021parameter}%
  \BibitemOpen
  \bibfield  {author} {\bibinfo {author} {\bibfnamefont {V.}~\bibnamefont
  {Akshay}}, \bibinfo {author} {\bibfnamefont {D.}~\bibnamefont {Rabinovich}},
  \bibinfo {author} {\bibfnamefont {E.}~\bibnamefont {Campos}},\ and\ \bibinfo
  {author} {\bibfnamefont {J.}~\bibnamefont {Biamonte}},\ }\href
  {https://doi.org/10.1103/PhysRevA.104.L010401} {\bibfield  {journal}
  {\bibinfo  {journal} {Physical Review A}\ }\textbf {\bibinfo {volume}
  {104}},\ \bibinfo {pages} {L010401} (\bibinfo {year} {2021})}\BibitemShut
  {NoStop}%
\bibitem [{\citenamefont {Coppersmith}\ \emph {et~al.}(2004)\citenamefont
  {Coppersmith}, \citenamefont {Gamarnik}, \citenamefont {Hajiaghayi},\ and\
  \citenamefont {Sorkin}}]{coppersmith2004random}%
  \BibitemOpen
  \bibfield  {author} {\bibinfo {author} {\bibfnamefont {D.}~\bibnamefont
  {Coppersmith}}, \bibinfo {author} {\bibfnamefont {D.}~\bibnamefont
  {Gamarnik}}, \bibinfo {author} {\bibfnamefont {M.}~\bibnamefont
  {Hajiaghayi}},\ and\ \bibinfo {author} {\bibfnamefont {G.~B.}\ \bibnamefont
  {Sorkin}},\ }\href {https://doi.org/https://doi.org/10.1002/rsa.20015}
  {\bibfield  {journal} {\bibinfo  {journal} {Random Structures \& Algorithms}\
  }\textbf {\bibinfo {volume} {24}},\ \bibinfo {pages} {502} (\bibinfo {year}
  {2004})}\BibitemShut {NoStop}%
\bibitem [{\citenamefont {Farhi}\ and\ \citenamefont
  {Harrow}(2019)}]{farhi2019quantum}%
  \BibitemOpen
  \bibfield  {author} {\bibinfo {author} {\bibfnamefont {E.}~\bibnamefont
  {Farhi}}\ and\ \bibinfo {author} {\bibfnamefont {A.~W.}\ \bibnamefont
  {Harrow}},\ }\href {http://arxiv.org/abs/1602.07674} {\bibfield  {journal}
  {\bibinfo  {journal} {arXiv:1602.07674}\ } (\bibinfo {year}
  {2019})}\BibitemShut {NoStop}%
\bibitem [{\citenamefont {Qiang}\ \emph {et~al.}(2018)\citenamefont {Qiang},
  \citenamefont {Zhou}, \citenamefont {Wang}, \citenamefont {Wilkes},
  \citenamefont {Loke}, \citenamefont {O'Gara}, \citenamefont {Kling},
  \citenamefont {Marshall}, \citenamefont {Santagati}, \citenamefont {Ralph},
  \citenamefont {Wang}, \citenamefont {O'Brien}, \citenamefont {Thompson},\
  and\ \citenamefont {Matthews}}]{qiang2018large-scale}%
  \BibitemOpen
  \bibfield  {author} {\bibinfo {author} {\bibfnamefont {X.}~\bibnamefont
  {Qiang}}, \bibinfo {author} {\bibfnamefont {X.}~\bibnamefont {Zhou}},
  \bibinfo {author} {\bibfnamefont {J.}~\bibnamefont {Wang}}, \bibinfo {author}
  {\bibfnamefont {C.~M.}\ \bibnamefont {Wilkes}}, \bibinfo {author}
  {\bibfnamefont {T.}~\bibnamefont {Loke}}, \bibinfo {author} {\bibfnamefont
  {S.}~\bibnamefont {O'Gara}}, \bibinfo {author} {\bibfnamefont
  {L.}~\bibnamefont {Kling}}, \bibinfo {author} {\bibfnamefont {G.~D.}\
  \bibnamefont {Marshall}}, \bibinfo {author} {\bibfnamefont {R.}~\bibnamefont
  {Santagati}}, \bibinfo {author} {\bibfnamefont {T.~C.}\ \bibnamefont
  {Ralph}}, \bibinfo {author} {\bibfnamefont {J.~B.}\ \bibnamefont {Wang}},
  \bibinfo {author} {\bibfnamefont {J.~L.}\ \bibnamefont {O'Brien}}, \bibinfo
  {author} {\bibfnamefont {M.~G.}\ \bibnamefont {Thompson}},\ and\ \bibinfo
  {author} {\bibfnamefont {J.~C.~F.}\ \bibnamefont {Matthews}},\ }\href
  {https://doi.org/10.1038/s41566-018-0236-y} {\bibfield  {journal} {\bibinfo
  {journal} {Nature Photonics}\ }\textbf {\bibinfo {volume} {12}},\ \bibinfo
  {pages} {534} (\bibinfo {year} {2018})}\BibitemShut {NoStop}%
\bibitem [{\citenamefont {Willsch}\ \emph {et~al.}(2020)\citenamefont
  {Willsch}, \citenamefont {Willsch}, \citenamefont {Jin}, \citenamefont
  {De~Raedt},\ and\ \citenamefont {Michielsen}}]{willsch2020benchmarking}%
  \BibitemOpen
  \bibfield  {author} {\bibinfo {author} {\bibfnamefont {M.}~\bibnamefont
  {Willsch}}, \bibinfo {author} {\bibfnamefont {D.}~\bibnamefont {Willsch}},
  \bibinfo {author} {\bibfnamefont {F.}~\bibnamefont {Jin}}, \bibinfo {author}
  {\bibfnamefont {H.}~\bibnamefont {De~Raedt}},\ and\ \bibinfo {author}
  {\bibfnamefont {K.}~\bibnamefont {Michielsen}},\ }\href
  {https://doi.org/10.1007/s11128-020-02692-8} {\bibfield  {journal} {\bibinfo
  {journal} {Quantum Information Processing}\ }\textbf {\bibinfo {volume}
  {19}},\ \bibinfo {pages} {197} (\bibinfo {year} {2020})}\BibitemShut
  {NoStop}%
\bibitem [{\citenamefont {Bengtsson}\ \emph {et~al.}(2020)\citenamefont
  {Bengtsson}, \citenamefont {Vikst\aa{}l}, \citenamefont {Warren},
  \citenamefont {Svensson}, \citenamefont {Gu}, \citenamefont {Kockum},
  \citenamefont {Krantz}, \citenamefont {Kri\ifmmode~\check{z}\else
  \v{z}\fi{}an}, \citenamefont {Shiri}, \citenamefont {Svensson}, \citenamefont
  {Tancredi}, \citenamefont {Johansson}, \citenamefont {Delsing}, \citenamefont
  {Ferrini},\ and\ \citenamefont {Bylander}}]{bengtsson2020improved}%
  \BibitemOpen
  \bibfield  {author} {\bibinfo {author} {\bibfnamefont {A.}~\bibnamefont
  {Bengtsson}}, \bibinfo {author} {\bibfnamefont {P.}~\bibnamefont
  {Vikst\aa{}l}}, \bibinfo {author} {\bibfnamefont {C.}~\bibnamefont {Warren}},
  \bibinfo {author} {\bibfnamefont {M.}~\bibnamefont {Svensson}}, \bibinfo
  {author} {\bibfnamefont {X.}~\bibnamefont {Gu}}, \bibinfo {author}
  {\bibfnamefont {A.~F.}\ \bibnamefont {Kockum}}, \bibinfo {author}
  {\bibfnamefont {P.}~\bibnamefont {Krantz}}, \bibinfo {author} {\bibfnamefont
  {C.}~\bibnamefont {Kri\ifmmode~\check{z}\else \v{z}\fi{}an}}, \bibinfo
  {author} {\bibfnamefont {D.}~\bibnamefont {Shiri}}, \bibinfo {author}
  {\bibfnamefont {I.-M.}\ \bibnamefont {Svensson}}, \bibinfo {author}
  {\bibfnamefont {G.}~\bibnamefont {Tancredi}}, \bibinfo {author}
  {\bibfnamefont {G.}~\bibnamefont {Johansson}}, \bibinfo {author}
  {\bibfnamefont {P.}~\bibnamefont {Delsing}}, \bibinfo {author} {\bibfnamefont
  {G.}~\bibnamefont {Ferrini}},\ and\ \bibinfo {author} {\bibfnamefont
  {J.}~\bibnamefont {Bylander}},\ }\href
  {https://doi.org/10.1103/PhysRevApplied.14.034010} {\bibfield  {journal}
  {\bibinfo  {journal} {Phys. Rev. Applied}\ }\textbf {\bibinfo {volume}
  {14}},\ \bibinfo {pages} {034010} (\bibinfo {year} {2020})}\BibitemShut
  {NoStop}%
\bibitem [{\citenamefont {Pagano}\ \emph {et~al.}(2020)\citenamefont {Pagano},
  \citenamefont {Bapat}, \citenamefont {Becker}, \citenamefont {Collins},
  \citenamefont {De}, \citenamefont {Hess}, \citenamefont {Kaplan},
  \citenamefont {Kyprianidis}, \citenamefont {Tan}, \citenamefont {Baldwin},
  \citenamefont {Brady}, \citenamefont {Deshpande}, \citenamefont {Liu},
  \citenamefont {Jordan}, \citenamefont {Gorshkov},\ and\ \citenamefont
  {Monroe}}]{pagano2020quantum}%
  \BibitemOpen
  \bibfield  {author} {\bibinfo {author} {\bibfnamefont {G.}~\bibnamefont
  {Pagano}}, \bibinfo {author} {\bibfnamefont {A.}~\bibnamefont {Bapat}},
  \bibinfo {author} {\bibfnamefont {P.}~\bibnamefont {Becker}}, \bibinfo
  {author} {\bibfnamefont {K.~S.}\ \bibnamefont {Collins}}, \bibinfo {author}
  {\bibfnamefont {A.}~\bibnamefont {De}}, \bibinfo {author} {\bibfnamefont
  {P.~W.}\ \bibnamefont {Hess}}, \bibinfo {author} {\bibfnamefont {H.~B.}\
  \bibnamefont {Kaplan}}, \bibinfo {author} {\bibfnamefont {A.}~\bibnamefont
  {Kyprianidis}}, \bibinfo {author} {\bibfnamefont {W.~L.}\ \bibnamefont
  {Tan}}, \bibinfo {author} {\bibfnamefont {C.}~\bibnamefont {Baldwin}},
  \bibinfo {author} {\bibfnamefont {L.~T.}\ \bibnamefont {Brady}}, \bibinfo
  {author} {\bibfnamefont {A.}~\bibnamefont {Deshpande}}, \bibinfo {author}
  {\bibfnamefont {F.}~\bibnamefont {Liu}}, \bibinfo {author} {\bibfnamefont
  {S.}~\bibnamefont {Jordan}}, \bibinfo {author} {\bibfnamefont {A.~V.}\
  \bibnamefont {Gorshkov}},\ and\ \bibinfo {author} {\bibfnamefont
  {C.}~\bibnamefont {Monroe}},\ }\href
  {https://doi.org/10.1073/pnas.2006373117} {\bibfield  {journal} {\bibinfo
  {journal} {Proceedings of the National Academy of Sciences}\ }\textbf
  {\bibinfo {volume} {117}},\ \bibinfo {pages} {25396} (\bibinfo {year}
  {2020})}\BibitemShut {NoStop}%
\bibitem [{\citenamefont {Abrams}\ \emph {et~al.}(2020)\citenamefont {Abrams},
  \citenamefont {Didier}, \citenamefont {Johnson}, \citenamefont {Silva},\ and\
  \citenamefont {Ryan}}]{abrams2020implementation}%
  \BibitemOpen
  \bibfield  {author} {\bibinfo {author} {\bibfnamefont {D.~M.}\ \bibnamefont
  {Abrams}}, \bibinfo {author} {\bibfnamefont {N.}~\bibnamefont {Didier}},
  \bibinfo {author} {\bibfnamefont {B.~R.}\ \bibnamefont {Johnson}}, \bibinfo
  {author} {\bibfnamefont {M.~P.~d.}\ \bibnamefont {Silva}},\ and\ \bibinfo
  {author} {\bibfnamefont {C.~A.}\ \bibnamefont {Ryan}},\ }\href
  {https://doi.org/10.1038/s41928-020-00498-1} {\bibfield  {journal} {\bibinfo
  {journal} {Nature Electronics}\ }\textbf {\bibinfo {volume} {3}},\ \bibinfo
  {pages} {744} (\bibinfo {year} {2020})}\BibitemShut {NoStop}%
\bibitem [{\citenamefont {Sreedhar}\ \emph {et~al.}(2022)\citenamefont
  {Sreedhar}, \citenamefont {Wikst{\aa}hl}, \citenamefont {Svensson},
  \citenamefont {Ask},\ and\ \citenamefont
  {Garc{\'\i}a-{\'A}lvarez}}]{sreedhar2022}%
  \BibitemOpen
  \bibfield  {author} {\bibinfo {author} {\bibfnamefont {R.}~\bibnamefont
  {Sreedhar}}, \bibinfo {author} {\bibfnamefont {P.}~\bibnamefont
  {Wikst{\aa}hl}}, \bibinfo {author} {\bibfnamefont {M.}~\bibnamefont
  {Svensson}}, \bibinfo {author} {\bibfnamefont {A.}~\bibnamefont {Ask}},\ and\
  \bibinfo {author} {\bibfnamefont {L.}~\bibnamefont
  {Garc{\'\i}a-{\'A}lvarez}},\ }\href {https://doi.org/10.5281/zenodo.6807307}
  {\bibinfo {title} {sr33dhar/mps-qaoa: mps-qaoa v1.0}},\ \bibinfo
  {howpublished} {{Z}enodo} (\bibinfo {year} {2022})\BibitemShut {NoStop}%
\bibitem [{\citenamefont {Hagberg}\ \emph {et~al.}(2008)\citenamefont
  {Hagberg}, \citenamefont {Schult},\ and\ \citenamefont
  {Swart}}]{hagberg2008exploring}%
  \BibitemOpen
  \bibfield  {author} {\bibinfo {author} {\bibfnamefont {A.~A.}\ \bibnamefont
  {Hagberg}}, \bibinfo {author} {\bibfnamefont {D.~A.}\ \bibnamefont
  {Schult}},\ and\ \bibinfo {author} {\bibfnamefont {P.~J.}\ \bibnamefont
  {Swart}},\ }in\ \href {https://www.osti.gov/biblio/960616} {\emph {\bibinfo
  {booktitle} {Proceedings of the 7th Python in Science Conference}}},\
  \bibinfo {editor} {edited by\ \bibinfo {editor} {\bibfnamefont
  {G.}~\bibnamefont {Varoquaux}}, \bibinfo {editor} {\bibfnamefont
  {T.}~\bibnamefont {Vaught}},\ and\ \bibinfo {editor} {\bibfnamefont
  {J.}~\bibnamefont {Millman}}}\ (\bibinfo {address} {Pasadena, CA USA},\
  \bibinfo {year} {2008})\ pp.\ \bibinfo {pages} {11 -- 15}\BibitemShut
  {NoStop}%
\bibitem [{loc()}]{localsolver}%
  \BibitemOpen
  \href {https://www.localsolver.com/} {\bibinfo {title}
  {Localsolver}}\BibitemShut {NoStop}%
\bibitem [{\citenamefont {Vikst\aa{}l}\ \emph {et~al.}(2020)\citenamefont
  {Vikst\aa{}l}, \citenamefont {Gr\"onkvist}, \citenamefont {Svensson},
  \citenamefont {Andersson}, \citenamefont {Johansson},\ and\ \citenamefont
  {Ferrini}}]{vikstaal2020applying}%
  \BibitemOpen
  \bibfield  {author} {\bibinfo {author} {\bibfnamefont {P.}~\bibnamefont
  {Vikst\aa{}l}}, \bibinfo {author} {\bibfnamefont {M.}~\bibnamefont
  {Gr\"onkvist}}, \bibinfo {author} {\bibfnamefont {M.}~\bibnamefont
  {Svensson}}, \bibinfo {author} {\bibfnamefont {M.}~\bibnamefont {Andersson}},
  \bibinfo {author} {\bibfnamefont {G.}~\bibnamefont {Johansson}},\ and\
  \bibinfo {author} {\bibfnamefont {G.}~\bibnamefont {Ferrini}},\ }\href
  {https://doi.org/10.1103/PhysRevApplied.14.034009} {\bibfield  {journal}
  {\bibinfo  {journal} {Phys. Rev. Applied}\ }\textbf {\bibinfo {volume}
  {14}},\ \bibinfo {pages} {034009} (\bibinfo {year} {2020})}\BibitemShut
  {NoStop}%
\bibitem [{\citenamefont {authors}(2016)}]{gpyopt2016}%
  \BibitemOpen
  \bibfield  {author} {\bibinfo {author} {\bibfnamefont {T.~G.}\ \bibnamefont
  {authors}},\ }\href@noop {} {\bibinfo {title} {Gpyopt: A bayesian
  optimization framework in python}},\ \bibinfo {howpublished}
  {\url{http://github.com/SheffieldML/GPyOpt}} (\bibinfo {year}
  {2016})\BibitemShut {NoStop}%
\end{thebibliography}%

\end{document}